\documentclass[journal,10pt]{IEEEtran}
\usepackage{amsmath}
\usepackage{bm}
\usepackage{graphicx,epsfig,balance}
\usepackage{cite}
\usepackage{color}
\usepackage{multirow}
\usepackage{amsfonts,amssymb}
\usepackage{booktabs}
\usepackage{threeparttable}
\usepackage{soul}
\soulregister\cite7 
\soulregister\citep7 
\soulregister\citet7 
\soulregister\ref7 
\soulregister\pageref7 
\makeatletter
\newif\if@restonecol
\makeatother

\usepackage[linesnumbered,ruled,lined]{algorithm2e}
\usepackage{algpseudocode}
\usepackage{array}
\usepackage{makecell}

\usepackage{verbatim}
\usepackage{graphicx}
\usepackage{float}
\usepackage{subfig}
\begin{document}

\title{Attention Mechanism Based Intelligent Channel Feedback for mmWave Massive MIMO Systems}
\author{Yibin Zhang, 
	Jinlong Sun, \emph{Member}, \emph{IEEE},
	Guan Gui, \emph{Senior Member}, \emph{IEEE},
	Yun Lin, \emph{Member}, \emph{IEEE},
	Haris Gacanin, \emph{Fellow}, \emph{IEEE}, Hikmet Sari, \emph{Life Fellow}, \emph{IEEE}, and Fumiyuki Adachi, \emph{Life Fellow}, \emph{IEEE}	
	\thanks{}
	\IEEEcompsocitemizethanks{
		\IEEEcompsocthanksitem Yibin Zhang, Jinlong Sun, Guan Gui, and Hikmet Sari are with the College of Telecommunications and Information Engineering, Nanjing University of Posts and Telecommunications, Nanjing 210003, China (e-mail: 2021010208@njupt.edu.cn, sunjinlong@njupt.edu.cn, guiguan@njupt.edu.cn, hsari@ieee.org).
		\IEEEcompsocthanksitem Yun Lin is with the College of Information and Communication Engineering, Harbin Engineering University, Harbin 150009, China (e-mail: linyun@hrbeu.edu.cn)
		\IEEEcompsocthanksitem Haris Gacanin is with the Institute for Communication Technologies and Embedded Systems, RWTH Aachen University, Aachen 52062, Germany (e-mail: harisg@ice.rwth-aachen.de).
		\IEEEcompsocthanksitem Fumiyuki Adachi is with the International Research Institute of Disaster Science (IRIDeS), Tohoku University, Sendai 980-8572 Japan (e-mail: adachi@ecei.tohoku.ac.jp).
}}

\markboth{IEEE Transactions on Cognitive Communications and Networking,~Vol.~XX, No.~XX, Month~2022}
{}

\maketitle

\begin{abstract}
The potential advantages of intelligent wireless communications with millimeter wave (mmWave) and massive multiple-input multiple-output (MIMO) are based on the availability of instantaneous channel state information (CSI) at the base station (BS). However, no existence of channel reciprocity leads to the difficult acquisition of accurate CSI at the BS in frequency division duplex (FDD) systems. Many researchers explored effective architectures based on deep learning (DL) to solve this problem and proved the success of DL-based solutions. However, existing schemes focused on the acquisition of complete CSI while ignoring the beamforming and precoding operations. In this paper, we propose an intelligent channel feedback architecture using eigenmatrix and eigenvector feedback neural network (EMEVNet). With the help of the attention mechanism, the proposed EMEVNet can be considered as a dual channel auto-encoder, which is able to jointly encode the eigenmatrix and eigenvector into codewords. Simulation results show great performance improvement and robustness with extremely low overhead of the proposed EMEVNet method compared with the traditional DL-based CSI feedback methods.
\end{abstract}


\IEEEpeerreviewmaketitle

\section{Introduction}

Intelligent wireless communications with millimeter wave (mmWave) band and massive multiple-input multiple-output (MIMO) are considered as the key technology of future communication \cite{Qichenhao2021,Liye2021_2,gui_6G}.  In  sixth-generation (6G) of mobile communication, mmWave will play an indispensable and important role \cite{Kato_6G1,Kato_6G2,Letaief6G,Huang6G}. In addition, massive MIMO  combined with the ultra-large bandwidth of mmWave will become a key technical for Internet of Everything (IoE) \cite{LiYe_IoT,LiYe2019,Liye2021_3}. However, all these potential advantages, such as beamforming, power allocation and antenna selection techniques,  will be achieved only when the base station (BS) can obtain accurate channel state information (CSI). In time-division duplexing (TDD) systems, the BS can infer the downlink CSI from the uplink CSI with the help of channel reciprocity.  Unfortunately, channel reciprocity does not exist in frequency-division duplex (FDD) systems. Therefore, many scholars have been exploring how the BS obtains accurate downlink CSI in the mmWave FDD system in recent years. Compared with the traditional codebook feedback scheme, existing works try to utilize either machine learning (ML) or deep learning (DL) algorithms to obtain  CSI at the BS.

Different from the existing codebook feedback scheme, many researches \cite{wen_csinet,guo_zongshu,guo_multirate,wang2019,sun_ancinet,sun_light,zeng_transfer,Gaofeifei2021, Liye2021_1} devoted to exploring more accurate CSI feedback schemes.
C. Wen \emph{et al.} \cite{wen_csinet} innovatively proposed a DL based CSI compression and feedback method, i.e., CsiNet. The proposed CsiNet method compresses the downlink CSI estimated by the user equipment (UE), and then transmits the compressed codewords to the BS. And the BS  obtains the accurate downlink CSI after decoding the codewords.
J. Guo  \emph{et al.} \cite{guo_multirate} further explored the solution of multiple compression ratios on the basis of CsiNet, which can adapt to different channel environments.
T. Wang  \emph{et al.} \cite{wang2019} explored the time correlation between CSI matrix and proposed to use  long short-term memory (LSTM) algorithm to improve the performance of CsiNet.
J. Guo  \emph{et al.} \cite{guo_zongshu} summarized the researches using DL methods for CSI feedback in recent years, and claimed that the overhead of DL is too large to be deployed with conventional BS.
Therefore, Y. Sun \emph{et al.} \cite{sun_ancinet,sun_light} explored an efficient algorithm for lightweight design, aiming to reduce the overhead of CsiNet.
M. Chen \emph{et al.} \cite{Liye2021_1}  proposed a DL-based implicit feedback architecture to inherit the low-overhead characteristic for  wideband systems.
Different from the CsiNet, J. Zeng \emph{et al.} \cite{zeng_transfer} explored a transfer learning-based fully convolutional network designed to satisfy  different channel environments.

Although the DL-based CSI feedback schemes can help the BS to obtain more accurate downlink CSI, they still face the challenge of  transmission overhead and  utilization of spectrum resources.
Furthermore, Z. Zhong  \emph{et al.} \cite{Zhong2020} pointed out  that there is no full channel reciprocity in FDD system but partial channel reciprocity still exists.
Hence, many researches \cite{Yang_pre,Yang_transfer,Safari_UL2DL,CV3DCNN,Yangyuwen_JSAC} still insisted on exploring the solution of predicting downlink CSI from uplink CSI.
Y. Yang  \emph{et al.} \cite{Yang_pre} proposed a  sparse complex-valued neural network (SCNet) to approximate the mapping function between uplink and downlink CSI so as to reduce the transmission overhead.
Meanwhile, they  proposed intelligent algorithms based on meta-learning and transfer learning for multiple different wireless communication environments \cite{Yang_transfer}, in order to address the problem of limited datasets in new scenarios.
Considering that the CSI matrix can be seen as in-phase/quadrature (I/Q) signal, Y. Zhang \emph{et al.} \cite{CV3DCNN}  introduced a complex network to excavate the implicit information between the two channels of I/Q signal and to improve the overall performance.
Y. Yang \emph{et al.} \cite{Yangyuwen_JSAC} developed a systematic framework based on deep multimodal learning to predict CSI by multi-source sensing information.

Neither of the feedback and prediction solutions described above is  perfect. The CSI feedback scheme will cause extra spectrum overhead, and the accuracy of the CSI prediction scheme is not very good. Hence, J. Wang \emph{et al.} \cite{WangJie2} proposed a compromise solution called SampleDL. The SampleDL requires the user equipment (UE) to transmit sampled downlink CSI  to assist the BS in  improving the prediction accuracy. The sampleDL aims to combine the advantages of feedback and prediction, which may reduce the feedback overhead and improve the system performance.
Recently, some researches have  paid attention to the application of downlink CSI. They conduct more specific research for the following beamforming module at the BS, instead of improving the  acquiring accuracy of CSI \cite{Liu_EV, Gaofeifei2021_2,Guo_beam, Gaofeifei2022, Gaofeifei2022_2}.
W. Liu \emph{et al.} \cite{Liu_EV} focused on the application of  eigenvectors and proposed  EVCsiNet  to compress and feedback eigenvectors.
J. Guo \emph{et al.} \cite{Guo_beam} explored the  feedback schemes designed for beamforming (CsiFBnet) in both single-cell and multi-cell scenarios.
Z. Liu \emph{et al.} \cite{Gaofeifei2022} proposed  a novel deep unsupervised learning-based approach to optimize antenna selection and hybrid beamforming.

In this paper, we pay more attention on  the eigenvector and eigenmatrix obtained by singular value decomposition (SVD) transformation. 
This paper proposes an attention mechanism based  intelligent channel feedback method designed for beamforming  at the BS.  Considering the applications of downlink CSI at the BS, each UE is required to transmit useful and effective information to the BS rather than the downlink CSI. The main contributions of this paper are summarized as follows:
\begin{itemize}
\item We propose a CSI feedback architecture designed for beamforming, where SVD transformation is utilized as a pre-processing module for CSI matrix.
\item We propose a two-channel compressed feedback  network using residual attention mechanism, which is suitable for the joint coding of multi-channel heterogeneous data.
\item We improve the  reconstruction performance of codewords at the BS with respect to switching different auto-encoders for different channel types.
\item Comparing with classical methods, the proposed method obtains better reconstruction performance with extremely low feedback overhead,  to verify the robustness of our proposed architecture.
\end{itemize}

\section{System Model And Problem Formulation}
This section  introduces the system model researched in this paper. First, The link-level channel model is introduced, which is based on 3rd Generation Partnership Project (3GPP) technical report. Then, we  introduce SVD transform and  its application to beamforming technology and precoding matrix acquisition. Finally, the scientific issues to be addressed in this paper are described in detail.

\subsection{Link-level Channel Model}
Considering a typical mmWave FDD MIMO communication system, we assume that the BS  is equipped with $N_t$  antennas in the form of uniform linear array (ULA)\footnote{We adopt the ULA model here for simpler illustration, nevertheless, the proposed approach does not restrict to the specifical array shape.} and the UE is equipped with $N_r$ antennas $(N_t \gg N_r)$. Meanwhile, the orthogonal frequency division multiplexing (OFDM) technique is applied to the link-level channel model. Then, the received  signal at the UE can be expressed as,
\begin{equation}
	\boldsymbol{y} = \mathbf{H}\boldsymbol{x}+ \boldsymbol{n}
	\label{system1}
\end{equation}
where $\mathbf{H} \in \mathbb{C}^{N_{RB} \times N_r \times N_t}$ is the downlink CSI between the BS and the UE, $n$ denotes the noise vector. For an OFDM  system, it is necessary to consider multiple subcarriers and OFDM symbols. In this paper, resource blocks (RBs) are used as the channel matrix resolution. Thus, $N_{RB}$ represents the number of RBs used in link-level channel model. Considering a single RB and a pair of transmit and receive antennas, a common multi-path fading channel model \cite{3GPP36873} is used and can be expressed as,
\begin{equation}
	\begin{aligned}
		\mathbf{H}  = \sum_{n=1}^{N} \sum_{m=1}^{M}\sqrt{P_{n,m}} [ c_{n,m} \cdot e^{(j2\pi v_{n,m}t) }
		  \boldsymbol{\alpha}(\theta_{n,m})]
	\end{aligned}
	\label{channel1}
\end{equation}
where $N$ and $M$ denotes the number of scattering clusters and ray pathes, respectively,  $P_{m,n}$ represents the power of $m$-th ray  in the $n$-th scattering cluster,  $c_{n,m}$ is the coefficient calculated by field patterns and initial random phases, $\theta_{n,m}$ is the corresponding azimuth angle-of-departure (AoD) of  ray path, $v_{n,m}$ stands for the speed and can be understood as Doppler shift parameter. Then, the the steering vector $\boldsymbol{\alpha}(\theta_{n,m}) \in \mathbb{C}^{N_t \times 1}$ can be formulated as,
\begin{equation}
	  \boldsymbol{\alpha}(\theta) = \left[ 1, e^{-j2\pi \frac{d}{\lambda}\sin(\theta)}, \dots, e^{-j2\pi \frac{(N_t -1)d}{\lambda}\sin(\theta)}   \right]
	  \label{steering}
\end{equation}
where $d$ and $\lambda$ are the antenna element spacing and carrier wavelength, respectively.

After channel model, we further discuss the probability distribution of LOS channel.  Considering an urban macro (UMa) scenario defined by 3GPP TR38.901 \cite{3GPP38901}, we assume that the plane straight-line distance from the UE to the BS is $d_{2D}$ and the LOS probability is $\mathrm{Pr}_{LOS}$. If  $d_{2D} \leq 18 ~{\rm m}$, then $\mathrm{Pr}_{LOS} = 1$, else the $\mathrm{Pr}_{LOS}$ can be calculated via
\begin{equation}
	\begin{aligned}
		\mathrm{Pr}_{LOS}  =&  \left[ \frac{18}{d_{2D}}+\exp\left(-\frac{d_{2D}}{63}\right)\left(1-\frac{18}{d_{2D}}\right) \right] \\
		& \cdot \left[    1+ 0.8 \cdot C(h_{UT}) \left(\frac{d_{2D}}{100}\right)^3 \exp\left(-\frac{d_{2D}}{150}\right) \right]
	\end{aligned}
	\label{PrLOS}
\end{equation}
where the $C(h_{UT})$ can be found in (\ref{chut}), and the $h_{UT}$ denotes the antenna height for the UE.
\begin{equation}
	C(h_{UT})=\left\{
	\begin{aligned}
		0 & , & h_{UT} \leq 13 ~{\rm m} \\
		\left(\frac{h_{UT-13}}{10}\right)^{1.5} & , & 13 ~{\rm m} \leq h_{UT} \leq 28 ~{\rm m}
	\end{aligned}
	\right.
	\label{chut}
\end{equation}

To sum up, NLOS channel is a more common scenario with the popularization of mmWave systems.

\subsection{Applications of SVD Transformation}
This subsection shows the advantages of SVD transformation and its application in wireless communication.  In order to reduce the conflict between multi-rays and increase the channel capacity in massive MIMO system, the transmitter needs to use the beamforming technology to precode the  data flow according to the quality of channel.  A conventional precoding matrix is based on SVD transformation of CSI matrix.

Considering channel model mentioned above with CSI matrix as $\mathbf{H} \in \mathbb{C}^{N_{RB} \times N_r \times N_t}$.  For simplicity of description, we  discuss only one RB here\footnote{This assumption is only for the brief illustration of SVD, and this solution is also applicable to OFDM systems.}, i.e. $RB = 1$ and $\mathbf{H} \in  \mathbb{C}^{N_r \times N_t}$. First, the CSI matrix $\mathbf{H}$ should carry on SVD transformation as
\begin{equation}
	\mathbf{H} = \mathbf{U} \cdot \mathbf{\Sigma} \cdot \mathbf{V}^*
	\label{svd}
\end{equation}
where $\mathbf{U} \in \mathbb{C}^{N_r \times N_r}$ and $\mathbf{V} \in \mathbb{C}^{N_t \times N_t}$ are the left-singular and the right-singular matrices\footnote{Both $\mathbf{U}$ and   $\mathbf{V}$ will be called as eigenmatrix in the follows.}, respectively. What is more,  $\mathbf{U}\mathbf{U}^*=\mathbf{I}_{N_r}, \mathbf{V}\mathbf{V}^*=\mathbf{I}_{N_t}$ \footnote{$\mathbf{X}^{*}$ denotes conjugate transpose matrix of $\mathbf{X}$.}. Note that $\mathbf{\Sigma} = (\Lambda,0)$ and $\Lambda$ can be expressed as follows:
\begin{equation}
	\Lambda = \left(
	\begin{array}{ccc}
		\sqrt{\lambda_1} & \cdots & 0 \\
		\vdots & \ddots & \vdots \\
		0 & \cdots & \sqrt{\lambda_{N_r}}
	\end{array}
	\right)_{N_r \times N_r}
	\label{lambda}
\end{equation}
which represents the singular value matrix. And we define the eigenvalues of $\mathbf{H}\mathbf{H}^{*}$ as  $\mathbf{s} = \left[ \lambda_1, \lambda_2, \cdots, \lambda_{N_r}  \right]$. Next, the application of SVD transformation is introduced in detail. The unitary  matrices $\mathbf{V}$ and   $\mathbf{U}$ are used as precoding matrices for transmitter and receiver, respectively.  When BS needs to sent the parallel data flow $\boldsymbol{x} = [x_1, x_2, \cdots, x_{N_t}]^T$ to multiuser, right-singular matrix $\mathbf{V}$ will be used for precoding: $\boldsymbol{x}_t = \mathbf{V} \cdot \boldsymbol{x}$. Thirdly, we consider a typical signal transmission model as
\begin{equation}
	\boldsymbol{y} = \mathbf{H}\boldsymbol{x}_t + \boldsymbol{n}
	\label{channel}
\end{equation}
where $\boldsymbol{y}$ is the received data flow and $\boldsymbol{n}$ denotes the noise vector. The channel matrix $\mathbf{H}$  can be expressed by (\ref{svd}), and we can obtain
\begin{equation}
	\begin{aligned}
		\boldsymbol{y} & = \mathbf{U}  \mathbf{\Sigma} \mathbf{V}^*  \mathbf{V} \cdot \boldsymbol{x}+ \boldsymbol{n} \\
		& =  \mathbf{U}  \mathbf{\Sigma} \cdot \boldsymbol{x} + \boldsymbol{n}
	\end{aligned}
	\label{channel2}
\end{equation}
Finally, the receiver will use   $\mathbf{U}^*$ for receiver combining, which can be expressed as
\begin{equation}
	\begin{aligned}
		\mathbf{U}^{*} \boldsymbol{y} & =  \mathbf{U}^{*} (\mathbf{U}  \mathbf{\Sigma} \cdot \boldsymbol{x}+\boldsymbol{n})  \\
		& =  \mathbf{\Sigma} \boldsymbol{x} + \mathbf{U}^* \boldsymbol{n}
	\end{aligned}
	\label{depreco}
\end{equation}
The noise component in (\ref{depreco})  will be  filtered out by the receiver. Therefore, the receiver can recover the data flow $\boldsymbol{x}$ by $\mathbf{\Sigma}$.

In summary, as a transmitter, the BS should pay more attention to eigenmatrix $\mathbf{V}$, which can support the following precoding module. And eigenvector $\mathbf{\Sigma}$ should be applied to receiver combining the data flow when the BS is a receiver.

\subsection{Problem Formulation}
As analyzed above, the  CSI matrix $\mathbf{H}$ of a massive MIMO system is too complex  to compress  and reconstruct at the BS accurately. Meanwhile, considering the specific application of downlink CSI at the BS, we find that the BS prefers to obtain a perfect  eigenmatrix $\mathbf{V}$ (right-singular matrix) and eigenvector $\mathbf{s}$. What's more, the eigenmatrix  $\mathbf{V}$  is unitary matrix which is symmetric and easily compressible, and the eigenvector $\mathbf{s}$ is a simple real-value vector. Therefore, this paper proposes to jointly compress the unitary matrix $\mathbf{V}$ and the corresponding eigenvector $\mathbf{s}$, and feed back the codeword to the BS.

Although downlink channel estimation is challenging, this topic is beyond the scope of this paper.We assume that perfect CSI has been acquired and focus on the feedback scheme. Cosidering a classical mmWave massive MIMO FDD system described above, we focus on the eigenmatrix $\mathbf{V} \in \mathbb{C}^{N_{RB} \times N_t \times N_t}$ and the eigenvector $\mathbf{S} \in \mathbb{R}^{N_{RB} \times N_r}$. The UE needs to deploy an encoder to jointly encode $\mathbf{V}$ and $\mathbf{S}$, which can be formulated as,
\begin{equation}
	\begin{aligned}
	\varepsilon = f_{en}\left( \mathbf{V}, \mathbf{S}, \Theta_{en} \right)
	\end{aligned}
	\label{fen}
\end{equation}
where $\varepsilon$ represents the codewords encoded by the UE, $\Theta_{en}$ is the weight parameter of the encoder, and $f_{en}(\cdot)$ stands for the framework of encoder. The role of the encoder is to extract high-dimensional features from $\mathbf{V}$ and $\mathbf{S}$ respectively, and match a suitable mapping function to convert them into codewords. When received the codewords $\varepsilon$, the BS needs to  switch the corresponding decoder to interpret and obtain the required $\mathbf{V}$ and $\mathbf{S}$. The decoder can be expressed as,
\begin{equation}
	\begin{aligned}
		\widehat{\mathbf{V}}, \widehat{\mathbf{S}} = f_{de}\left(  f_{en}\left( \mathbf{V}, \mathbf{S}, \Theta_{en} \right), \Theta_{de} \right)
	\end{aligned}
	\label{fdn}
\end{equation}
where $\widehat{\mathbf{V}}, \widehat{\mathbf{S}}$ are the reconstructed eigenmatrix and eigenvector at the BS,  $f_{de}(\cdot)$ denotes the framework of decoder, and $\Theta_{de}$ is the corresponding weight value.

\begin{figure*}[htbp]
\centering
\includegraphics[width=7 in,trim=0 0 0 0,clip]{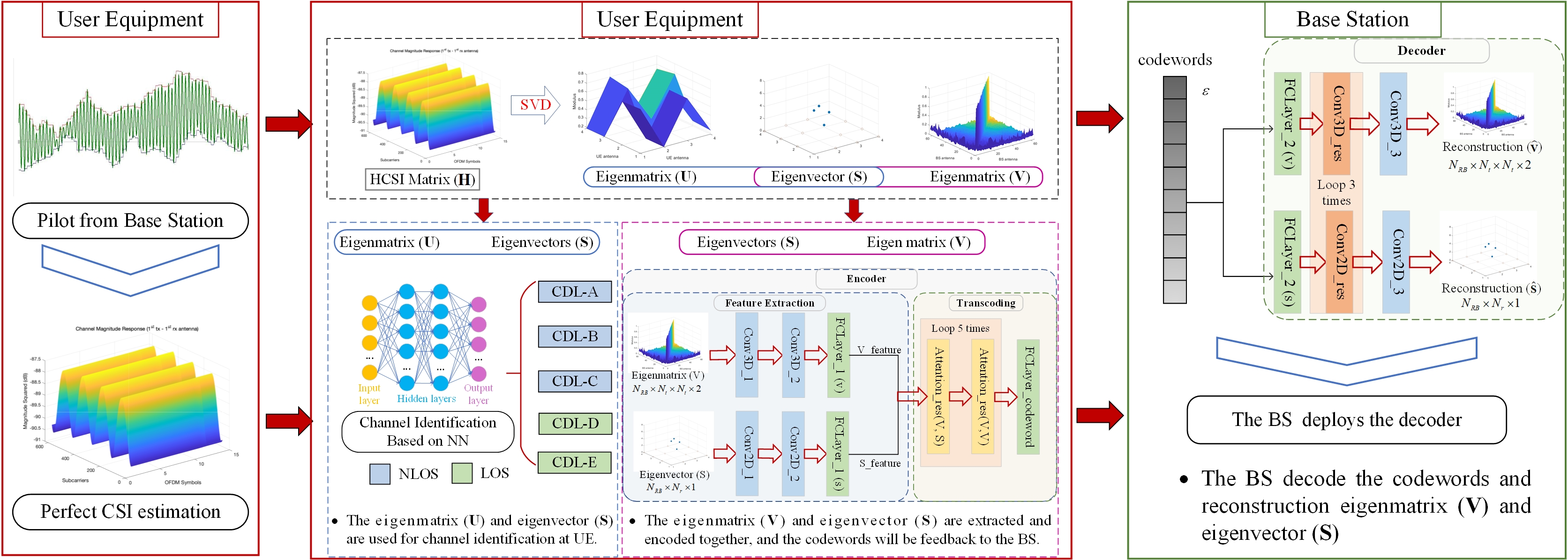}
\caption{Illustration of the overview of proposed EMEV feedback architecture. CSI matrix $\mathbf{H}$ is estimated by pilot, and then be divided into eigenmatrix $\mathbf{U},\mathbf{V}$ and eigenvector $\mathbf{S}$ by SVD transformation. $\mathbf{U}$  and $\mathbf{S}$ are used  for channel identification, and  the  encoder is used to jointly encode $\mathbf{V}$ and $\mathbf{S}$ into  codewords $\varepsilon$. The BS deploys the decoder to reconstruct $\widehat{\mathbf{V}}, \widehat{\mathbf{S}}$.}
\label{fig:zongtu}
\end{figure*}

This paper utilizes neural network (NN) based encoder and decoder to complete the compression and feedback of $\mathbf{V}$ and $\mathbf{S}$.  When training NN based auto-encoder, the loss function used by the optimizer is  mean squared error (MSE)  and can be expressed as,
\begin{equation}
	\begin{aligned}
		MSE = \mathbb{E}\left[  \Gamma\left( \Vert \mathbf{V}-\widehat{\mathbf{V}} \Vert^2_2, \Vert \mathbf{S} - \widehat{\mathbf{S}} \Vert^2_2 \right) \right]
	\end{aligned}
	\label{mse}
\end{equation}
where $\mathbb{E}(\cdot)$ stands for mathematical expectation, $\Vert \cdot \Vert^2_2$ denotes the Euclidean norm, and $\Gamma(\cdot)$  is a joint loss estimation function, weighted average function in general. The main problem explored  is to solve the optimal weights  of the NN-based encoder and decoder, which can be formulated by,
\begin{equation}
	\begin{aligned}
		\left(\Theta_{en}^*, \Theta_{de}^* \right) = \mathop{\arg\min}\limits_{\Theta_{en},\Theta_{de}} \mathbb{E}\left[  \Gamma\left( \Vert \mathbf{V}-\widehat{\mathbf{V}} \Vert^2_2, \Vert \mathbf{S} - \widehat{\mathbf{S}} \Vert^2_2 \right) \right]
	\end{aligned}
	\label{arg}
\end{equation}
where $\Theta_{en}^*$ and $\Theta_{de}$ are the optimal weights of encoder and decoder, respectively.

\section{DL-based EMEV Feedback Architecture}
This section describes the proposed DL-based eigenmatrix and eigenvector (EMEV) feedback architecture in detail. Based on the SVD transformation and its application for beamforming, we pay more attention to the eigenmatrix $\mathbf{V}$ and eigenvector $\mathbf{S}$ in this paper.  First, the overview of proposed EMEV feedback architecture is shown. Then, the NN desigend for the EMEV auto-encoder is displayed and analyzed by different modules.

\subsection{Overview of The Proposed Architecture}
This part is the overview of DL-based EMEV feedback architecture. We aim to explore efficient feedback schemes for beamforming.  As is shown in Fig. \ref{fig:zongtu}, the whole process starts when the UE estimates the real-time downlink CSI through pilot.  And then the CSI matrix $\mathbf{H}$ is divided into eigenmatrix $\mathbf{U}, \mathbf{V}$ and eigenvector $\mathbf{S}$ by SVD transformation. From the figure we can find that $\mathbf{H}$ is complex and irregular, but $\mathbf{U}$ and $\mathbf{V}$ are unitary matrices and $\mathbf{S}$  exhibits a scatter distribution. Furthermore, the power distributions of  $\mathbf{U}$ and $\mathbf{V}$ are symmetric.  Thirdly, the UE inputs $\mathbf{U}$ and $\mathbf{S}$ to the NN-based channel identification to obtain the exact channel type. The  detailed NN-based channel identification is shown in our conference paper \cite{ZYB_arXiv},  called  EMEV-IdNet. Considering the clustered delay line (CDL) channel model compliant with 5G new radio (NR) standards\cite{3GPP38901}, we explores five common channel types, composed of three none line of sight (NLOS) channels and two line of sight (LOS) channels. Since the eigenmatrix distributions of five channel types are quite different \cite{Mantoro2017}, we cascade the channel identification before EMEV encoder to improve the system performance. After channel identification, the UE will joint-encode $\mathbf{V}$ and $\mathbf{S}$ into codewords and feedback to the BS. As is analyzed above, $\mathbf{V}$ and $\mathbf{S}$ will be able to meet the requirements of beamforming and communication for BS. Finally, the BS receives and decodes the codewords $\varepsilon$ and reconstructs eigenmatrix $\widehat{\mathbf{V}}$ and eigenvector $\widehat{\mathbf{S}}$. The algorithm flow of proposed EMEV feedback architecture is described in \textbf{Algorithm \ref{svdalg}}.
\begin{figure*}[htbp]
	\centering
	\includegraphics[width=7 in,trim=0 0 0 0,clip]{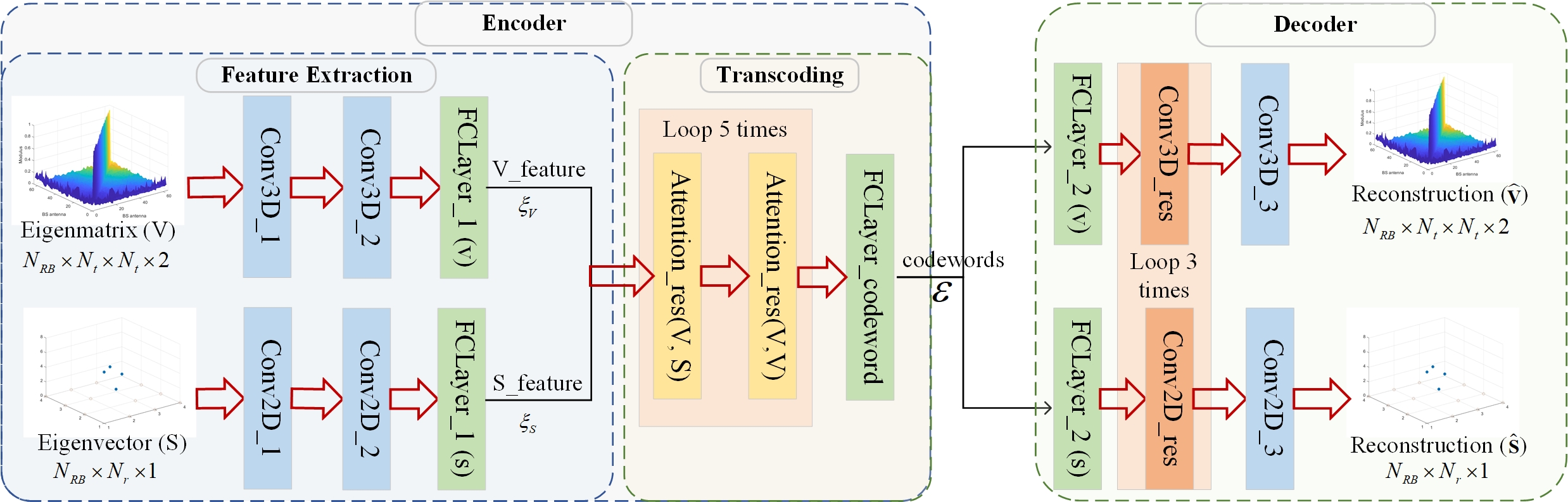}
	\caption{Illustration of the overall framework of DL-based EMEVNet.  Feature extraction module: Input are $\mathbf{V}$ and $\mathbf{S}$; Output $\xi_{V}$ and $\xi_{S}$ are high-dimensional feature of $\mathbf{V}$ and $\mathbf{S}$ respectively. Transcoding module: Input are $\xi_{V}$ and $\xi_{S}$; Output is codewords $\varepsilon$. Decoder module: Input is received codewords $\varepsilon$ at the BS;  Output are reconstructed $\widehat{\mathbf{V}}$ and $\widehat{\mathbf{S}}$. }
	\label{fig:network}
\end{figure*}
\begin{algorithm}
	\caption{The algorithm of proposed channel feedback architecture based on EMEV feature.}
	\label{svdalg}
	\KwIn{$\mathbf{H} \in \mathbb{C}^{N_{RB} \times N_r \times N_t} \gets$  CSI matrix; }
	\KwOut{$\widehat{\mathbf{S}} \in \mathbb{R}^{N_{RB}\times N_r} \gets$ Reconstructed eigenvector;
	$\widehat{\mathbf{V}} \in \mathbb{C}^{N_{RB \times N_t \times N_t}} \gets$ Reconstructed eigenmatirx;
	$id \gets$ Channel type;}
	\textbf{Stage \uppercase\expandafter{\romannumeral1}: UE operations:}\\
	\textbf{SVD transformation: }
	{Initialize $ \mathbf{U} \in \mathbb{C}^{N_{RB} \times N_r \times N_r}, \mathbf{S} \in \mathbb{R}^{N_{RB} \times N_r},\mathbf{V} \in \mathbb{C}^{N_{RB \times N_t \times N_t}}$};\\
	\For{$i = 1, \cdots N_{RB}$ }{
		$\mathbf{U_t}, \boldsymbol{S_t}, \mathbf{V_t} = f_{svd}(\mathbf{H}(i, :, :)) $ \\
		\If{$\mathbf{U_t} \cdot \boldsymbol{S_t} \cdot \mathbf{V_t} == \mathbf{H}(i, :, :)$}{
			$\mathbf{U}(i, ;, :) = \mathbf{U_t}$;
			$\mathbf{S}(i, :) = \boldsymbol{S_t}$;
			$\mathbf{V}(i, ;, :) = \mathbf{V_t}$.
		}
	}
	{Save $\mathbf{U}, \mathbf{S},\mathbf{V}$.}\\
	{\textbf{Channel identification: }Load trained EMEV-IdNet and identify the channel type $id \gets f_{id}(\mathbf{U},\mathbf{S})$;} \\
	{\textbf{Encoder: }Switch appropriate encoder by $id$ and generate feedback codewords: $\varepsilon=f_{en}(\mathbf{V},\mathbf{V})$}\\
	\textbf{Stage \uppercase\expandafter{\romannumeral2}: BS operations:}\\
	{\textbf{Decoder: }Switch appropriate decoder by $id$ and reconstruct precoding matrix: $\left(\widehat{\mathbf{V}},\widehat{\mathbf{S}} \right)=f_{de}(\varepsilon)$}
\end{algorithm}

\subsection{The Proposed DL-based EMEV Feedback Network}
This subsection  shows the overall framework of proposed DL-based EMEV feedback neural network, called EMEVNet. Fig. \ref{fig:network} is the illustration of its framework.

As is described in Fig. \ref{fig:network}, the EMEVNet is an auto-encoder which is combined with an encoder at the UE and a decoder at the BS. And the encoder can be further divided into feature extraction  and transcoding modules. We design a  feature extraction module with dual-channel input layer.  Different convolution layers are used for different inputs, i.e. three-dimensional convolution layer (Conv3D) is for $\mathbf{V} \in \mathbb{C}^{N_{RB} \times N_t \times N_t}$ and two-dimensional convolution layer (Conv2D) for $\mathbf{S} \in \mathbb{R}^{N_{RB} \times N_r}$.  The  high-dimensional feature maps after convolution layers will be compressed to one dimensional feature $\xi_{V}$ and $\xi_{S}$ by fully-connected layers.  Thus, the feature extraction module can be described as,
\begin{equation}
	\begin{aligned}
		\left( \xi_{V}, \xi_{S}\right) = \mathcal{L} _{fc} \left[ \mathcal{L}_{conv}\left(\mathbf{V}, \mathbf{S}, \Omega_{conv} \right), \Omega_{fc} \right]
	\end{aligned}
	\label{feature ex}
\end{equation}
where $\mathcal{L}_{fc}(\cdot), \mathcal{L}_{conv}(\cdot)$ represent fully-connected layer and convolution layer respectively, and $\Omega_{fc}, \Omega_{conv}$ are their corresponding weight values. Then, $\xi_{V}$ and $\xi_{S}$ pass through the transcoding module and output codewords $\varepsilon$ at specified system compression ratio. The attention mechanism\footnote{The attention mechanism will be described separately when analyzing the transcoding module.} based attention residual block and fully-connected layer  are combined into a transcoding module, which is formulated as,
\begin{equation}
	\begin{aligned}
	\varepsilon = \mathcal{L} _{fc} \left[\mathcal{L}_{att}^{(5)}\left(\xi_{V}, \xi_{S}, \Omega_{att} \right), \beta_{CR}, \Omega_{fc}  \right]
	\end{aligned}
	\label{trans co}
\end{equation}
where $\mathcal{L}_{att}^{(5)}$ indicates 5 loops of attention residual block, $\beta_{CR}$ is the input system compression ratio,  and $\Omega_{att}$ is the corresponding weight value. Therefore, the length of $\varepsilon$ is determined by system $\beta_{CR}$, which can be defined as,
\begin{equation}
	\begin{aligned}
		L_{\varepsilon} = \frac{L[\Re({\mathbf{V}})]+L[\Im({\mathbf{V}})]+L[\mathbf{S}] }{\beta_{CR}} +id
	\end{aligned}
	\label{trans cr}
\end{equation}
where $L[\cdot]$ denotes the length of the variable, $\Re(\cdot)$ and $\Im(\cdot)$ are the  real and imaginary parts of complex numbers, $L_{\varepsilon}$ is the length of codewords $\varepsilon$, and $id$ represents the control symbol of channel identification result.
Finally, the BS can reconstruct eigenmatrix $\widehat{\mathbf{V}}$ and eigenvector $\widehat{\mathbf{S}}$ from received $\varepsilon$ by utilizing decoder.  The decoder is composed of fully-connected layers, convolutional residual blocks and convolution layers, which can be written as,
\begin{equation}
	\begin{aligned}
		\left(\widehat{\mathbf{V}}, \widehat{\mathbf{S}}\right) = \mathcal{L} _{conv} \left\lbrace  \mathcal{L}_{res}^{(3)}\left[ \mathcal{L}_{fc}\left(\varepsilon, \Omega_{fc} \right) \Omega_{res} \right], \Omega_{conv} \right\rbrace
	\end{aligned}
	\label{deco}
\end{equation}
where $\mathcal{L}_{res}^{(3)}$ stands for 3 loops of  convolutional residual block and  $\Omega_{res}$ is its weight values.  In summary, the  concrete steps of  EMEVNet algorithm are given in the  \textbf{Algorithm \ref{emevnetalg}}.
\begin{algorithm}
	\caption{The proposed EMEVNet algorithm for eigenmatrix and eigenvector feedback architecture.}
	\label{emevnetalg}
	\KwIn{$\mathbf{V} \in \mathbb{C}^{N_{RB} \times N_t \times N_t} \gets$ Eigenmatrix;
		$\mathbf{S} \in \mathbb{R}^{N_{RB} \times N_r} \gets$ Eigenvector;
		$\eta \gets$ Initial learning rate;  $\tau \gets$ Maximum epoch number; $\beta_{CR} \gets$ System compression ratio;}
	\KwOut{$\widehat{\mathbf{V}} \to$ Reconstructed eigenmatrix;
		$\widehat{\mathbf{S}} \to$ Reconstructed eigenvector;
		$\left( \Theta_{en}, \Theta_{de}\right) \to$ Trained autoencoder parameters;}
	\textbf{Training stage:} \\
	{Load $\mathbf{V} \in \mathbb{C}^{N_{RB} \times N_t \times N_t}, \mathbf{S} \in \mathbb{R}^{N_{RB} \times N_r}$};\\
	{Randomly initialize  NN weight parameters $\Theta_{en}, \Theta_{de}$};\\
	\For{$t = 1, \cdots, \tau$}{
		$\left( \xi_{V}, \xi_{S}\right) = \mathcal{L} _{fc} \left[ \mathcal{L}_{conv}\left(\mathbf{V}, \mathbf{S}, \Omega_{conv} \right), \Omega_{fc} \right] $\\
		$\varepsilon = \mathcal{L} _{fc} \left[\mathcal{L}_{att}^{(5)}\left(\xi_{V}, \xi_{S}, \Omega_{att} \right), \beta_{CR}, \Omega_{fc} \right]$\\
		$\widehat{\mathbf{V}}, \widehat{\mathbf{S}} = \mathcal{L} _{conv} \left\lbrace  \mathcal{L}_{res}^{(3)}\left[ \mathcal{L}_{fc}\left(\varepsilon, \Omega_{fc} \right) \Omega_{res} \right], \Omega_{conv} \right\rbrace $ \\
		$loss_{t} =\mathbb{E}\left[  \Gamma\left( \Vert \mathbf{V}-\widehat{\mathbf{V}} \Vert^2_2, \Vert \mathbf{S} - \widehat{\mathbf{S}} \Vert^2_2 \right) \right]$ \\
		\If{$loss_{t}$ converges to $loss^*$}{break;}
		\If{$loss_{t} $ is not updated after 20 loops}{$\eta = \eta \times 0.7$;}
		$\Omega \gets {\rm Adam}(\Omega, \eta,  \nabla loss_{t})$ \\
	}
	{$\left[\Theta_{en}^*, \Theta_{de}^* \right] \gets \left[\Omega_{fc}, \Omega_{att}, \Omega_{conv}, \Omega_{res}\right]$};\\
	{Save $f_{en}(\beta_{CR}, \Theta_{en}^*), f_{de}(\beta_{CR}, \Theta_{de}^*)$.}
\end{algorithm}

\begin{table*}[htbp]
	\centering
	\caption{The hyper-parameters setting  and analysis of parameters and FLOPs for feature extraction module. }
	\begin{tabular}{cccccc}
		\toprule
		{\bf Layer name} & {\bf Hyper-parameters} & {\bf Activation} & {\bf Output shape} & {\bf Parameter size} & {\bf FLOPs} \\ \hline
		Input$(\mathbf{V})$  &    --     &     --    &    $N_{RB} \times N_t \times N_t \times 2$    &    --    &    --   \\
		Input$(\mathbf{S})$    &       --     &   --      &     $N_{RB} \times N_r \times 1$       &      --         &   --    \\
		Conv3D\_1 &  \multirow{2}{*}{\begin{tabular}[c]{@{}c@{}}Filter = 2,\\ Kernel = 3.\end{tabular}}  &  Leaky Relu  &  $N_{RB} \times N_t \times N_t \times 2$ &  $2 \times 2 \times 3^2$  & $(N_{RB} \times N_t \times N_t \times 2) \times (2 \times 3^2)$    \\
		Conv2D\_1 &      &   Leaky Relu  &    $N_{RB} \times N_t  \times 2$  & $2 \times 2 \times 3^2$   &  $(N_{RB} \times N_r \times 2) \times (2 \times 3^2)$      \\
		Conv3D\_2 &  \multirow{2}{*}{\begin{tabular}[c]{@{}c@{}}Filter = 8,\\ Kernel = 3.\end{tabular}}   &  Leaky Relu  &   $N_{RB} \times N_t \times N_t \times 8$  &    $8 \times 2 \times 3^2$  &     $(N_{RB} \times N_t \times N_t \times 8) \times (2 \times 3^2)$     \\
		Conv2D\_2 &       &    Leaky Relu &   $N_{RB} \times N_r  \times 8$  &  $8 \times 2 \times 3^2$&   $(N_{RB} \times N_r \times 8) \times (2 \times 3^2)$     \\
		FCLayer\_1$(\mathbf{V})$&   Units = $L_{\xi_{V}}$    &    Relu    &   $L_{\xi_{V}} \times 1$   & \scriptsize   $[N_{RB} \times N_t \times N_t \times 8] \times L_{\xi_{V}} $  &  \scriptsize  $2 \times[N_{RB} \times N_t \times N_t \times 8] \times L_{\xi_{V}} $       \\
		FCLayer\_1$(\mathbf{S})$&   Units = $L_{\xi_{S}}$ &   Relu  &   $L_{\xi_{S}} \times 1$  &    $[N_{RB} \times N_r \times 8] \times L_{\xi_{S}}$   & $2 \times [N_{RB} \times N_r \times 8] \times L_{\xi_{S}}$    \\
		\bottomrule   
	\end{tabular}
	\label{table1}
\end{table*}

\subsection{Analysis and Discussion of EMEVNet}
In this subsection, feature extraction, transcoding and decoding modules are discussed one by one. We  analyze the proposed NN from two aspects: time complexity and space complexity \cite{he2015convolutional}.  Time complexity refers to the number of operations of the model, which can be measured by FLOPs, i.e., the number of floating-point operations. The space complexity includes two parts: parameter quantity and output characteristic map. Meanwhile, the hyper-parameters and activation functions set for each layer are given in detail.

\subsubsection{Feature extraction module}
As is shown in Fig. \ref{fig:network}, the feature extraction module includes two 2D convolution layers, two 3D convolution layers and two fully-connected layers.  The time complexity of 2D and 3D convolution layers can be respectively expressed as,
\begin{equation}
	\begin{aligned}
		Time \left\lbrace Conv2D\right\rbrace  \sim O\left(S_{M} \cdot K^2 \cdot C_{in} \cdot C_{out} \right)
	\end{aligned}
	\label{conv2d}
\end{equation}
\begin{equation}
	\begin{aligned}
		Time \left\lbrace Conv3D\right\rbrace  \sim O\left(V_{M} \cdot K^3 \cdot C_{in} \cdot C_{out} \right)
	\end{aligned}
	\label{conv3d}
\end{equation}
where $S_M, V_M$ are the area and volume of the output feature map of the convolution layer, respectively, $K$ is the size of the convolution kernel, $C_{in}$  denotes the number of input channels from the upper layer,  and $C_{out}$  represents the number of output channels. Considering the fully-connected layer, its time complexity can be formulated as,
\begin{equation}
	\begin{aligned}
		Time \left\lbrace fc \right\rbrace  \sim O\left(2 \times L_{in} \times L_{out} \right)
	\end{aligned}
	\label{fc}
\end{equation}
where $L_{in}$ and $L_{out}$ are the input and output tensor shape of fully-connected layer.  After introducing the above time complexity, we continue to describe the space complexity. Eq. (\ref{conv2d_s}) to Eq. (\ref{fc_s}) are   the space complexity  expressions corresponding to 2D convolution layer, 3D convolution layer and fully-connected layer.
\begin{equation}
	\begin{aligned}
		Space \left\lbrace Conv2D \right\rbrace  \sim O\left(K^2\times C_{in} \times C_{out} \right)
	\end{aligned}
	\label{conv2d_s}
\end{equation}
\begin{equation}
	\begin{aligned}
		Space \left\lbrace Conv3D \right\rbrace  \sim O\left(K^3\times C_{in} \times C_{out} \right)
	\end{aligned}
	\label{convdd_s}
\end{equation}
\begin{equation}
	\begin{aligned}
		Space \left\lbrace fc \right\rbrace  \sim O\left(L_{in} \times L_{out} \right)
	\end{aligned}
	\label{fc_s}
\end{equation}
Through the above analysis we can find that the convolution layer requires more FLOPs,  but its space complexity is relatively low. On the contrary, the space complexity of fully-connected layer is larger, which means high demand for memory overhead. The hyper-parameters set in feature extraction module is listed in Tab. \ref{table1}, and the analysis of complexity is also included.

\begin{table*}[htbp]
	\centering
	\caption{The hyper-parameters setting  and analysis of parameters and FLOPs for transcoding module. }
	\begin{tabular}{cccccc}
		\toprule
		{\bf Layer name} & {\bf Hyper-parameters} & {\bf Activation} & {\bf Output shape} & {\bf Parameter size} & {\bf FLOPs} \\ \hline
		Input$(\xi_{V})$  &    --     &     --    &    $L_{\xi_{V}} \times 1$    &    --    &    --   \\
		Input$(\xi_{S})$    &       --     &   --      &     $L_{\xi_{S}} \times 1$       &      --         &   --    \\
		Attention\_res$(\mathbf{V},\mathbf{S})$ &  \multirow{2}{*}{\begin{tabular}[c]{@{}c@{}}Head\_num = 2,\\ Key\_dim = 3.\end{tabular}}  &  --  &  $L_{\xi_{V}} \times 1$ &  $2 \times (L^2_{\xi_{V}}+L^2_{\xi_{S}})$  & $8 \times (L^2_{\xi_{V}}+L^2_{\xi_{S}})$    \\
		Attention\_res$(\mathbf{V},\mathbf{V})$ &      &  -- &     $L_{\xi_{V}} \times 1$ & $2 \times (L^2_{\xi_{V}}+L^2_{\xi_{V}})$   &  $8 \times (L^2_{\xi_{V}}+L^2_{\xi_{V}})$      \\
		FCLayer\_codewords &   Units = $L_{\varepsilon}$ &   Linear  &   $L_{\varepsilon} \times 1$  &    $(L_{\varepsilon} \times 1) \times (L_{\varepsilon} \times 1)$  & $2 \times (L_{\varepsilon} \times 1) \times (L_{\varepsilon} \times 1)$    \\
		\bottomrule   
	\end{tabular}
	\label{table2}
\end{table*}
\begin{table*}[htbp]
	\centering
	\caption{The hyper-parameters setting  and analysis of parameters and FLOPs for decoder module. }
	\begin{tabular}{cccccc}
		\toprule
		{\bf Layer name} & {\bf Hyper-parameters} & {\bf Activation} & {\bf Output shape} & {\bf Parameter size} & {\bf FLOPs} \\ \hline
		Input$(\varepsilon)$  &    --     &     --    &    $L_{\varepsilon} \times 1$    &    --    &    --   \\
		FCLayer\_2$(\mathbf{V})$&    \scriptsize Units = [$N_{RB}\cdot N_t \cdot N_t \cdot 2]$    &    Linear    &     $[N_{RB}\cdot N_t \cdot N_t \cdot 2]$   & \scriptsize    $L_{\varepsilon}\times [N_{RB}\cdot N_t \cdot N_t \cdot 2]$   &  \scriptsize  $2 \times L_{\varepsilon}\times [N_{RB}\cdot N_t \cdot N_t \cdot 2] $       \\
		FCLayer\_2$(\mathbf{S})$&  \scriptsize Units = [$N_{RB}\cdot N_r \cdot 1]$    &    Linear    &     $[N_{RB}\cdot N_t ]$   &     $L_{\varepsilon}\times [N_{RB}\cdot N_r ]$   &    $2 \times L_{\varepsilon}\times [N_{RB}\cdot N_t ] $       \\
		Conv3D\_res &  \multirow{2}{*}{\begin{tabular}[c]{@{}c@{}}Filter = [2,8,2],\\ Kernel = 3.\end{tabular}}  &  -- &  $N_{RB} \times N_t \times N_t \times 2$ &  \scriptsize $(2+8+2) \times 2 \times 3^2$  &\scriptsize $(N_{RB} \times N_t \times N_t \times 12) \times (12 \times 3^2)$    \\
		Conv2D\_res &      &   --  &    $N_{RB} \times N_t  \times 2$  & \scriptsize $(1+8+2) \times 2 \times 3^2$   & \scriptsize $(N_{RB} \times N_r \times 11) \times (11 \times 3^2)$      \\
		Conv3D\_3 &  Filter=2, Kernel=3  &  Tanh  &   $N_{RB} \times N_t \times N_t \times 2$  &    $2 \times 2 \times 3^2$  &     $(N_{RB} \times N_t \times N_t \times 2) \times (2 \times 3^2)$     \\
		Conv2D\_3 &    Filter=2, Kernel=3   &    Linear &   $N_{RB} \times N_r $  &  $1 \times 2 \times 3^2$&   $(N_{RB} \times N_r \times 1) \times (2 \times 3^2)$     \\
		\bottomrule   
	\end{tabular}
	\label{table3}
\end{table*}

\subsubsection{Transcoding module}
\begin{figure}[htbp]
	\centering
	\includegraphics[width=3.5 in,trim=0 0 0 0,clip]{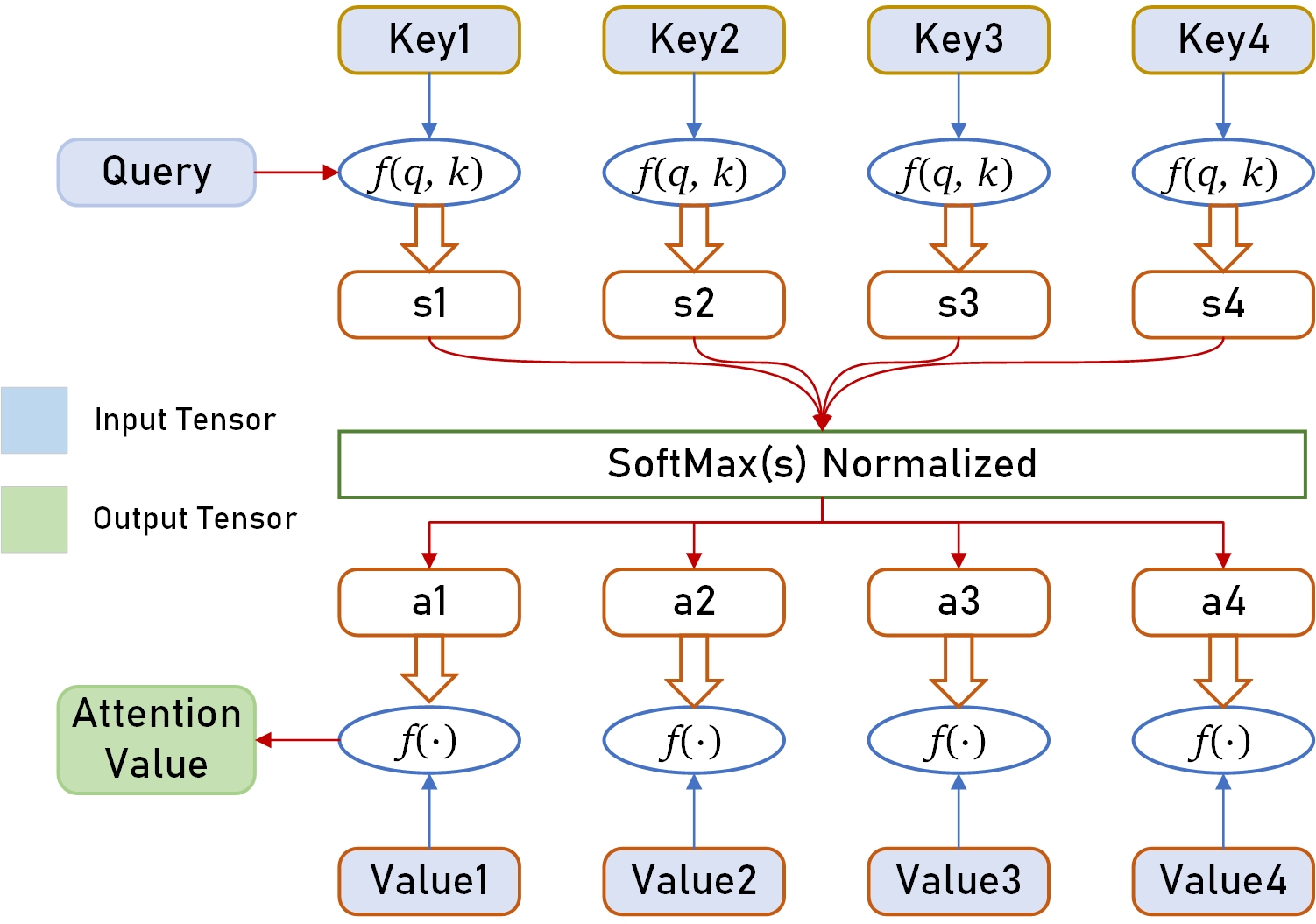}
	\caption{Illustration of attention mechanism. The query ($\boldsymbol{q}$), key ($\boldsymbol{k}$), and value ($\boldsymbol{v}$) are input tensors, and attention value ($\boldsymbol{z}$) is output tensors.}
	\label{fig:attention}
\end{figure}

Before the transcoding module, we want to briefly introduce the attention mechanism, which plays an important role in trascoding task. Attention mechanism \cite{attention2017} was proposed by the Bengio team  and has been widely used in various fields of deep learning in recent years, such as in computer vision for capturing the receptive field on images, or for positioning in natural language processing (NLP) key token or feature.  Fig. \ref{fig:attention} shows the detailed tensor flow of the attention mechanism.

First, attention distribution $\boldsymbol{s}$ between the input the query vector $\boldsymbol{q}$  and the keyword vector $\boldsymbol{k}$ need to  be known, which can be calculated via,
\begin{equation}
	\boldsymbol{s}_{i} = f(\boldsymbol{q}, \boldsymbol{k}_i) =
	\begin{cases}
		\boldsymbol{q}^T  \boldsymbol{k}_{i} \\
		\boldsymbol{q}^{T}  \boldsymbol{W}  \boldsymbol{k}_{i} \\
		\left[ \boldsymbol{q}^T \boldsymbol{k}_{i} \right]/ \sqrt{d} \\
		\boldsymbol{v} \cdot \tanh(\boldsymbol{W}\boldsymbol{q}+\boldsymbol{U}\boldsymbol{k}_{i})
	\end{cases}
	\label{att1}
\end{equation}
where $\boldsymbol{W}$, $\boldsymbol{U}$ and  $\boldsymbol{v}$ are trainable weight coefficients in neural network, and $d$ denotes the input dimension. Then,  the attention distribution $\boldsymbol{s}$ is normalized to attention score $\boldsymbol{a}$, which can be written as,
\begin{equation}
	\boldsymbol{a}_{i} = {\rm softmax} \left[ f(\boldsymbol{q}, \boldsymbol{k}_i) \right]  = \frac{e^{f(\boldsymbol{q}, \boldsymbol{k}_i)}}{\sum_{j} e^{f(\boldsymbol{q}, \boldsymbol{k}_j)}}
	\label{att2}
\end{equation}
Finally, the output $\boldsymbol{z}$ of  attention mechanism is the result of  weighted average of  vector $\boldsymbol{v}$, which is shown as,
\begin{equation}
	\boldsymbol{z} = {\rm Attention}(\boldsymbol{q},\boldsymbol{k},\boldsymbol{v}) = \sum_{i}\boldsymbol{a}_i \cdot \boldsymbol{v}_i
	\label{att3}
\end{equation}
In short, the attention mechanism is actually designed to give larger weight to the parts that need attention, highlighting important information and ignoring other content.  L. Chen  \emph{et al.} \cite{Chen_CVPR} explored the combination of CNN and attention mechanism, which gave excellent performance. Y. Cui  \emph{et al.} \cite{Cui2022} applied the attention mechanism to the CSI feedback solution and demonstrated its performance improvement.

\begin{figure}[htbp]
	\centering
	\includegraphics[width=3.5 in,trim=0 0 0 0,clip]{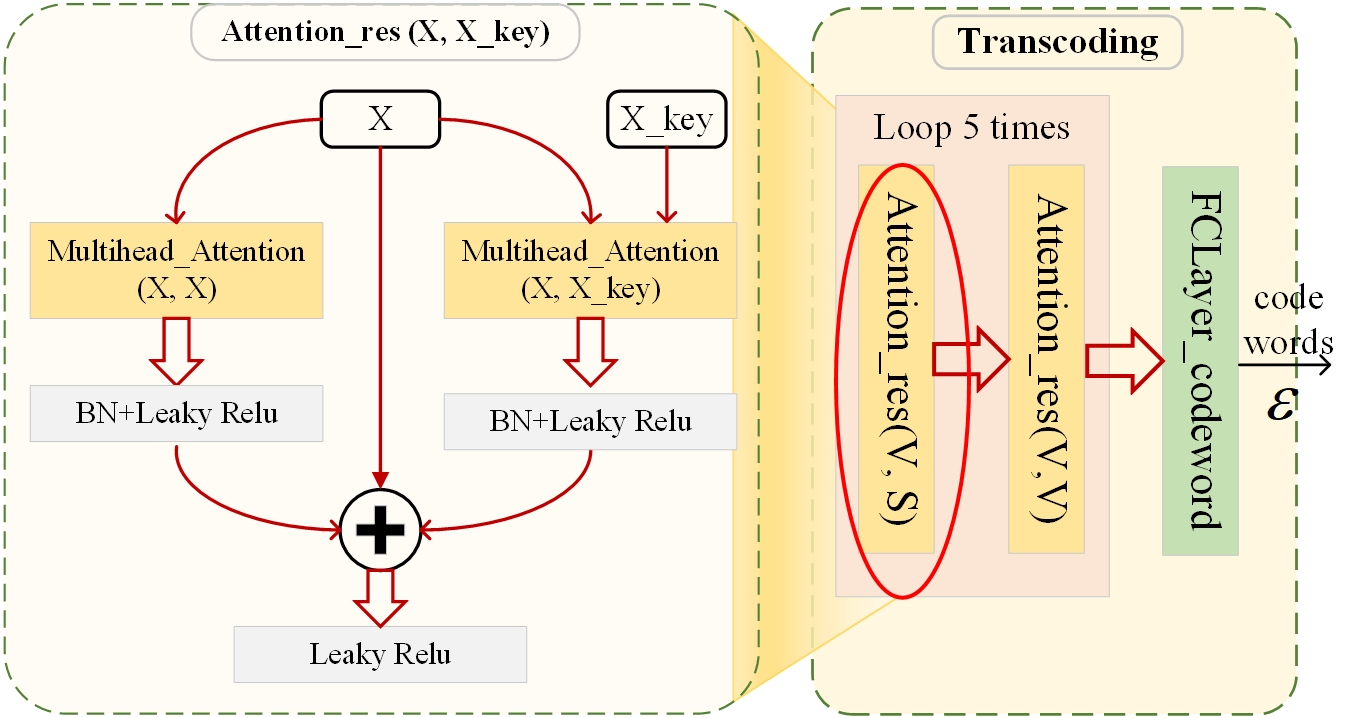}
	\caption{Illustration of the transcoding module and attention residual block. }
	\label{fig:transcoding}
\end{figure}

As is shown in Fig. \ref{fig:transcoding}, transcoding module is combined with two attention residual blocks and a fully-connected layer. The attention residual block receives two parallel input tensors $\mathbf{X}$ and $ \mathbf{X}_{key}$. If $\mathbf{X} = \mathbf{X}_{key}$, then we call it self-attention residual block, which pays more attention  to self hidden information. If $\mathbf{X} \neq \mathbf{X}_{key}$,  the attention residual block will embed the  $\mathbf{X}_{key}$'s feature information  in $\mathbf{X}$, which is called cross-attention residual block. First, we design a cross-attention residual block to  embed the information of $\xi_{S}$ into $\xi_{V}$. And then a self-attention residual block is followed to explore the self hidden features of $\xi_{V}$.
The time complexity and space complexity of multi-head attention layer can be respectively formulated as,
\begin{equation}
	\begin{aligned}
		Time \left\lbrace att \right\rbrace  \sim O\left( L_{in}^2 \times d_{in} \right)
	\end{aligned}
	\label{att_t}
\end{equation}
\begin{equation}
	\begin{aligned}
		Space \left\lbrace att \right\rbrace  \sim O\left( 4\times L_{in}^2 \times d_{in} \right)
	\end{aligned}
	\label{att_s}
\end{equation}
where $L_{in}$ is the length of input tensor, and $d_{in}$ denotes the value of hyper-parameter $head\_num$. The hyper-parameters set in transcoding module is listed in Tab. \ref{table2}, and the analysis of complexity is also included.

\subsubsection{Decoder module}
\begin{figure}[htbp]
	\centering
	\includegraphics[width=3.5 in,trim=0 0 0 0,clip]{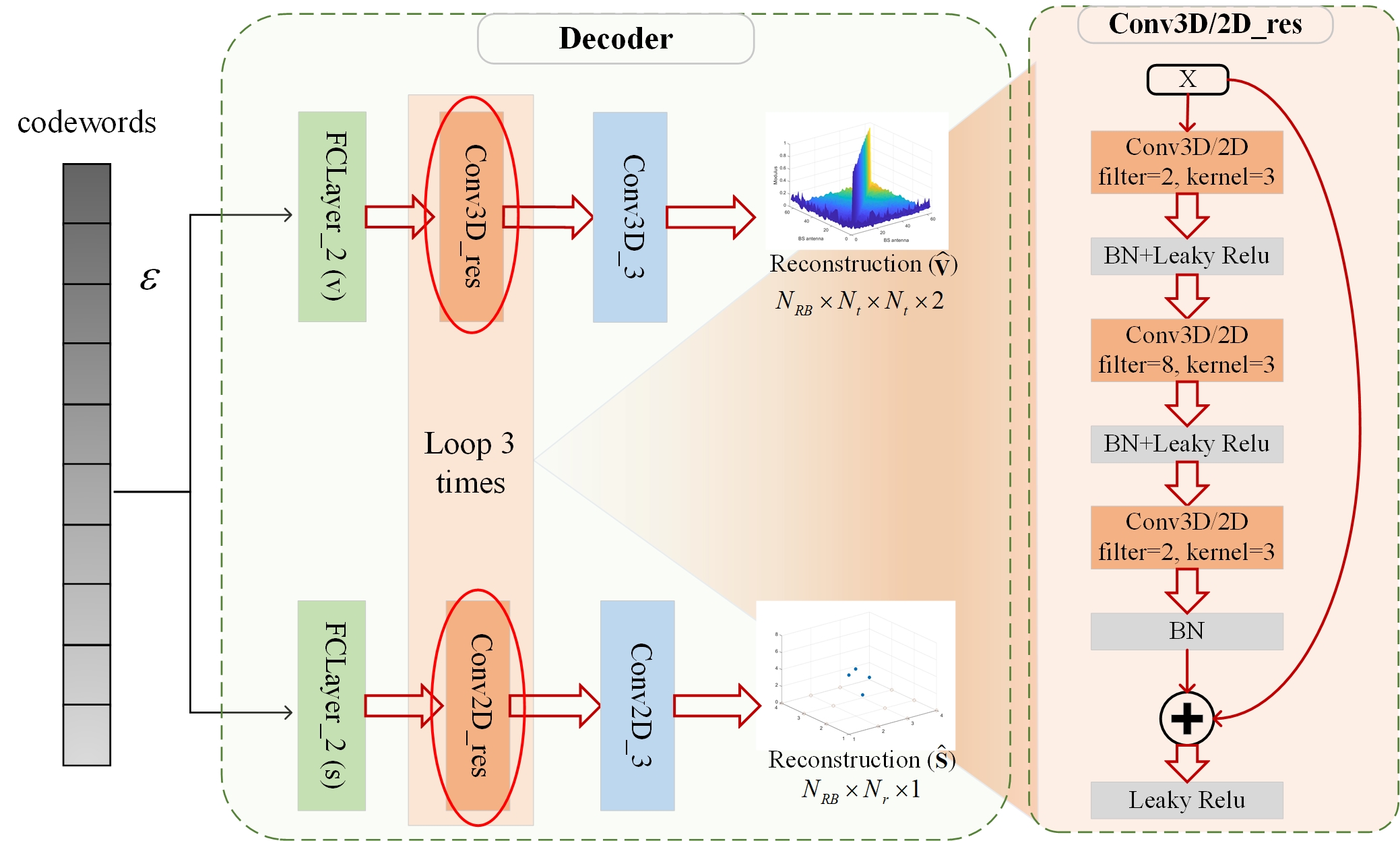}
	\caption{Illustration of the decoder module and convolutional residual block. }
	\label{fig:decoder}
\end{figure}

In this part we will discuss the decoder module in EMEVNet. The decoder will be deployed at the BS to reconstruct  eigenmatrix $\widehat{\mathbf{V}}$ and eigenvectors $\widehat{\mathbf{S}}$.  Fig. \ref{fig:decoder} shows the  detailed decoder module and the convolution residual block applied in it. We design two different branches to reconstruct $\widehat{\mathbf{V}}$ and $\widehat{\mathbf{S}}$ respectively. Two different fully-connected layers are utilized to extract the high-dimensional features of $\mathbf{V}$ and $\mathbf{S}$ from codewords $\varepsilon$.  And then we design convolutional residual blocks to reconstruct $\widehat{\mathbf{V}}$ and $\widehat{\mathbf{S}}$.
Since this part depends on convolution layer and fully-connected layer, and their complexity analysis is mentioned above.  The hyper-parameters set in transcoding module is listed in Tab. \ref{table3}, and the analysis of complexity is also included.

\section{Simulation Results and Discussions} \label{simulation results}
This section  shows the simulation experiments of proposed EMEV feedback NN (EMEVNet)  in detail, including simulation platform, datasets generation, parameter settings and performance evaluation. And then we  give some analysis and discussion about the results, including feasibility analysis, superiority analysis and robustness analysis.  Meanwhile, this section exhibits all numerical results corresponding to  simulation experiments.  The datasets  used in this paper and the simulation codes  can be found at Github\footnote{Github link: https://github.com/CodeDwan/EMEV-feedback}.

\subsection{Parameters Setting}
\subsubsection{Simulation platform}
All the simulations and experiments are carried out on the workstation with CentOS 7.0. The workstation is equipped with two Intel(R) Xeon(R) Silver 4210R CPU and four Nvidia RTX 2080Ti GPU, it also has 256GB Random Access Memory (RAM).

\subsubsection{Datasets generation}
With the help of MATLAB 5G Toolbox and Communication Toolbox, we define a standard CDL channel object and carry on the link-level simulation. The dataset used in our experiments are extracted from the link-level simulator. Tab. \ref{data} shows the alternative parameter and  default values in the data generator. Both UE and BS antennas follow uniform panel array (UPA) distribution.
\begin{table}[htpb]
	\caption{Simulation experiment parameter setting}
	\begin{tabular}{|c|ccccc|}
		\hline
		\multirow{2}{*}{\begin{tabular}[c]{@{}c@{}}Channel\\ environment\end{tabular}} & \multicolumn{3}{c|}{NLOS}                                                               & \multicolumn{2}{c|}{LOS}           \\ \cline{2-6}
		& \multicolumn{1}{c|}{CDL-A}  & \multicolumn{1}{c|}{CDL-B}  & \multicolumn{1}{c|}{CDL-C}  & \multicolumn{1}{c|}{CDL-D} & CDL-E \\ \hline
		NRB    & \multicolumn{5}{c|}{13}               \\ \hline
		Center frequency     & \multicolumn{5}{c|}{28 GHz}                \\ \hline
		Subcarrier spacing   & \multicolumn{5}{c|}{60 KHz}                  \\ \hline
		UE speeds    & \multicolumn{5}{c|}{\{4.8, 24, 40, 60\} km/h}           \\ \hline
		Delay spreads          & \multicolumn{1}{c|}{129 ns} & \multicolumn{1}{c|}{634 ns} & \multicolumn{1}{c|}{634 ns} & \multicolumn{1}{c|}{65 ns} & 65ns  \\ \hline
		BS antenna          & \multicolumn{5}{c|}{UPA $[8. 8] = 64$}               \\ \hline
		UE antenna            & \multicolumn{5}{c|}{UPA $[2, 2] = 4$}                \\ \hline
	\end{tabular}
	\label{data}
\end{table}

We totally generate $60,000$  data samples for each CDL channel using MATLAB.  The $60,000$ samples of each channel type are divided into $50,000$ and $10,000$. We utilize $50,000$ samples $\mathbb{D}_{sp}$ to train  specific network $\mathbb{N}_{sp}$ for each channel.  For comparison, we mix $10,000$ samples of  five CDL channels to obtain $\mathbb{D}_{mix}$, and train a general network $\mathbb{N}_{mix}$.  The operations of datasets and neural networks can be respectively expressed as,
\begin{equation}
\mathbb{N}_{sp}^* \gets \mathbb{D}_{sp}^* = \left\lbrace \mathbf{H}_{*}^{(50k)} \right\rbrace
\label{datann_1}
\end{equation}
\begin{equation}
\mathbb{N}_{mix} \gets \mathbb{D}_{mix} = \left\lbrace \mathbf{H}_{A}^{(10k)}, \mathbf{H}_{B}^{(10k)}, \cdots, \mathbf{H}_{E}^{(10k)} \right\rbrace
\label{datann_2}
\end{equation}
where $\mathbf{H}_{*}^{(50k)}$ denotes the $50,000$ samples of each channel environment represented by different subscripts, and $\mathbf{H}_{A}^{(10k)}$ means the $10,000$ samples of CDL-A channel.

\subsubsection{Setting of compression ratios}
This paper  explores the EMEV feedback solution. The UE needs to compress $\mathbf{V}$ and $\mathbf{S}$ first, and then feed back the compressed  codewords $\varepsilon$ to the BS. The length of codewords $L_{\varepsilon}$ is related to the overhead of feedback, which will affect the spectral efficiency of communication system. And the $L_{\varepsilon}$ is decided by compression ratio $\beta_{CR}$.  Since the size of channel matrix $\mathbf{H}$, eigenmatrix $\mathbf{V}$ and eigenvectors $\mathbf{S}$ are different, we detailed define the setting of compression ratio $\beta_{CR}$  in this part.  For unity and comparison, almost all researches utilize the compression ratio of $\mathbf{H}$ as the measurement standard. Hence, the length of codewords $L_{\varepsilon}$ can be expressed as,
\begin{equation}
	\begin{aligned}
		L_{\varepsilon} &=\frac{\Re(\mathbf{H})+\Im(\mathbf{H})}{\beta_{h}}+1\\
		&= \frac{N_{RB} \times N_r \times N_t \times 2}{\beta_{h}}+1
	\end{aligned}
	\label{lc}
\end{equation}
where $\beta_{h}$  is the compression ratio of $\mathbf{H}$.  In order to guarantee the effectiveness of the comparative simulation experiments,  fixed $L_{\varepsilon}$ is given in  the follow discussion. To ensure the same $L_{\varepsilon}$, the compression ratio of  EMEVNet can be defined as,
\begin{equation}
	\begin{aligned}
		\beta_{emev} &\approx \frac{\Re(\mathbf{V})+\Im(\mathbf{V})+\Re(\mathbf{S})}{\Re(\mathbf{H})+\Im(\mathbf{H})}\beta_{h} \\
		&= \frac{N_{RB} \times (N_t \times N_t \times 2 + N_r) }{N_{RB} \times N_r \times N_t \times 2} \beta_{h}
	\end{aligned}
	\label{cremev}
\end{equation}
where $\beta_{emev}$ represents the compression ratio of $\mathbf{V},\mathbf{S}$.
Referring to the  datasets parameters settings in Tab. \ref{data}, we can assume  $\beta_{emev}=16\beta_{h}$ in this paper.
\begin{figure*}[htbp]
	\centering
	\subfloat[Initial sample $\mathbf{V}$ at UE]{
		\label{vini}
		\includegraphics[width=1.6 in]{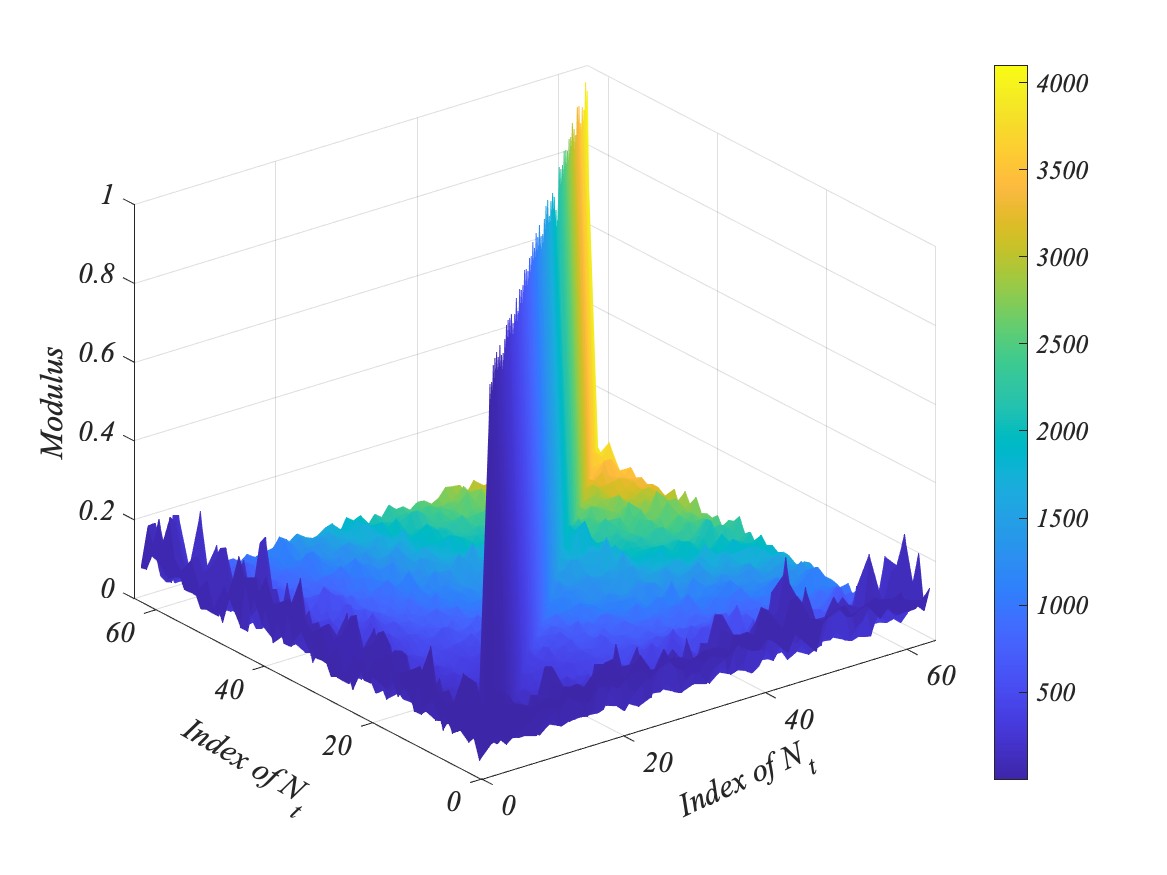}}
	\hspace{0pt}
	\subfloat[Reconstructed $\widehat{\mathbf{V}}: L_{\varepsilon}=416$]{
		\label{v16}
		\includegraphics[width=1.6 in]{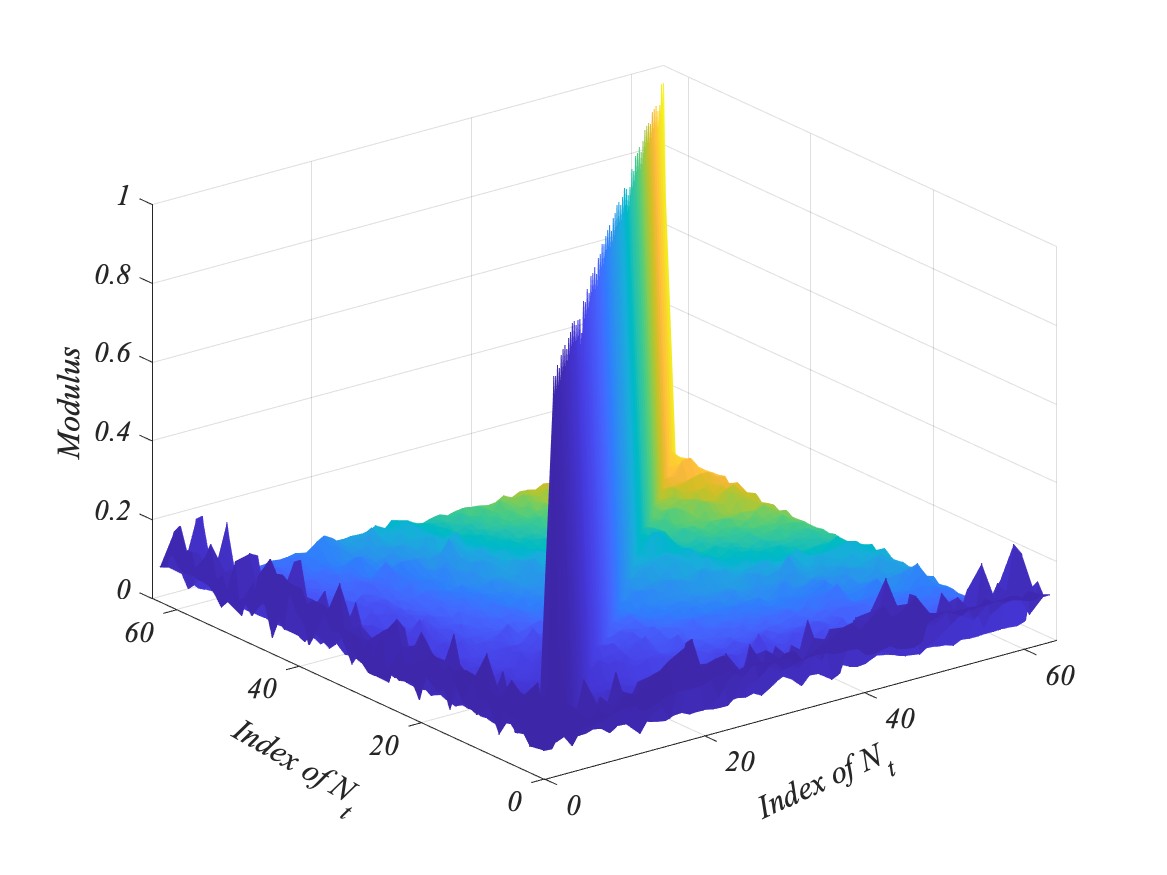}}
	\hspace{0pt}
	\subfloat[Reconstructed $\widehat{\mathbf{V}}: L_{\varepsilon}=208$]{
		\label{v32}
		\includegraphics[width=1.6 in]{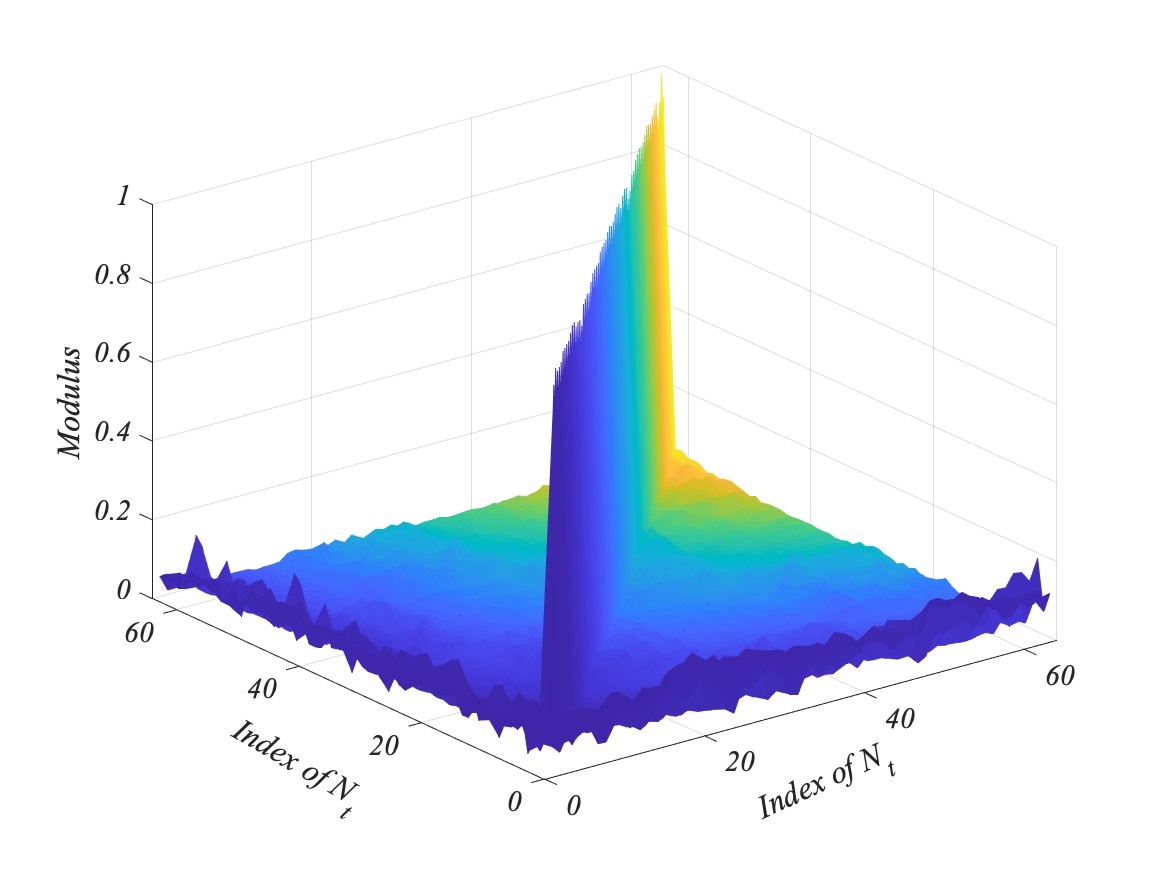}}
	\hspace{0pt}
	\subfloat[Reconstructed $\widehat{\mathbf{V}}: L_{\varepsilon}=104$]{
		\label{v64}
		\includegraphics[width=1.6 in]{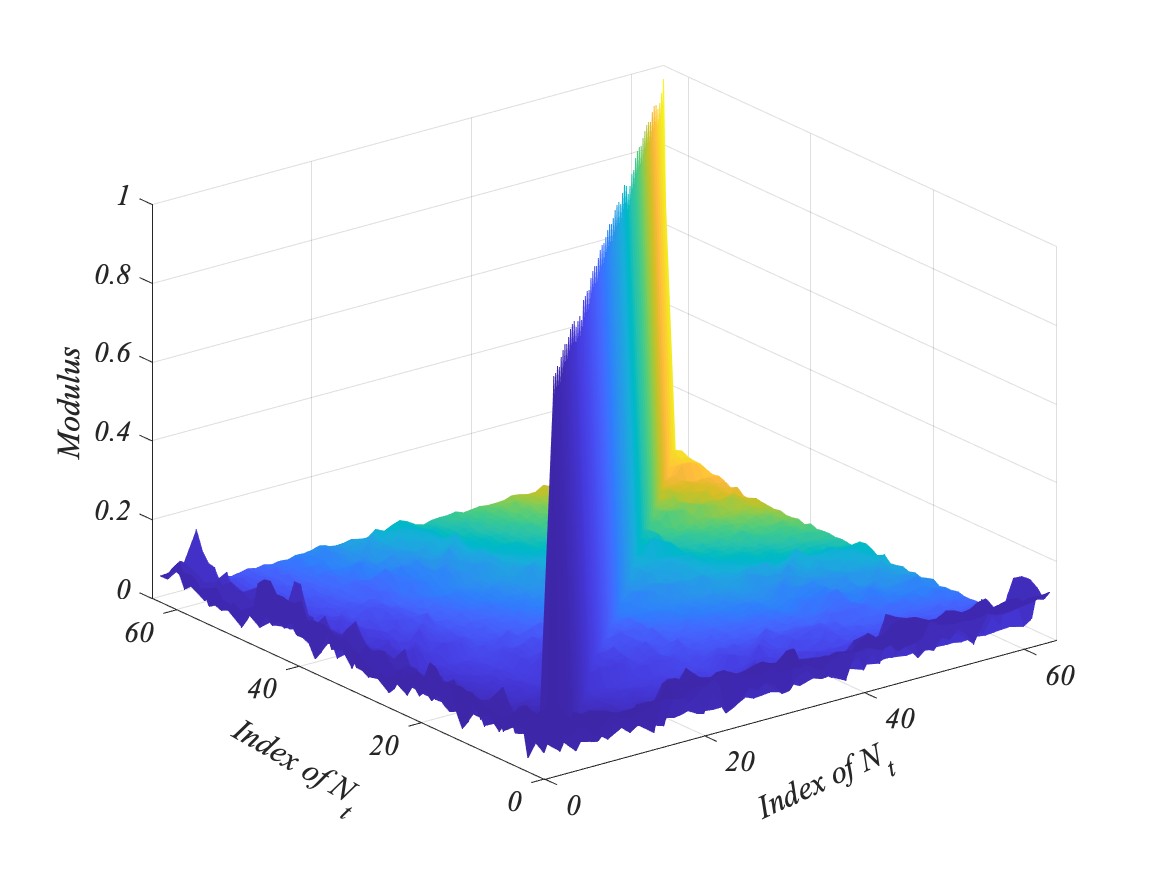}}
	\hspace{0pt}
	\subfloat[Reconstructed $\widehat{\mathbf{V}}: L_{\varepsilon}=52$]{
		\label{v128}
		\includegraphics[width=1.6 in]{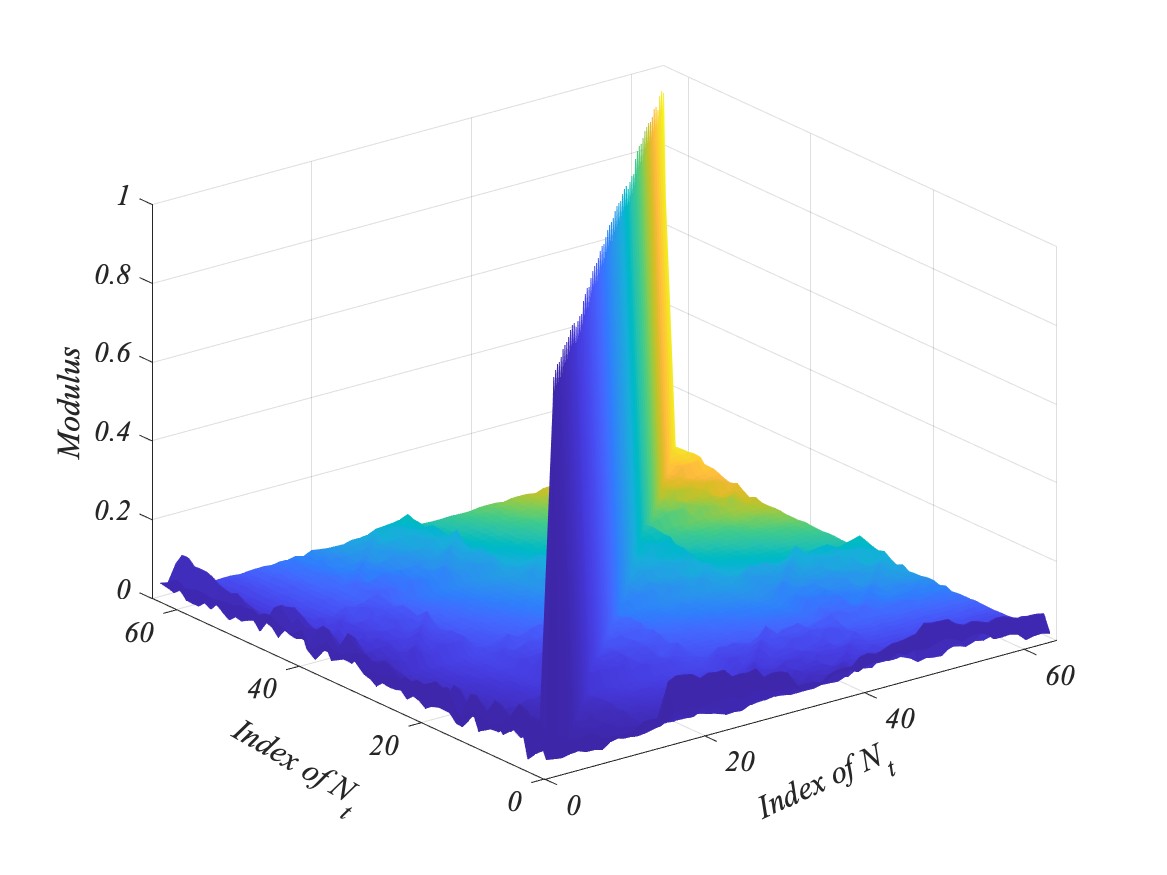}}
	\hspace{0pt}
	\subfloat[Reconstructed $\widehat{\mathbf{V}}: L_{\varepsilon}=26$]{
		\label{v256}
		\includegraphics[width=1.6 in]{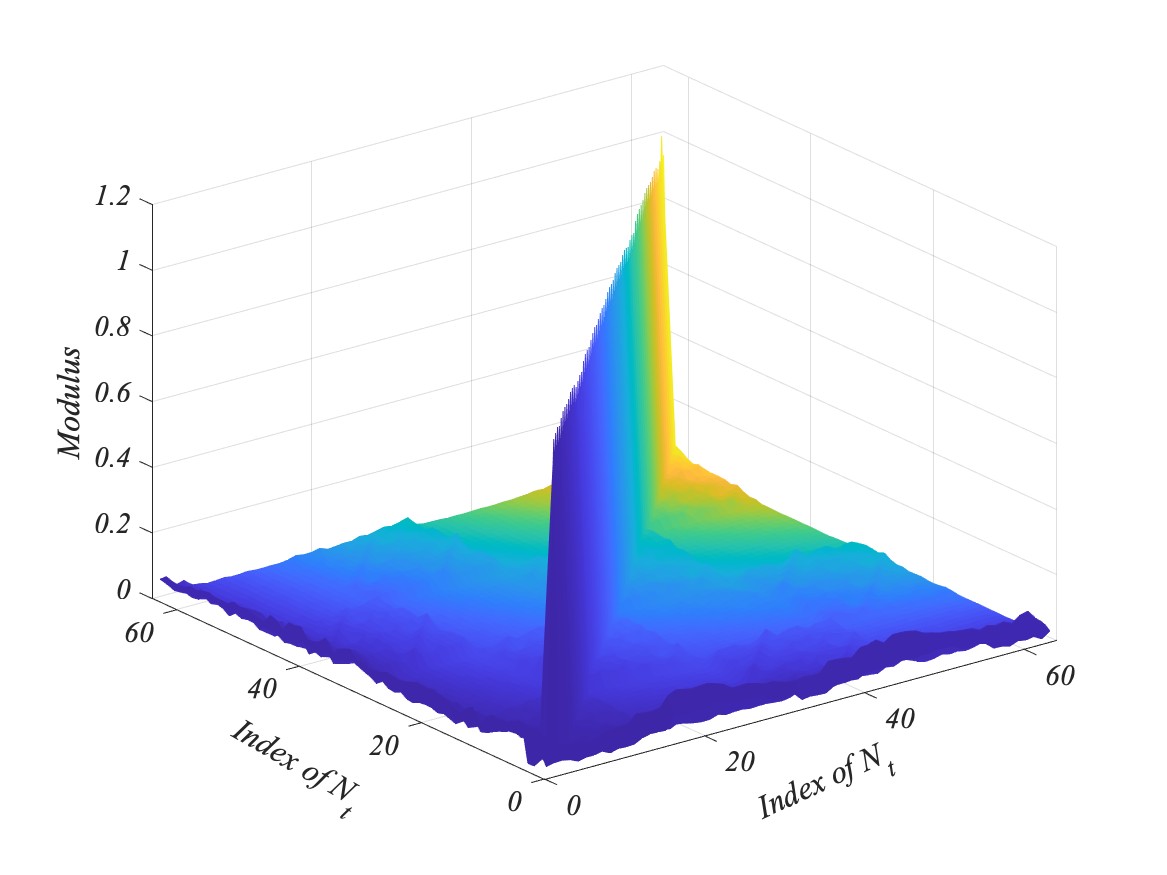}}
	\hspace{0pt}
	\subfloat[Reconstructed $\widehat{\mathbf{V}}: L_{\varepsilon}=13$]{
		\label{v512}
		\includegraphics[width=1.6 in]{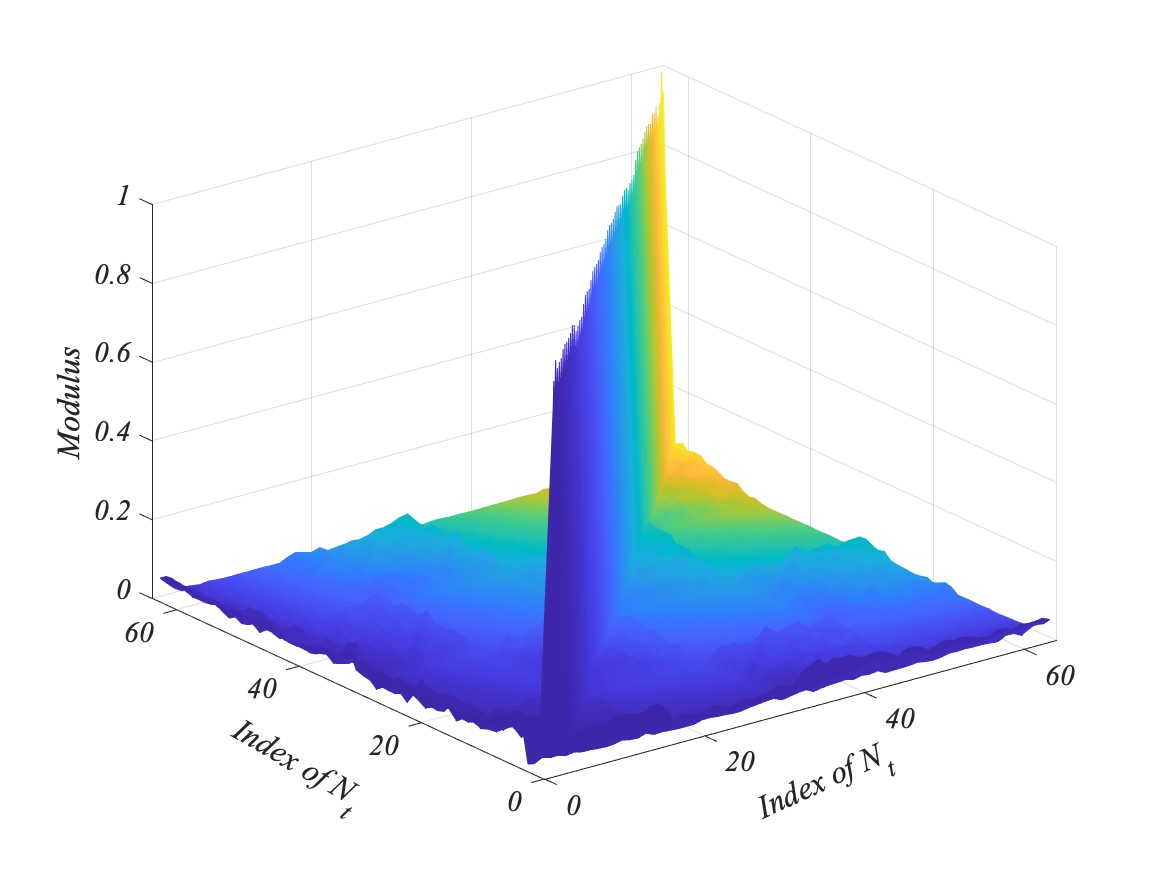}}
	\hspace{0pt}
	\subfloat[Reconstructed $\widehat{\mathbf{V}}: L_{\varepsilon}=6$]{
		\label{v1024}
		\includegraphics[width=1.6 in]{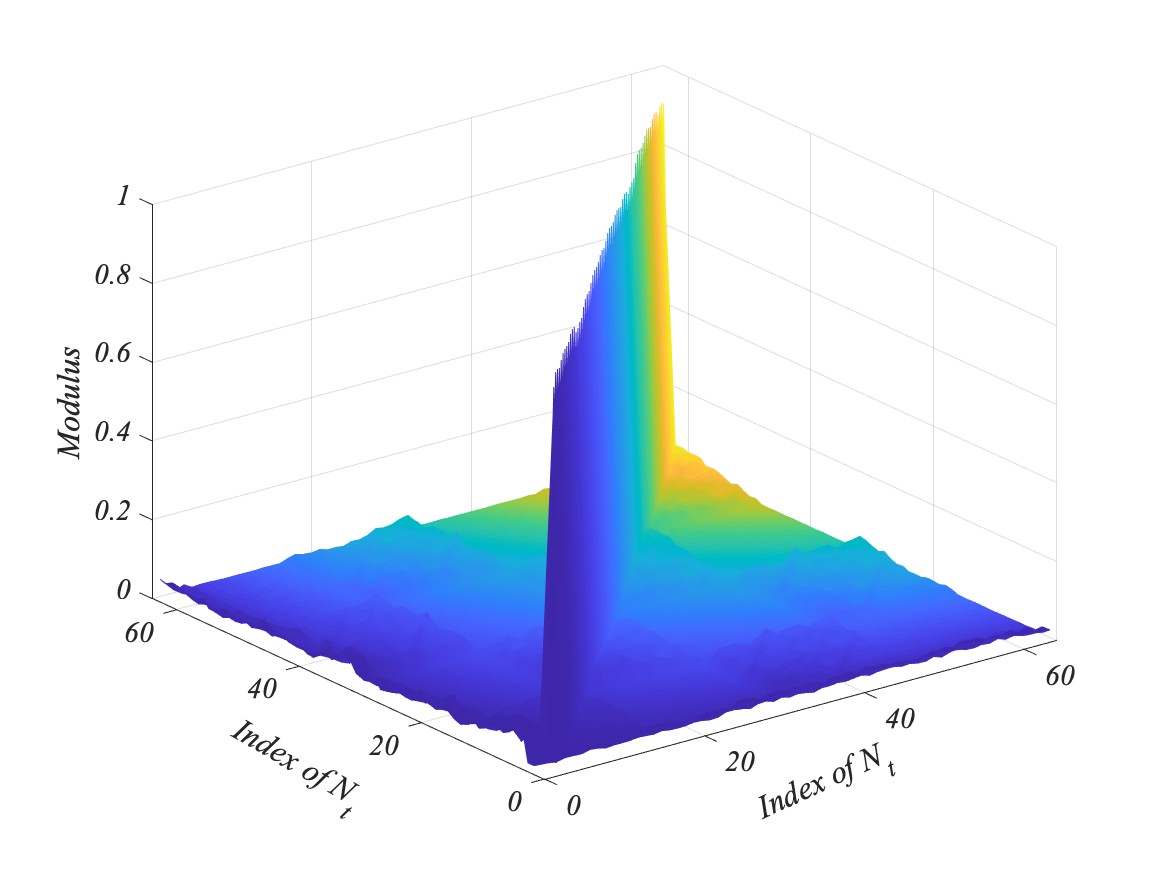}}
	\hspace{0pt}
	\subfloat[Initial sample $\mathbf{S}$ at UE]{
		\label{sini}
		\includegraphics[width=1.6 in]{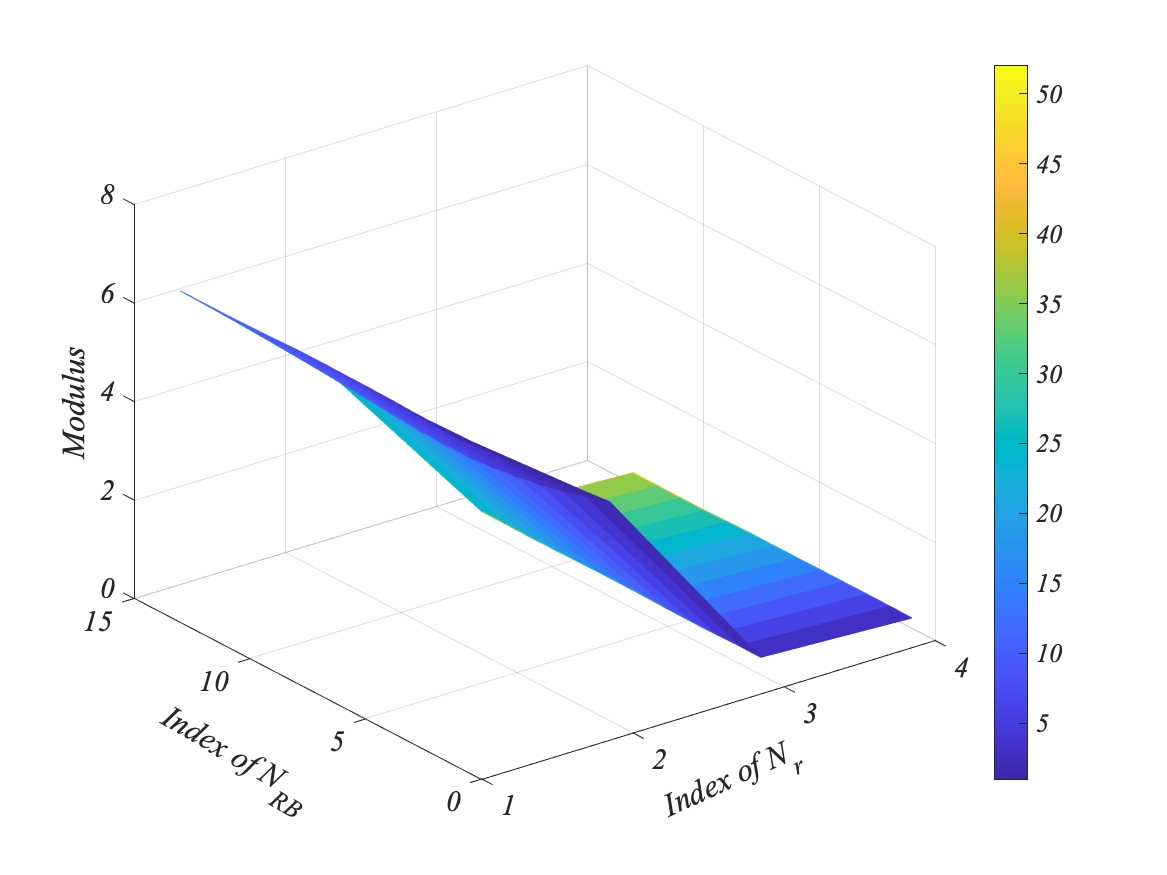}}
	\hspace{0pt}
	\subfloat[Reconstructed $\widehat{\mathbf{S}}: L_{\varepsilon}=416$]{
		\label{s16}
		\includegraphics[width=1.6 in]{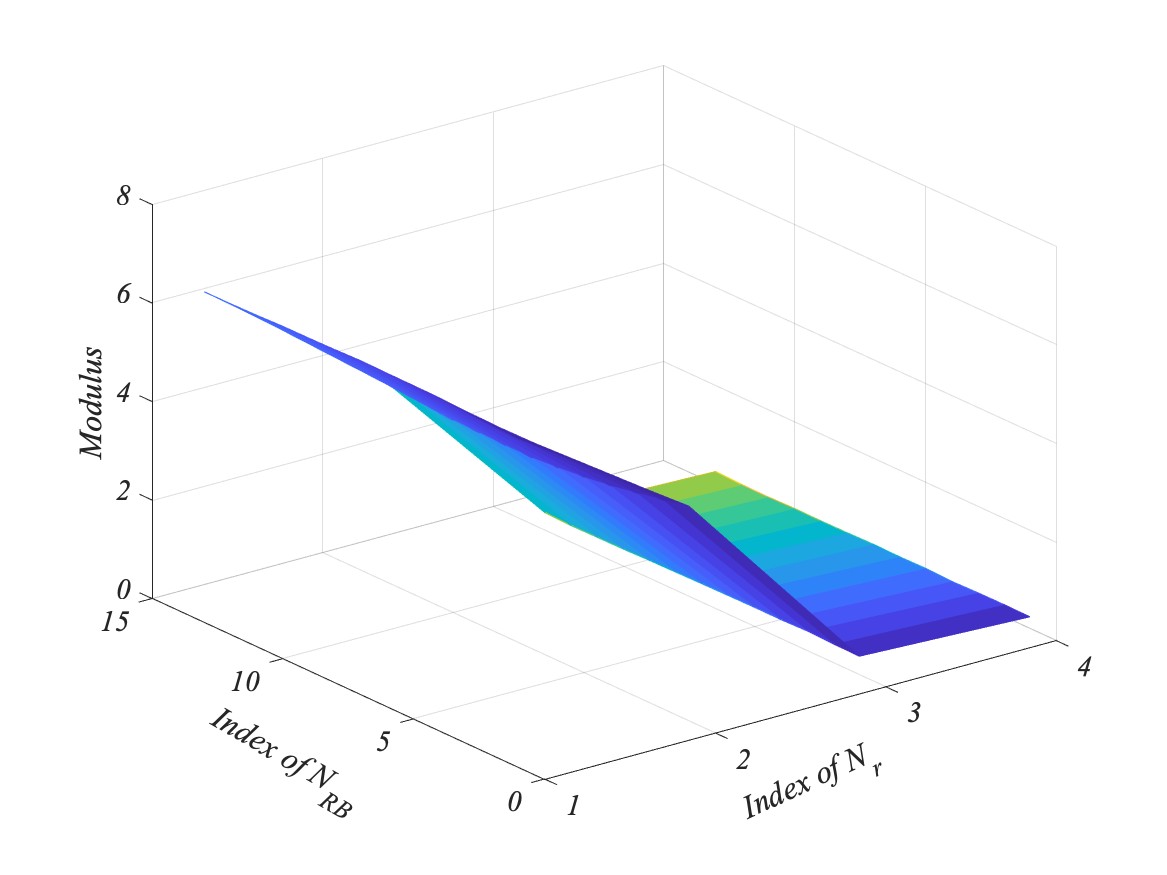}}
	\hspace{0pt}
	\subfloat[Reconstructed $\widehat{\mathbf{S}}: L_{\varepsilon}=208$]{
		\label{s32}
		\includegraphics[width=1.6 in]{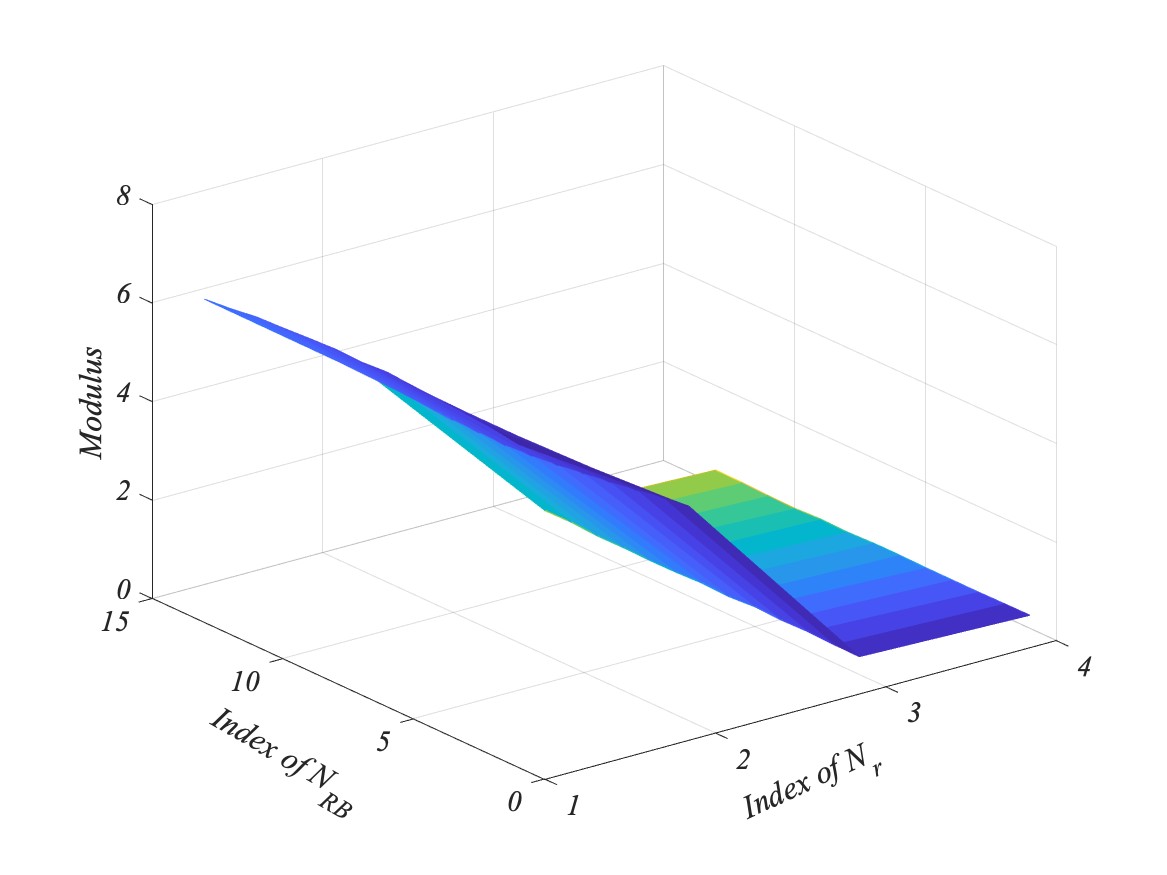}}
	\hspace{0pt}
	\subfloat[Reconstructed $\widehat{\mathbf{S}}: L_{\varepsilon}=104$]{
		\label{s64}
		\includegraphics[width=1.6 in]{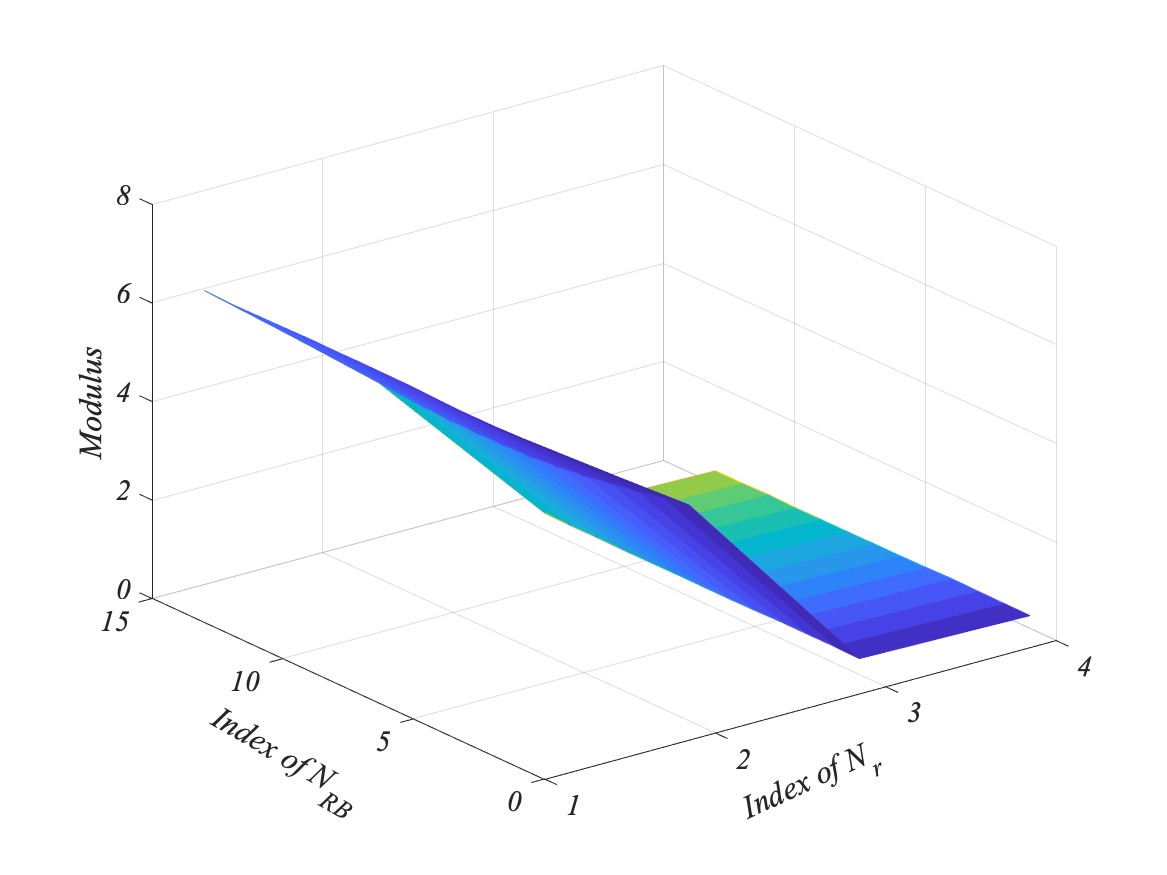}}
	\hspace{0pt}
	\subfloat[Reconstructed $\widehat{\mathbf{S}}: L_{\varepsilon}=52$]{
		\label{s128}
		\includegraphics[width=1.6 in]{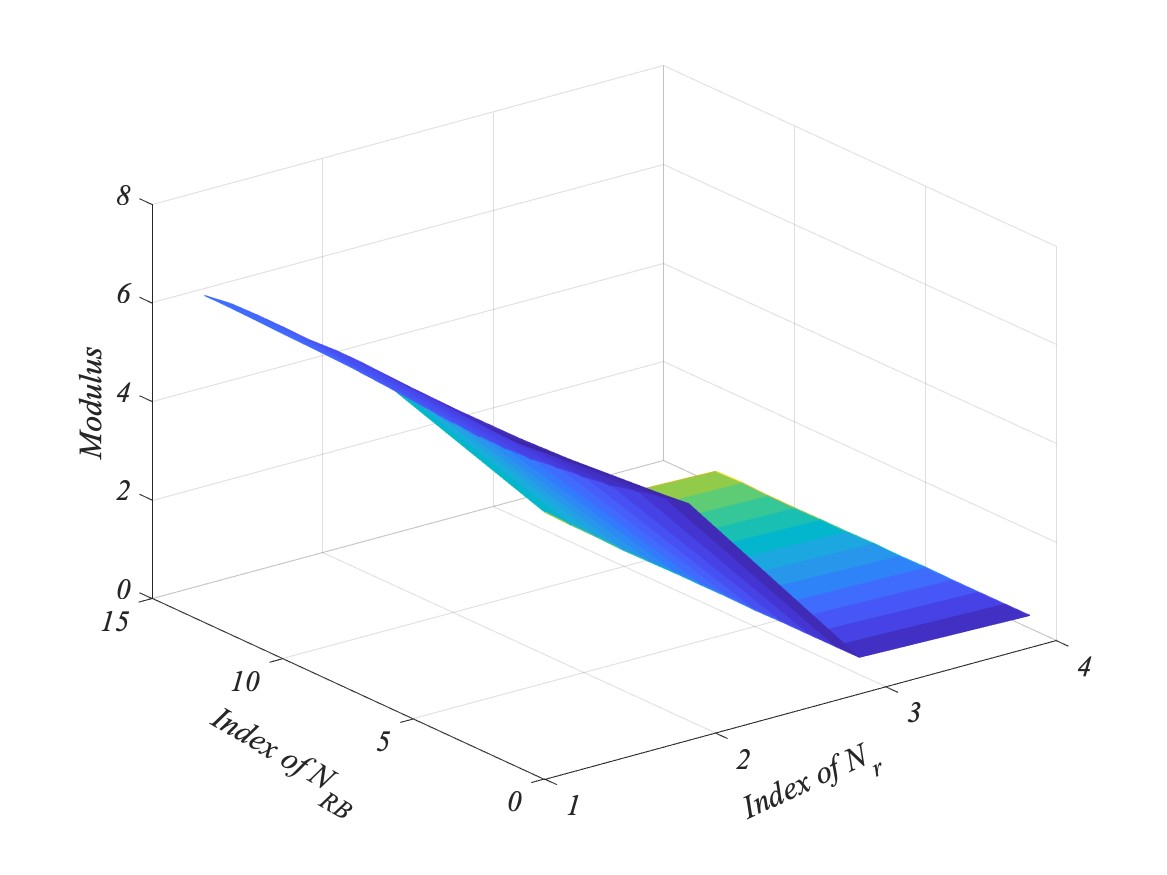}}
	\hspace{0pt}
	\subfloat[Reconstructed $\widehat{\mathbf{S}}: L_{\varepsilon}=26$]{
		\label{s256}
		\includegraphics[width=1.6 in]{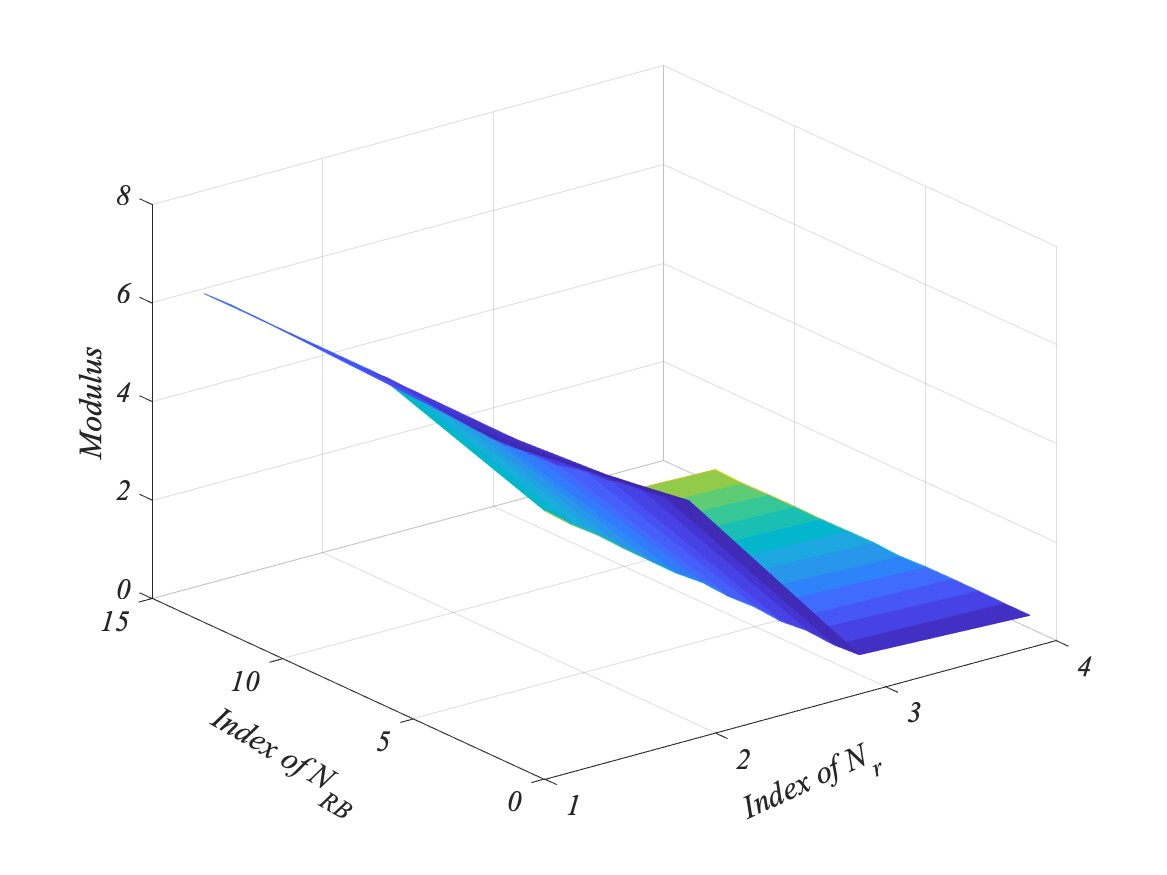}}
	\hspace{0pt}
	\subfloat[Reconstructed $\widehat{\mathbf{S}}: L_{\varepsilon}=13$]{
		\label{s512}
		\includegraphics[width=1.6 in]{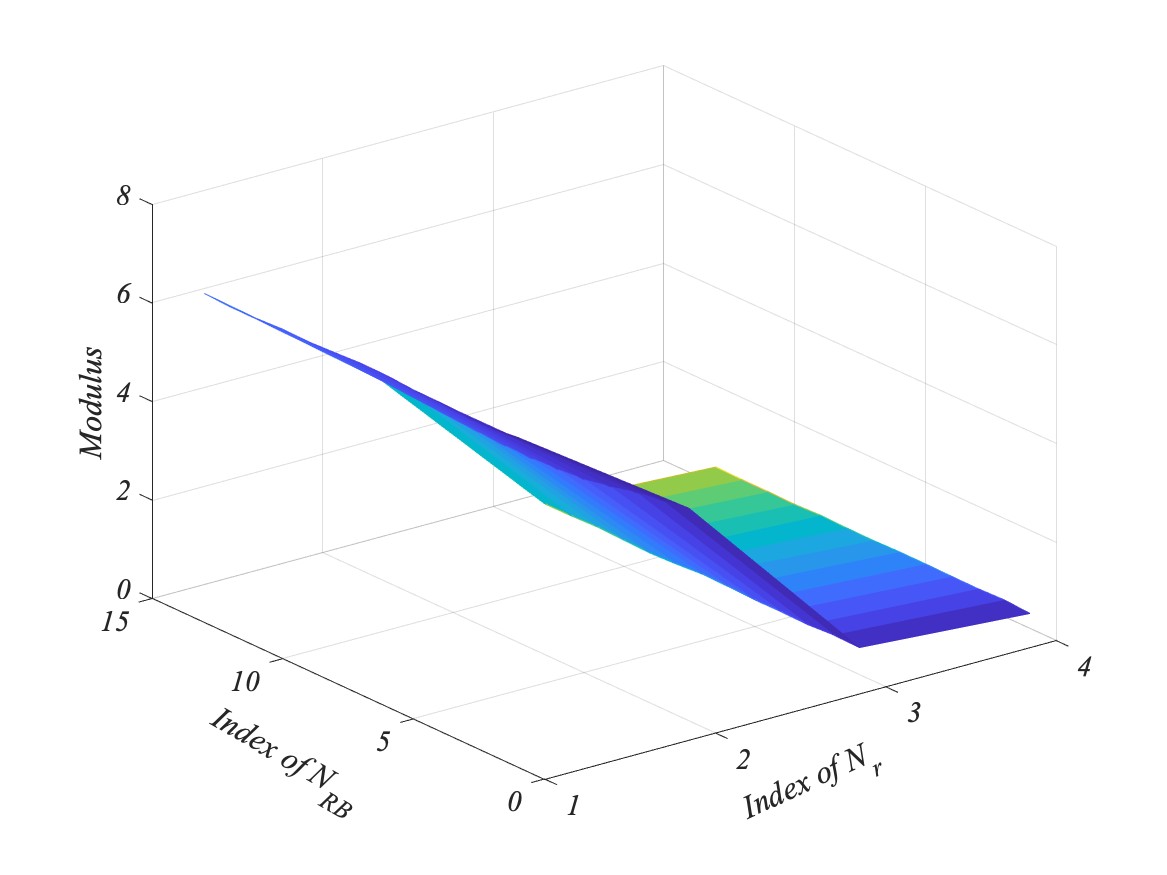}}
	\hspace{0pt}
	\subfloat[Reconstructed $\widehat{\mathbf{S}}: L_{\varepsilon}=6$]{
		\label{s1024}
		\includegraphics[width=1.6 in]{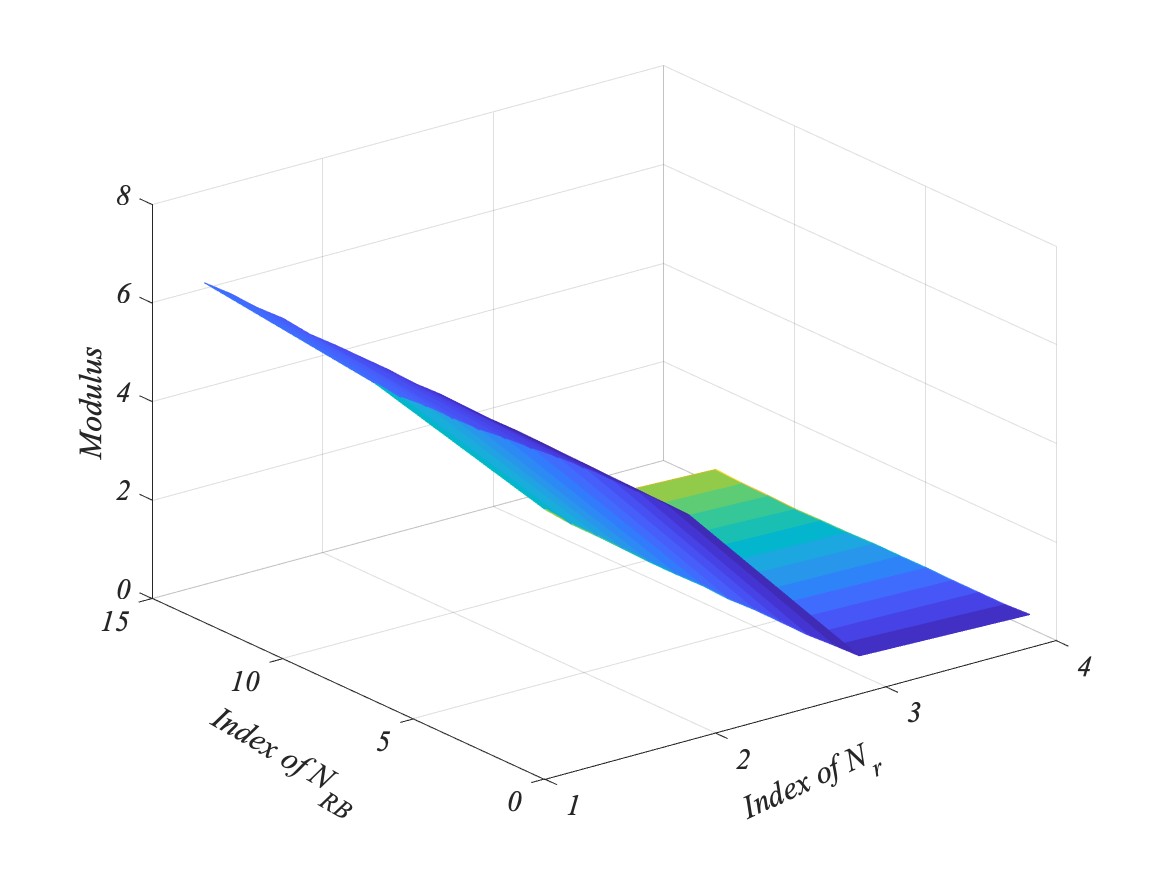}}
	\hspace{0pt}
	\caption{Illustration of partial visualization feasibility experimental results.  Fig. \ref{vini} and Fig. \ref{sini} show the perfect eigenmatrix $\mathbf{V}$ and eigenvector $\mathbf{S}$ at the UE. Fig. \ref{v16}  to Fig. \ref{v1024} describe the reconstructed $\widehat{\mathbf{V}}$ from $L_{\varepsilon}=416$ to $L_{\varepsilon}=6$.  Fig. \ref{s16} to Fig. \ref{s1024} show the reconstructed eigenvector $\widehat{\mathbf{S}}$ with the same $L_{\varepsilon}$. }
	\label{feasibility}
\end{figure*}

\subsubsection{Setting of NN training} \label{NNsetting}
To ensure that the NN can converge to the best performance, we have tried many times for the NN training parameters, and finally defined as follows: Maximum  epoch number $\tau=500$; Initial learning rate $\eta=1\times 10^{-3}$; Early stop patience is 50 epoches. For the fairness evaluation, above NN training parameters are consistent in different experiments. The length of  codewords and the corresponding compression ratio are set as follows:
\begin{itemize}
	\item  $L_{\varepsilon}=[416,208,104,52,26,13,6]$ means the length of codewords.
	\item $\beta_{h}=[16,32,64,128,256.512,1024]$ represents the compression ratio of CSI matrix $\mathbf{H}$.
	\item $\beta_{emev}=[256,512,1024,2048,4096,8192,16384]$ denotes the system compression ratio of EMEVNet.
\end{itemize}

\subsubsection{Performance evaluation} \label{performance eva}
The ultimate purpose of our architecture is to help BS reconstruct the eigenmatrix $\mathbf{V}$ and eigenvector $\mathbf{S}$.  Therefore, we utilize normalized mean square error (NMSE) and cosine similarity $(\rho)$ to measure the reconstruction accuracy, which can be respectively defined as,
\begin{equation}
	\begin{aligned}
		NMSE\left( \mathbf{V},\widehat{\mathbf{V}}\right)  &= \mathbb{E}\left\lbrace \lVert \mathbf{V} - \widehat{\mathbf{V}} \rVert_2^2/\lVert \mathbf{V} \rVert_2^2 \right\rbrace  \\
		& = 10\log\left(\mathbb{E}\left\lbrace \lVert \mathbf{V} - \widehat{\mathbf{V}} \rVert_2^2/\lVert \mathbf{V} \rVert_2^2 \right\rbrace \right)
	\end{aligned}
	\label{nmse}
\end{equation}

\begin{equation}
	\begin{aligned}
		NMSE\left( \mathbf{S},\widehat{\mathbf{S}}\right)  &= \mathbb{E}\left\lbrace \lVert \mathbf{S} - \widehat{\mathbf{S}} \rVert_2^2/\lVert \mathbf{S} \rVert_2^2 \right\rbrace \\
		& = 10 \log \left(\mathbb{E}\left\lbrace \lVert \mathbf{S} - \widehat{\mathbf{S}} \rVert_2^2/\lVert \mathbf{S} \rVert_2^2 \right\rbrace \right)
	\end{aligned}
	\label{nmse}
\end{equation}

\begin{equation}
	\begin{aligned}
		\rho\left( \mathbf{V},\widehat{\mathbf{V}}\right)  &= \mathbb{E}\left\lbrace \frac{ \langle \mathbf{V}^* , \widehat{\mathbf{V}} \rangle}{\lVert \mathbf{V} \rVert_2 \lVert\widehat{\mathbf{V}} \rVert_2}  \right\rbrace
	\end{aligned}
	\label{rho}
\end{equation}

\begin{equation}
	\begin{aligned}
		\rho\left( \mathbf{S},\widehat{\mathbf{S}}\right)  &= \mathbb{E}\left\lbrace \frac{ \langle \mathbf{S}^* , \widehat{\mathbf{S}}\rangle}{\lVert \mathbf{S} \rVert_2 \lVert\widehat{\mathbf{S}} \rVert_2}  \right\rbrace
	\end{aligned}
	\label{rho}
\end{equation}
where $\langle a,b \rangle$ is the scalar product in Euclidean space. Generally, $NMSE$ is converted to logarithmic domain and the smaller value represents the better performance, e.g. $NMSE=-20$ dB shows better performance than $-10$ dB.  The  range of cosine similarity is $\rho \in [-1,1]$, which can measure the similarity between reconstructed  and original matrix. $\rho$ close to 1 indicates better reconstruction performance.

\subsection{Feasibility Analysis} \label{Feasibility Analysis}
This subsection will show some simulation results and discuss the feasibility of our proposed architecture. In order to verify the feasibility of proposed EMEVNet, this part utilize $50,000$ samples of CDL-A channel, i.e. $\mathbb{D}_{sp}^{A} = \{ \mathbf{H}_{A}^{(50k)} \}$, as simulation datasets.  We set the training, validation  and testing datasets  obeying the ratio of $70:15:15$. Before training EMEVNet, the datasets $\mathbb{D}_{sp}^{A}$ need to carry out SVD transformation according to \textbf{Algorithm \ref{svdalg}}. And the generated $\mathbf{V},\mathbf{S}$ are the input of EMEVNet. After NN training stage according to \textbf{Algorithm \ref{emevnetalg}}, we can obtain the trained $\mathbb{N}_{sp}^{A}$, which is the specific EMEVNet  for CDL-A channel. In this part, we have trained and tested seven different compression ratios set as \textbf{Section \ref{NNsetting}}.

The experimental results can be found in Fig. \ref{feasibility}. Considering that $\mathbf{V} \in \mathbb{C}^{N_{RB}\times N_t\times N_t}$ is a 3D complex-valued matrix, so we split one RB for more intuitive exhibition.  The program randomly selects the third resource block, i.e. $\mathbf{V}(3) \in \mathbb{C}^{N_t \times N_t}$. Then, the Euclidean norm of $\mathbf{V}(3)$ is applied for convenient plotting,  which can be interpreted as the power distribution.  Fig. \ref{vini} shows  the initial $\mathbf{V}$ sample at UE and Fig. \ref{v16} to Fig. \ref{v1024} show the reconstructed $\widehat{\mathbf{V}}$ at BS with different compression ratio from $L_{\varepsilon}=416$ to $L_{\varepsilon}=6$. As for eigenvector $\mathbf{S} \in \mathbb{R}^{N_{RB}\times N_r}$, it is a 2D real-valued matrix. We directly plot the figure without any processing. Fig. \ref{sini} shows the initial $\mathbf{S}$ at UE and Fig. \ref{s16} to Fig. \ref{s1024} show the reconstructed $\widehat{\mathbf{S}}$ at BS with the same settings as mentioned above. After many experiments, the shape of $\mathbf{V}$ of different RBs is almost the same.  They all present  diagonal power distribution, and the edges are serrated. Moreover, the power distribution of $\mathbf{V}$ is a symmetrical image.

As shown in Fig. \ref{feasibility}, with the reduction of feedback codewords $L_{\varepsilon}$, the reconstructed $\widehat{\mathbf{V}}$ by the BS will lose more information. This loss is mainly reflected in the sawtooth energy at the edge of eigenmatrix, while the loss of power principal component (diagonal distribution) is limited. When the length of codewords comes to $L_{\varepsilon}=26$, i.e. $\beta_{h}=256$ and $\beta_{emev}=4096$, the reconstructed $\widehat{\mathbf{V}}$ losses almost all edge information.  It can be seen from the figure that when $L_{\varepsilon}\leq26$, the edge power of $\widehat{\mathbf{V}}$  cannot  be reconstructed. However, we can ensure that the  distribution of $\widehat{\mathbf{V}}$ remains unchanged and the diagonal power is hardly affected in any case. As for $\widehat{\mathbf{S}}$ at the BS,  it can be well reconstructed under almost any $\L_{\varepsilon}$. The detailed numerical results  of $\mathbb{N}_{sp}^A$, i.e. $NMSE,\rho$, will be shown and discussed in \textbf{Section \ref{num results}}. From the exhibition and analysis of this subsection,   we vividly  proved the feasibility of EMEVNet by visualizing some experimental results.

\begin{figure}[htbp]
	\centering
	\subfloat[CDL-A: $\widehat{\mathbf{V}}$]{
		\label{vNMSErho_A}
		\includegraphics[width=1.605 in]{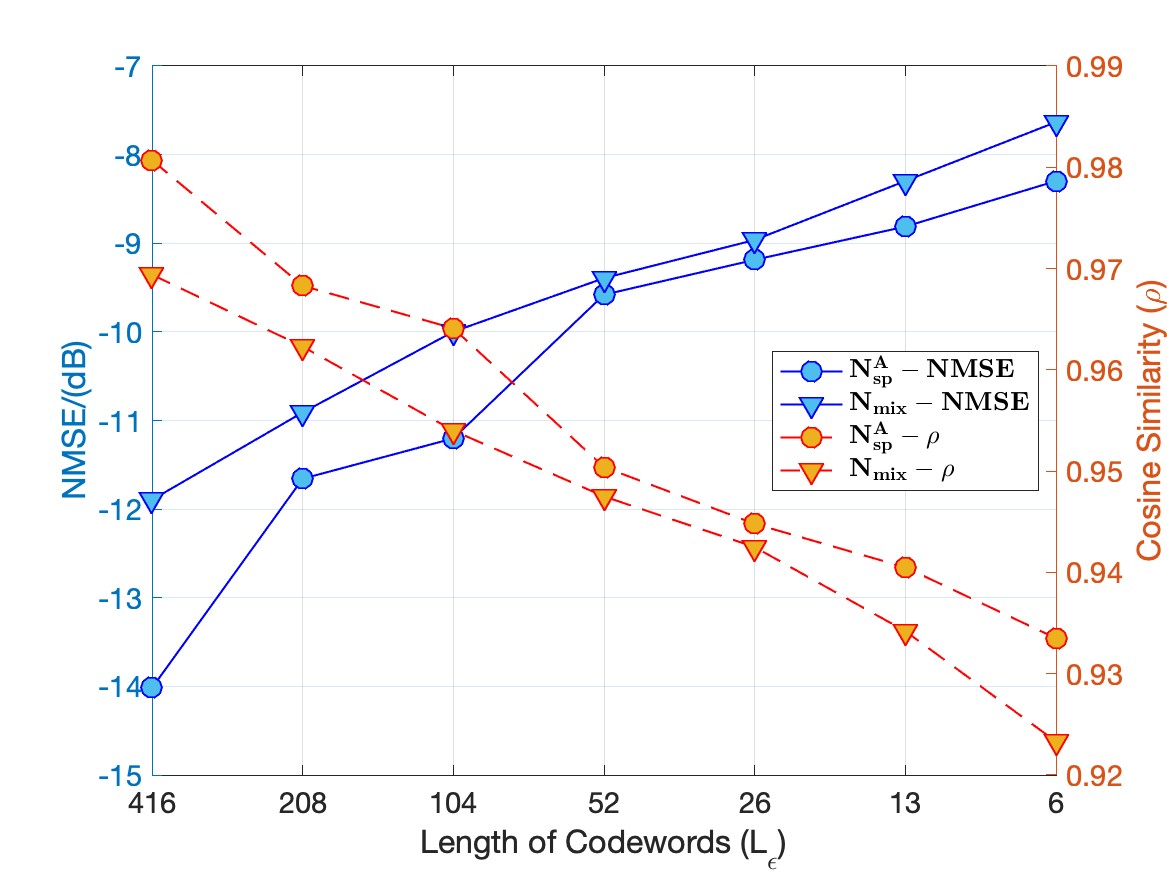}}
	\hspace{0pt}
	\subfloat[CDL-A: $\widehat{\mathbf{S}}$]{
		\label{sNMSErho_A}
		\includegraphics[width=1.60 in]{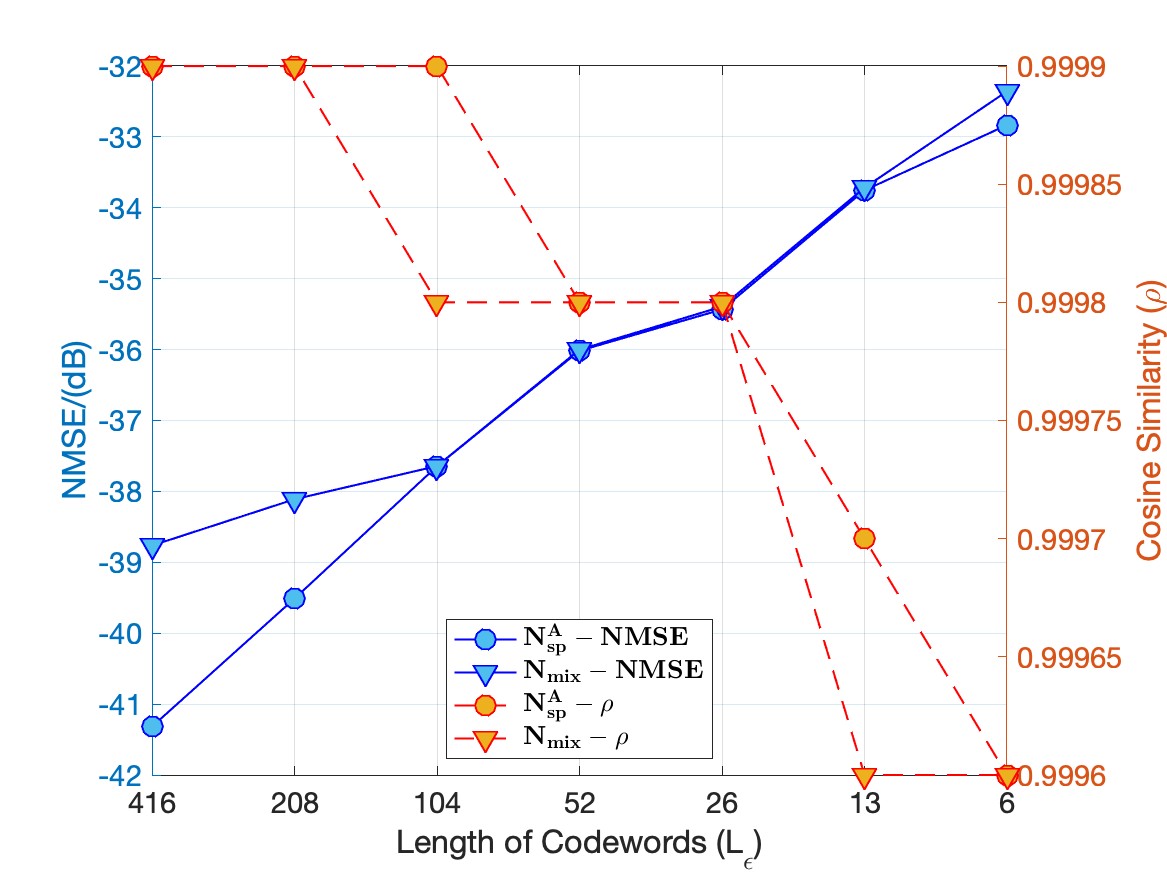}}
	\hspace{0pt}
	\subfloat[CDL-B: $\widehat{\mathbf{V}}$]{
		\label{vNMSErho_B}
		\includegraphics[width=1.60 in]{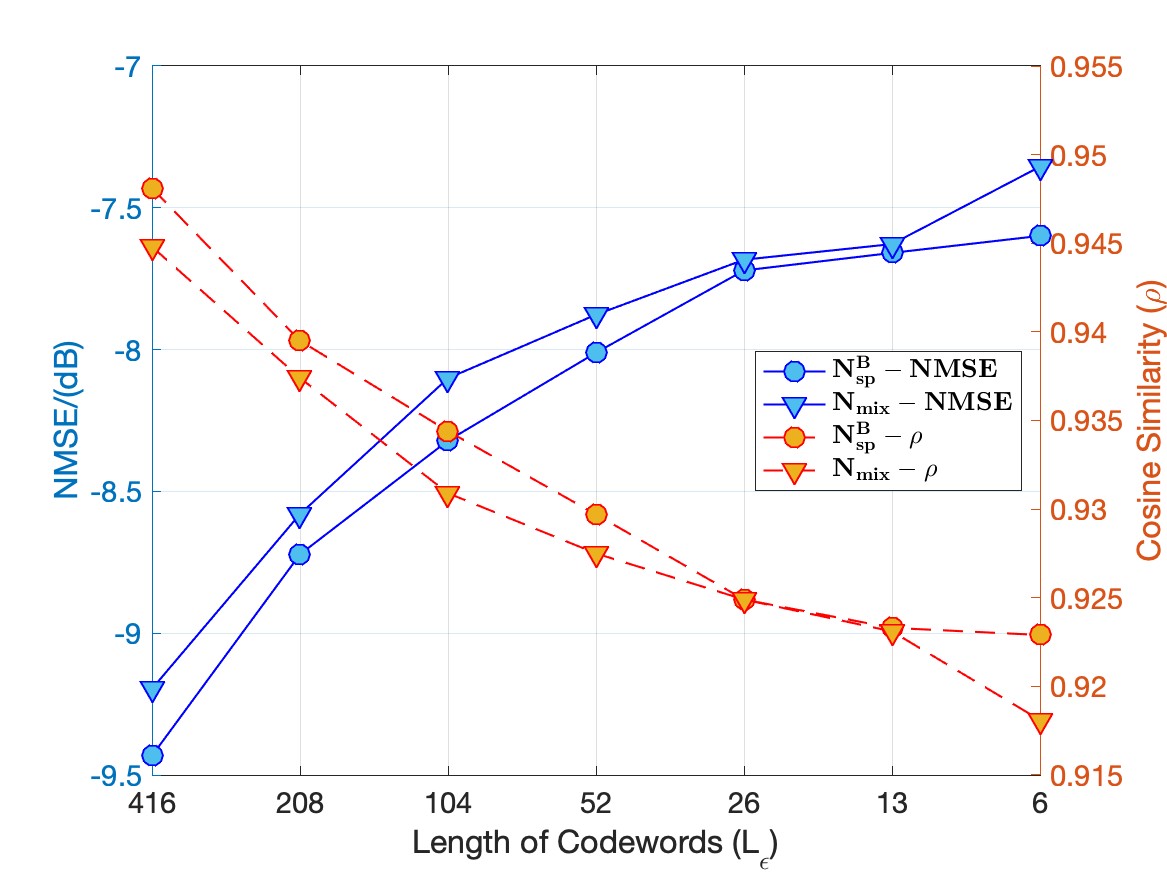}}
	\hspace{0pt}
	\subfloat[CDL-B: $\widehat{\mathbf{S}}$]{
		\label{sNMSErho_B}
		\includegraphics[width=1.60 in]{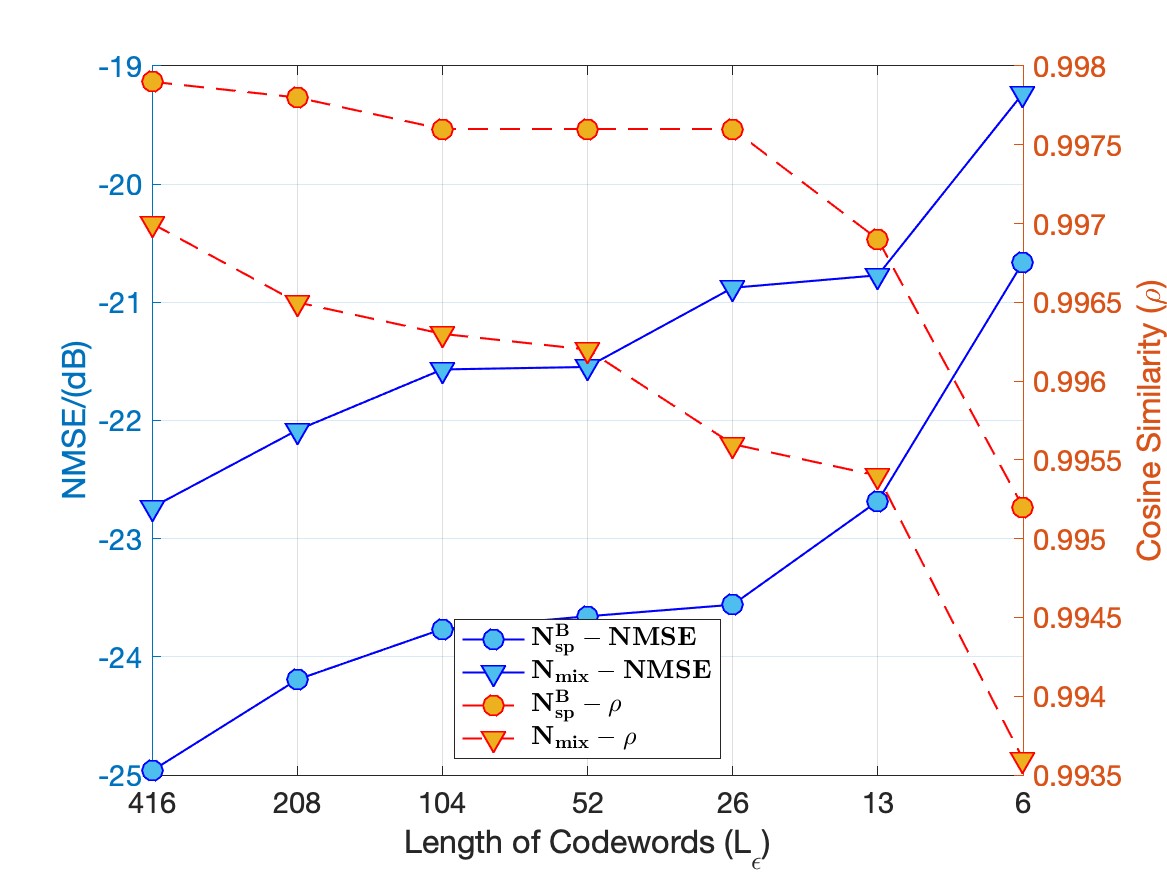}}
	\hspace{0pt}
	\subfloat[CDL-C: $\widehat{\mathbf{V}}$]{
		\label{vNMSErho_C}
		\includegraphics[width=1.60 in]{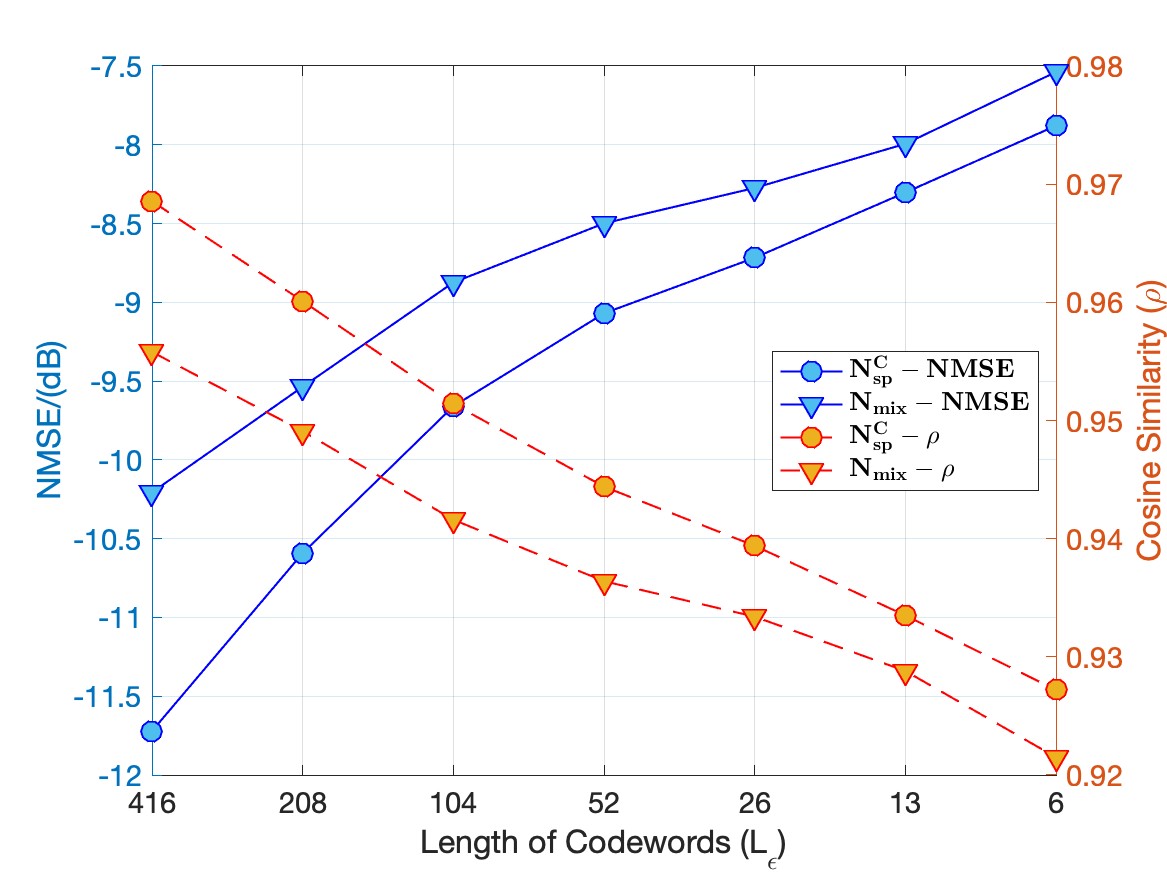}}
	\hspace{0pt}
	\subfloat[CDL-C: $\widehat{\mathbf{S}}$]{
		\label{sNMSErho_C}
		\includegraphics[width=1.60 in]{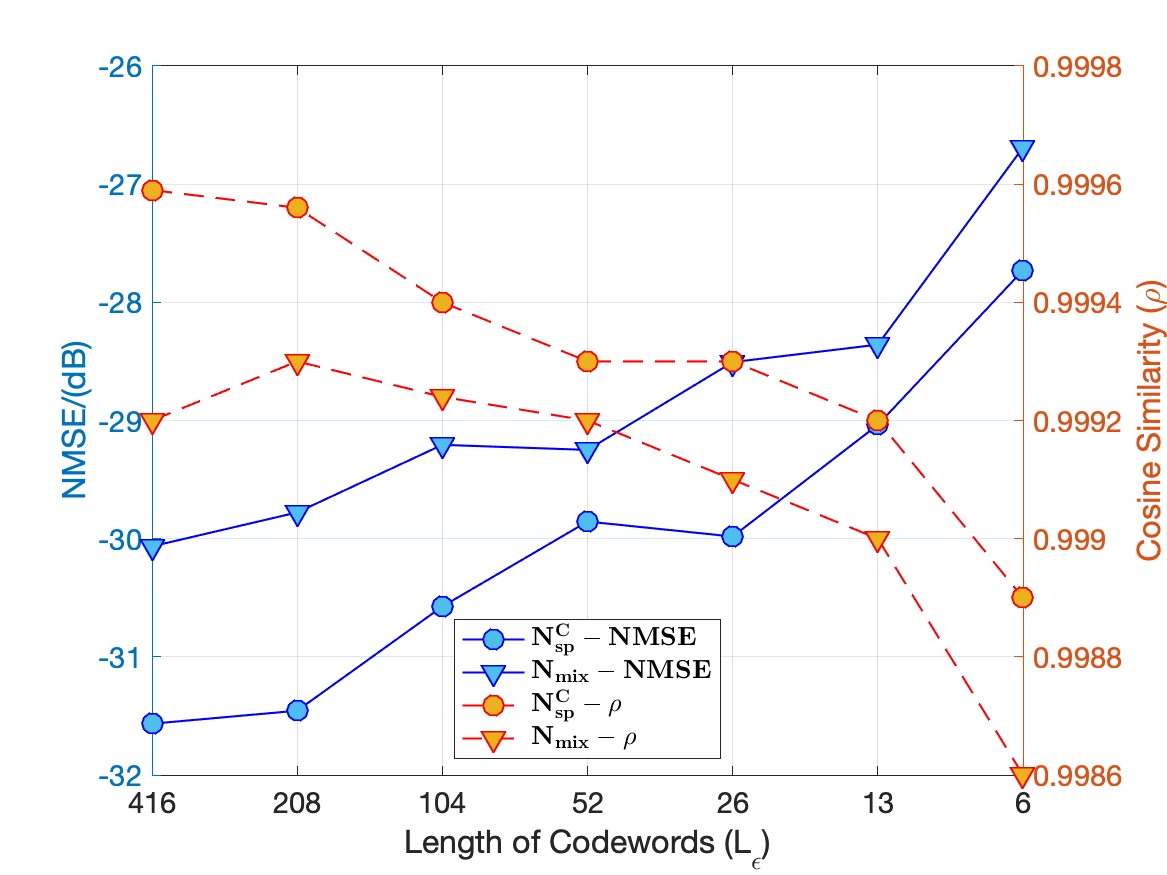}}
	\hspace{0pt}
	\subfloat[CDL-D: $\widehat{\mathbf{V}}$]{
		\label{vNMSErho_D}
		\includegraphics[width=1.60 in]{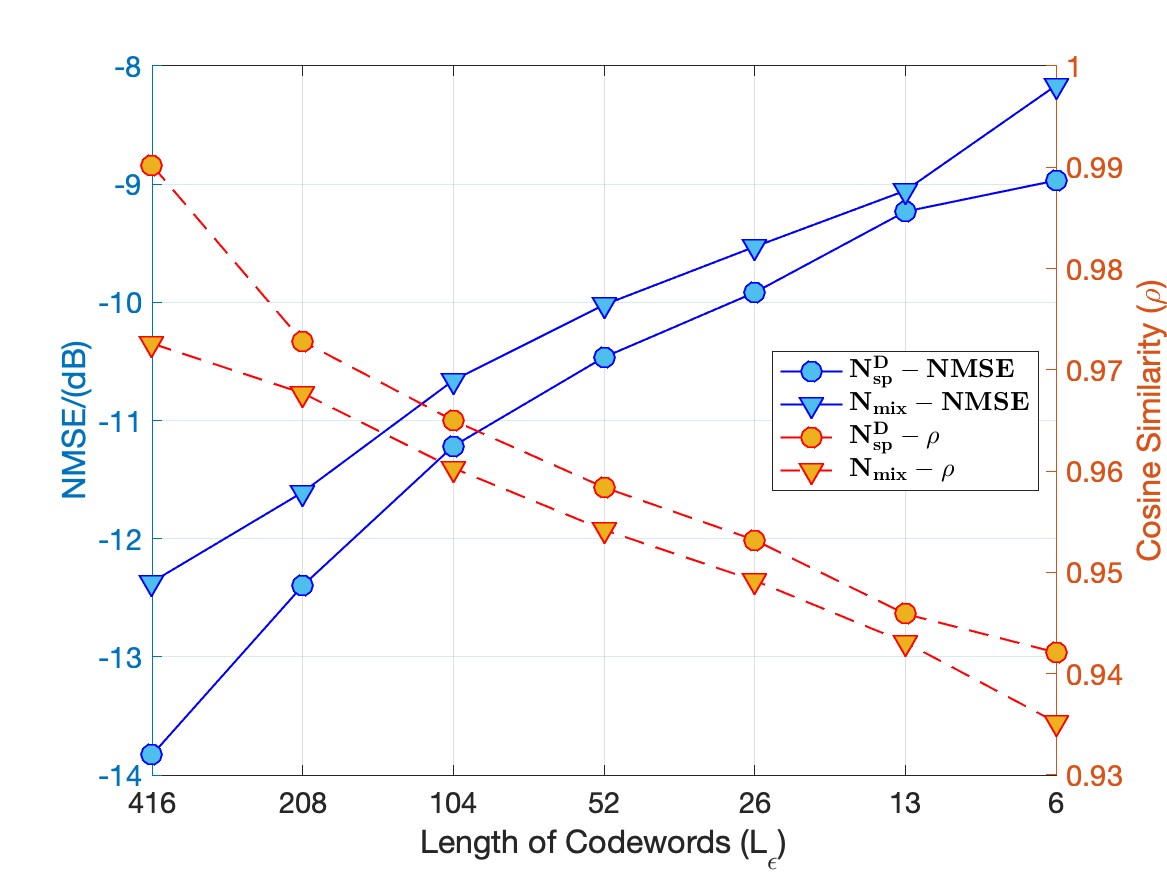}}
	\hspace{0pt}
	\subfloat[CDL-D: $\widehat{\mathbf{S}}$]{
		\label{sNMSErho_D}
		\includegraphics[width=1.60 in]{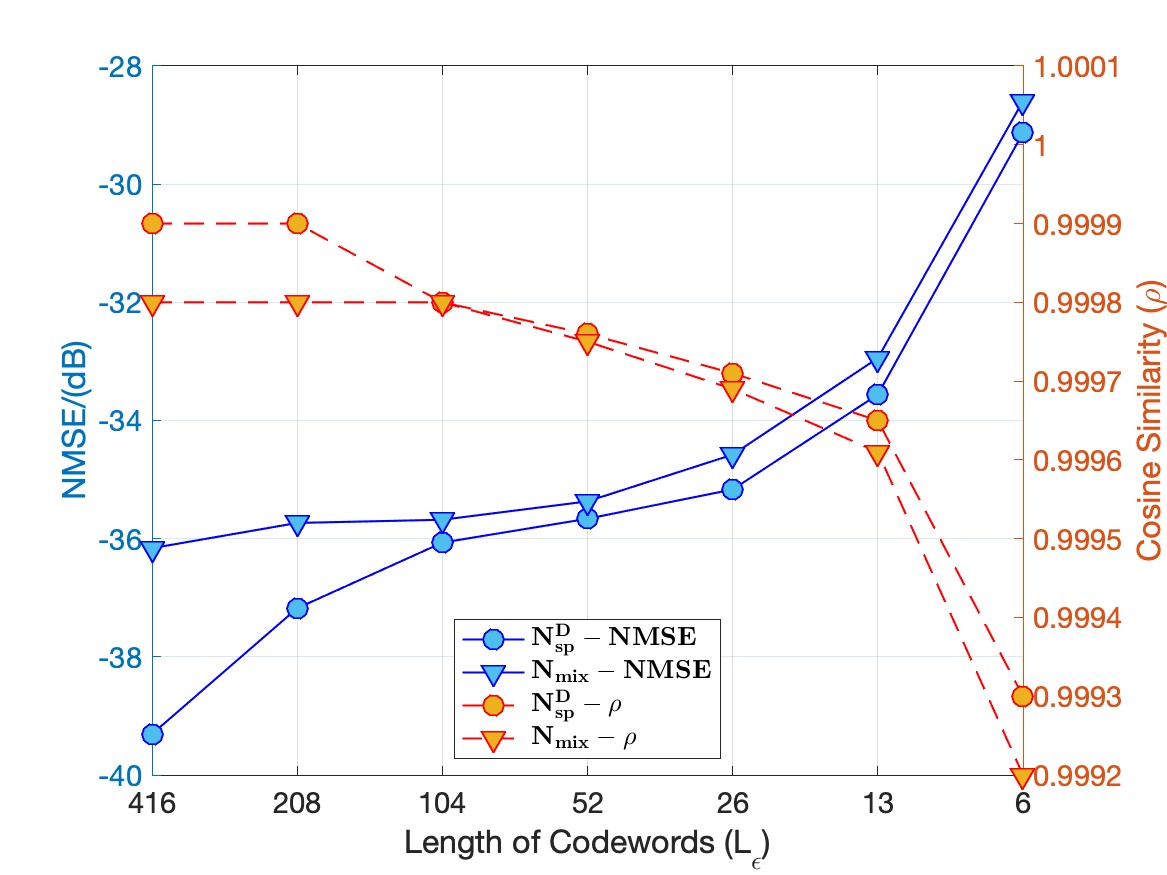}}
	\hspace{0pt}
	\subfloat[CDL-E: $\widehat{\mathbf{V}}$]{
		\label{vNMSErho_E}
		\includegraphics[width=1.60 in]{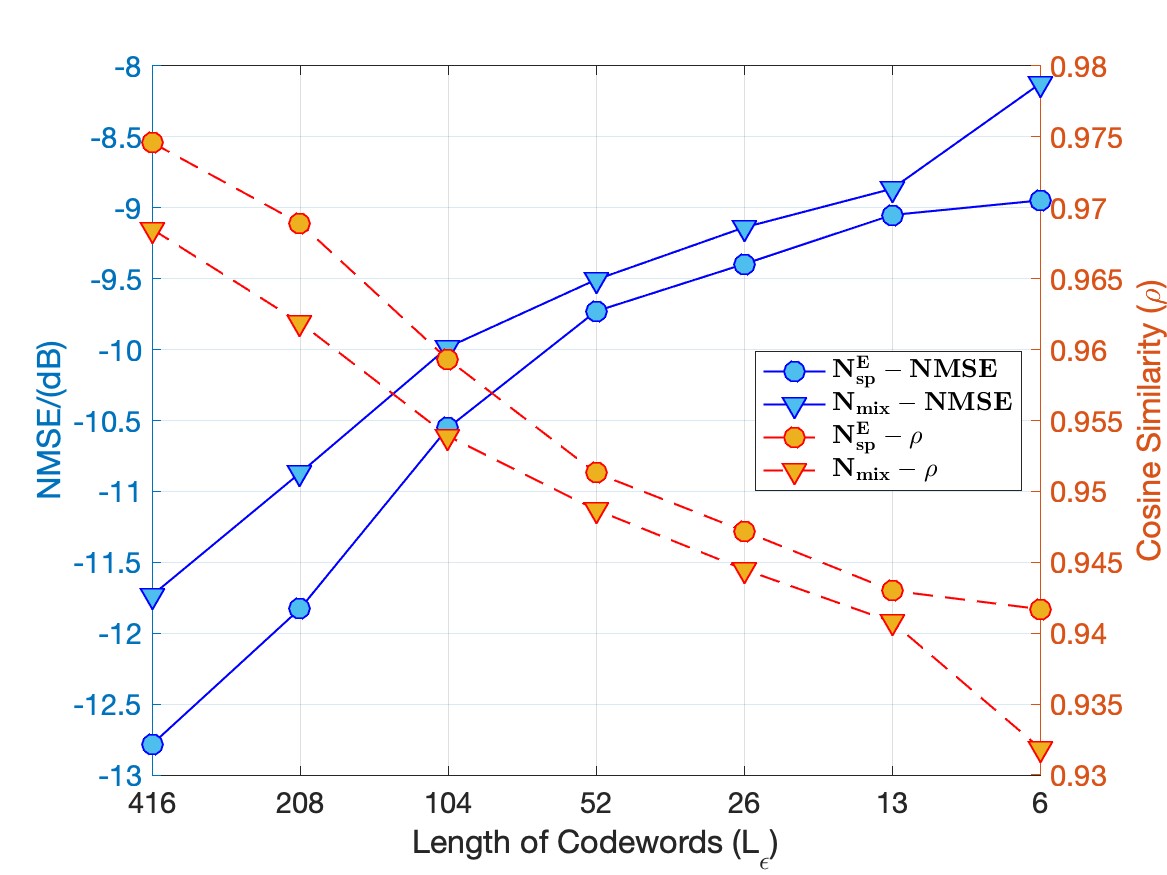}}
	\hspace{0pt}
	\subfloat[CDL-E: $\widehat{\mathbf{S}}$]{
		\label{sNMSErho_E}
		\includegraphics[width=1.60 in]{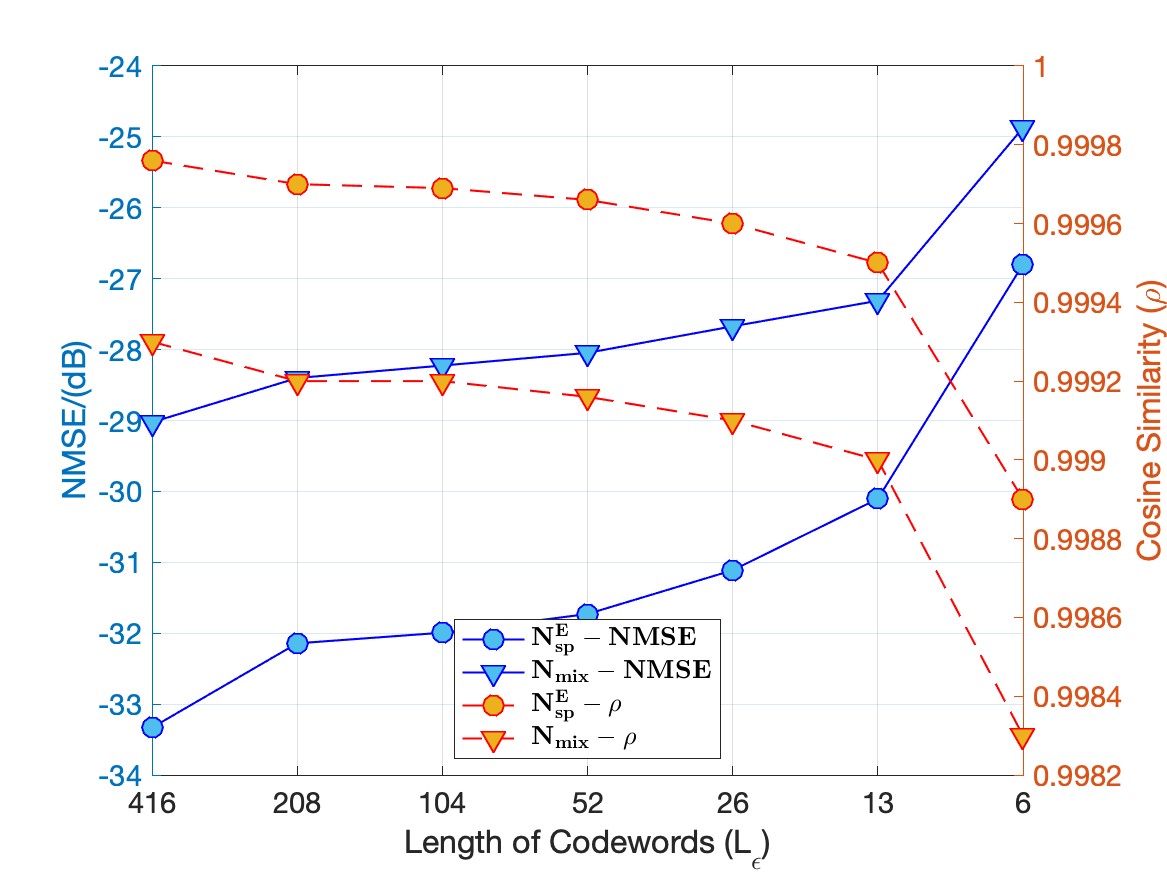}}
	\hspace{0pt}
	\caption{Illustration of superiority experimental results. The x-axis represents different lengths of codewords.  The blue and red y-axes  denotes NMSE (dB) and cosine similarity ($\rho$).}
	\label{super}
\end{figure}

\subsection{Superiority Analysis} \label{Superioity Analysis}
In this subsection, we hope to prove the superior performance of the proposed architecture. As is mentioned in the previous section, we assign a channel identification module before  encoding $\mathbf{V}$ and $\mathbf{S}$. Therefore, this part compares the  performance of specific networks $\mathbb{N}_{sp}^{*}$ trained with large datasets $\mathbb{D}_{sp}^{*}$ and general network $\mathbb{N}_{mix}$ trained with mixed datasets $\mathbb{D}_{mix}$. In order to completely verify the superiority of the proposed architecture, we have checked five  CDL channels defined by 3GPP. The testing objects are $\left\lbrace \mathbb{N}_{sp}^{A}, \mathbb{N}_{sp}^{B}, \mathbb{N}_{sp}^{C}, \mathbb{N}_{sp}^{D}, \mathbb{N}_{sp}^{E} \right\rbrace$, and the baseline is set as $\left\lbrace \mathbb{N}_{mix} \right\rbrace$.  For each experiment, we give four evaluation indexes defined in \textbf{Section \ref{performance eva}}, including  $NMSE(\mathbf{V},\widehat{\mathbf{V}}),NMSE(\mathbf{S},\widehat{\mathbf{S}}),\rho(\mathbf{V},\widehat{\mathbf{V}})$ and $\rho(\mathbf{S},\widehat{\mathbf{S}})$.

The experimental results can be seen in Fig. \ref{super}. Each channel type has been tested and verified by 7 different compression ratios mentioned in \textbf{Section \ref{NNsetting}}. The  horizontally adjacent figures respectively show the reconstruction performance curves of $\widehat{\mathbf{V}}$ and $\widehat{\mathbf{S}}$ under the same channel, e.g. Fig. \ref{vNMSErho_A} and Fig. \ref{sNMSErho_A} show the performance under CDL-A channel environment. In addition, the vertically adjacent figures show the performance  of the same feedback information with different channel types, e.g. Fig \ref{vNMSErho_A} and Fig. \ref{vNMSErho_B} show the performance with different channels. Each figure has two y-axes of different scales, corresponding to $NMSE (dB)$ and $\rho$ respectively.  In addition, there are four performance curves in each figure, among of which the solid blue lines show $NMSE$ performance, the dotted red lines represent $\rho$ performance, the circle marked lines represent specific network $\mathbb{N}_{sp}^{*}$, and the triangle marked lines represent general network $\mathbb{N}_{mix}$.

As can be seen from Fig. \ref{super}, the two blue solid lines of each figure show an upward trend, while the two red dotted lines show a downward trend. This result is the same as the analysis in \textbf{Section \ref{performance eva}}. The performance of reconstruction is getting worse with the decrease of $L_{\varepsilon}$, i.e., $NMSE (dB)$ tends to be larger and $\rho$ deviates from 1. What's more,  all the blue lines with circle mark are lower than the triangle mark, and all the red lines with circle mark are higher than the triangle mark.   These results mean that the performance of specific network $\mathbb{N}_{sp}^{*}$ is better than general network $\mathbb{N}_{mix}$. It can also be found that two lines with the same color are almost parallel, which shows that the performance improvement of  $\mathbb{N}_{sp}^{*}$ is relatively stable compared with $\mathbb{N}_{mix}$, and is less affected by $L_{\varepsilon}$. In summary, we can conclude that the channel identification module is necessary, and the performance of feedback and reconstruction can be improved by selecting a specific network $\mathbb{N}_{sp}^{*}$.

\subsection{Robustness Analysis} \label{Robustness Analysis}
\begin{figure}[htbp]
	\centering
	\subfloat[CDL-A: $\widehat{\mathbf{V}}$]{
		\label{vemev_A}
		\includegraphics[width=1.6 in]{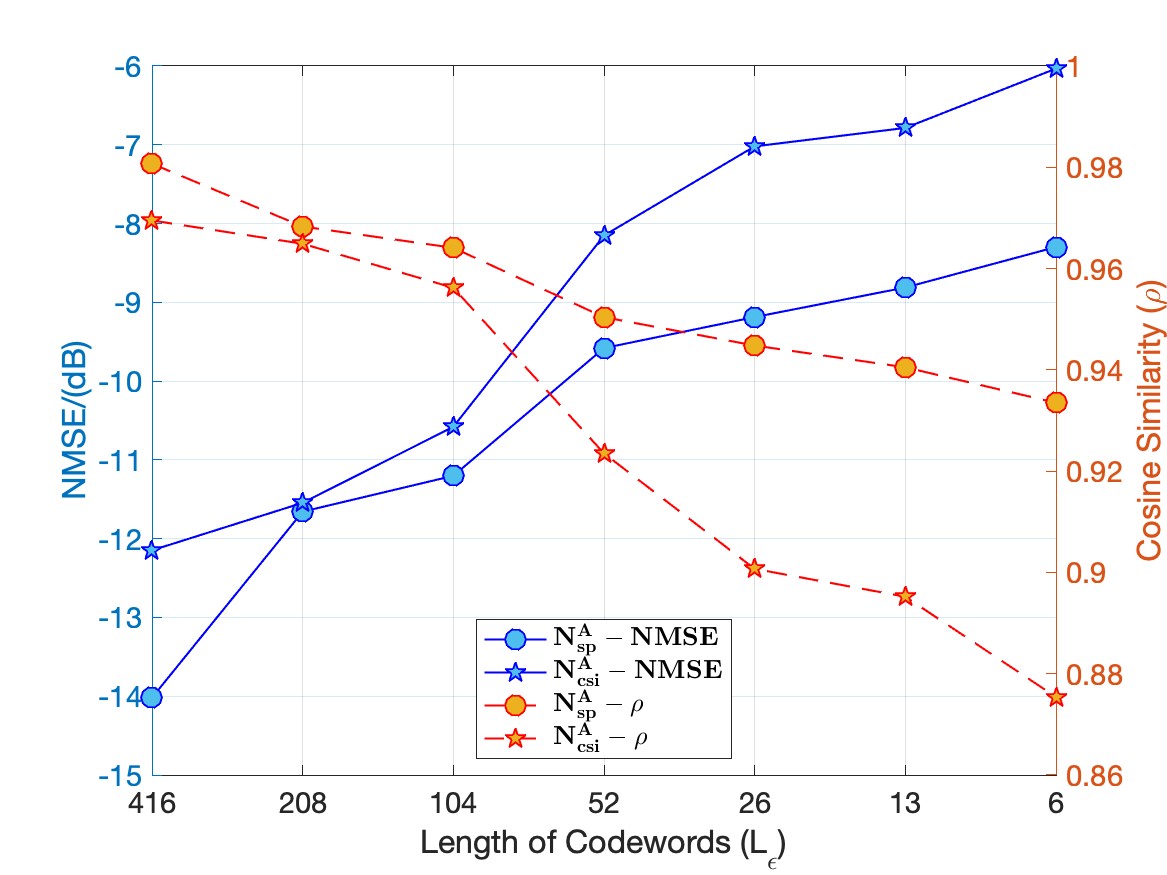}}
	\hspace{0pt}
	\subfloat[CDL-A: $\widehat{\mathbf{S}}$]{
		\label{semev_A}
		\includegraphics[width=1.6 in]{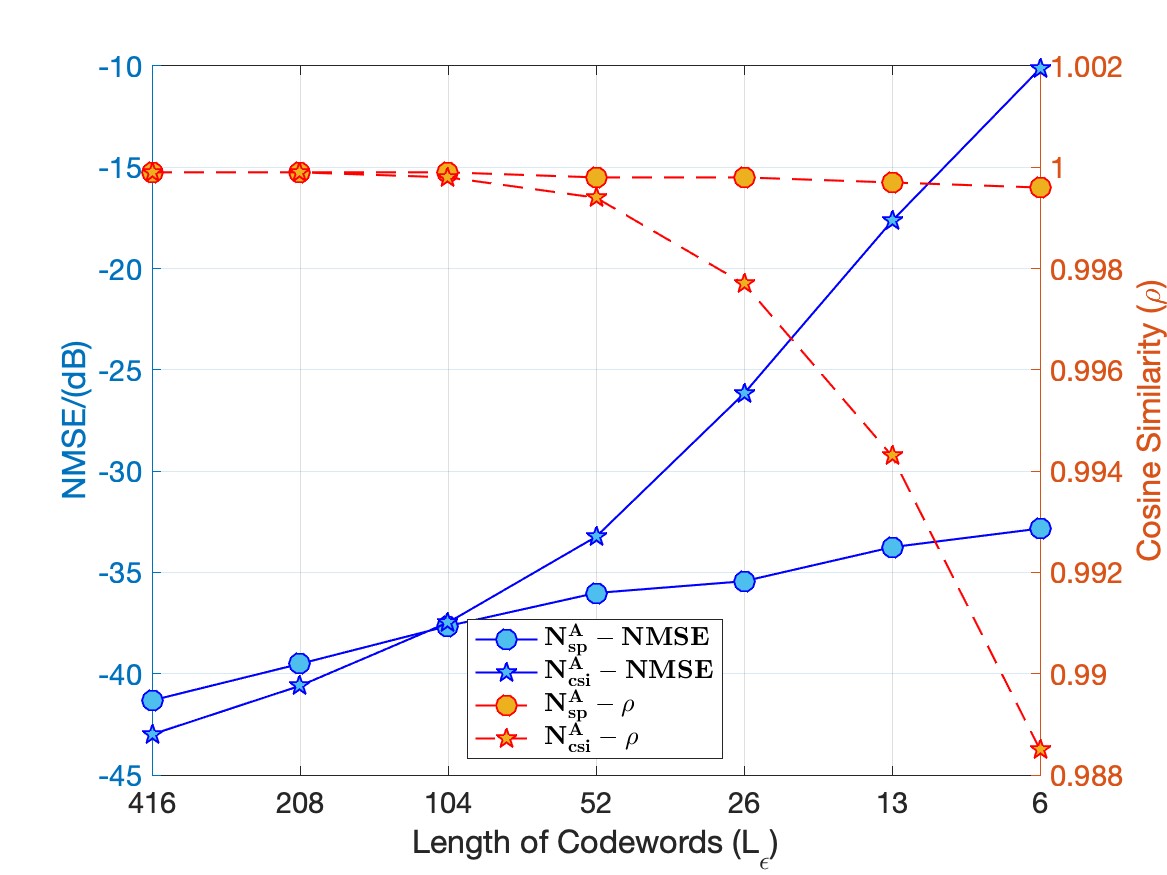}}
	\hspace{0pt}
	\subfloat[CDL-B: $\widehat{\mathbf{V}}$]{
		\label{vemev_B}
		\includegraphics[width=1.6 in]{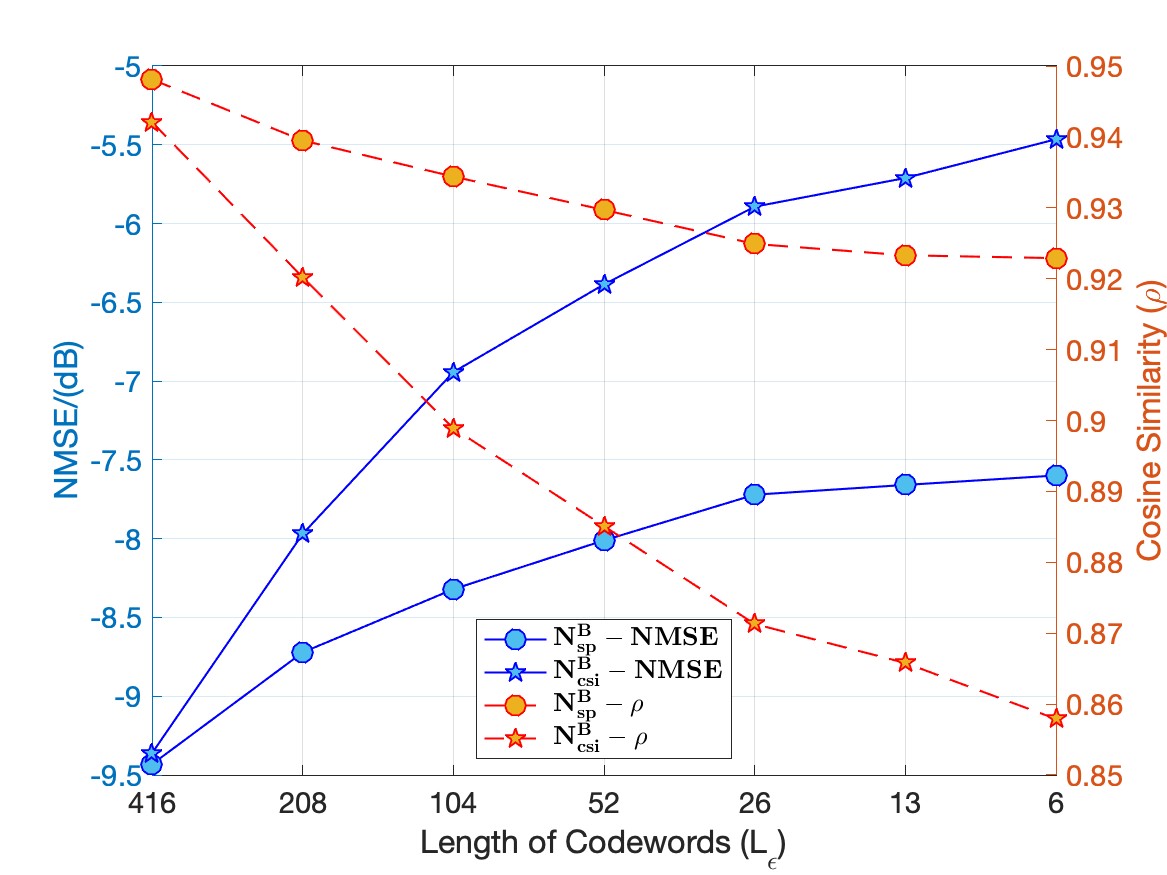}}
	\hspace{0pt}
	\subfloat[CDL-B: $\widehat{\mathbf{S}}$]{
		\label{semev_B}
		\includegraphics[width=1.6 in]{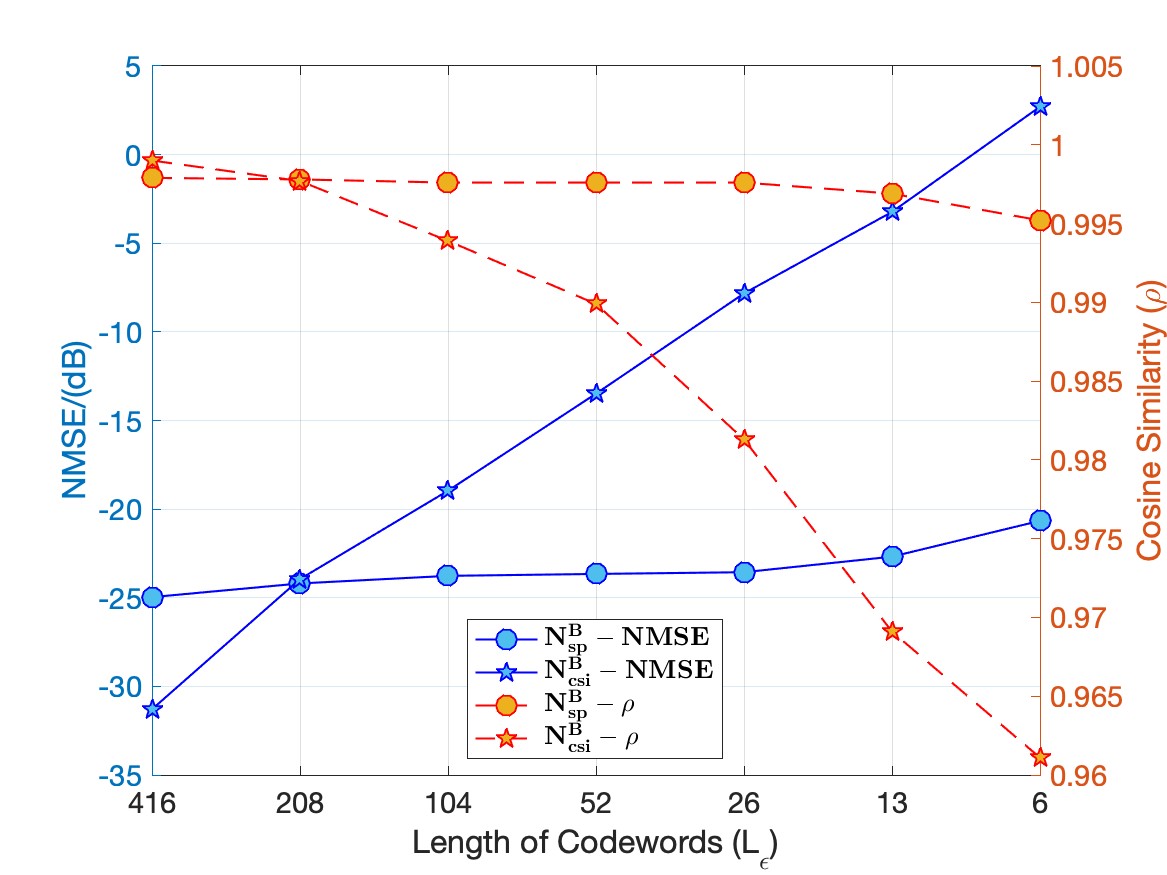}}
	\hspace{0pt}
	\subfloat[CDL-C: $\widehat{\mathbf{V}}$]{
		\label{vemev_C}
		\includegraphics[width=1.6 in]{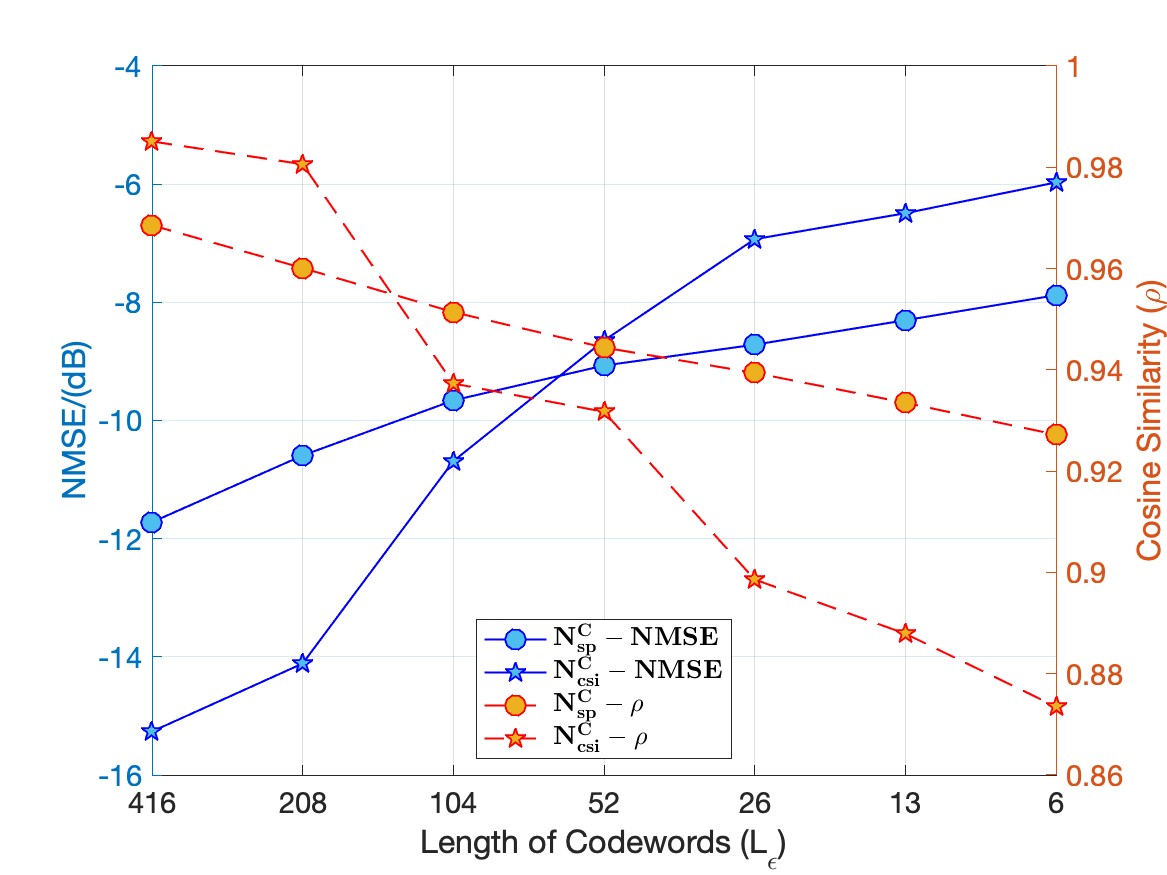}}
	\hspace{0pt}
	\subfloat[CDL-C: $\widehat{\mathbf{S}}$]{
		\label{semev_C}
		\includegraphics[width=1.6 in]{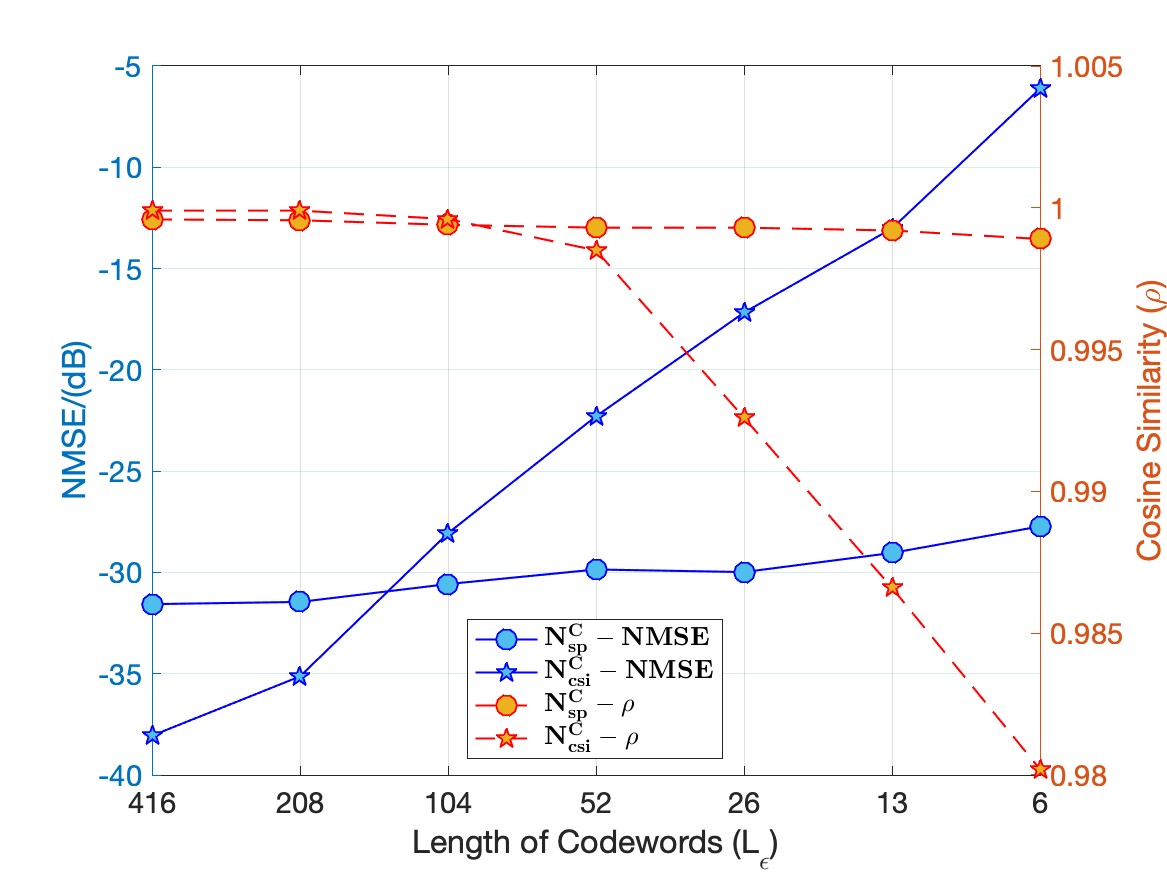}}
	\hspace{0pt}
	\subfloat[CDL-D: $\widehat{\mathbf{V}}$]{
		\label{vemev_D}
		\includegraphics[width=1.6 in]{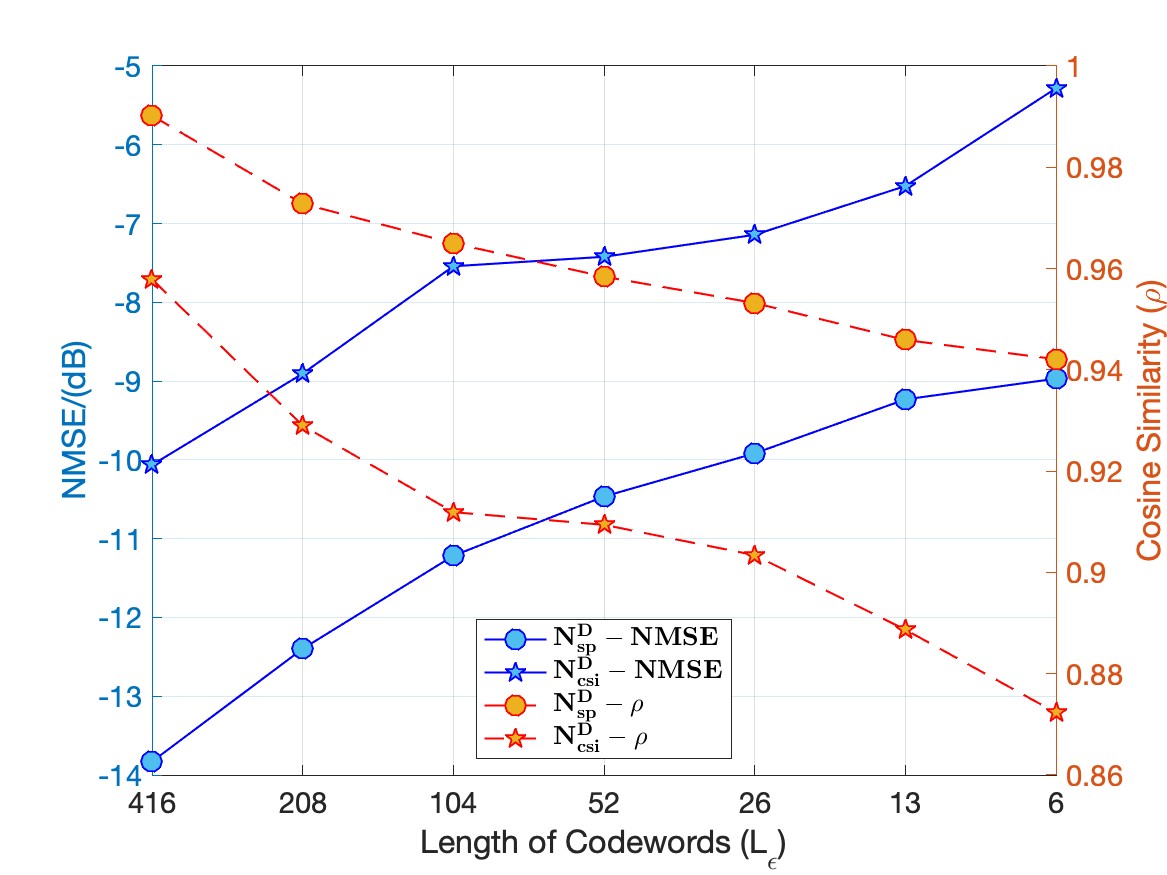}}
	\hspace{0pt}
	\subfloat[CDL-D: $\widehat{\mathbf{S}}$]{
		\label{semev_D}
		\includegraphics[width=1.6 in]{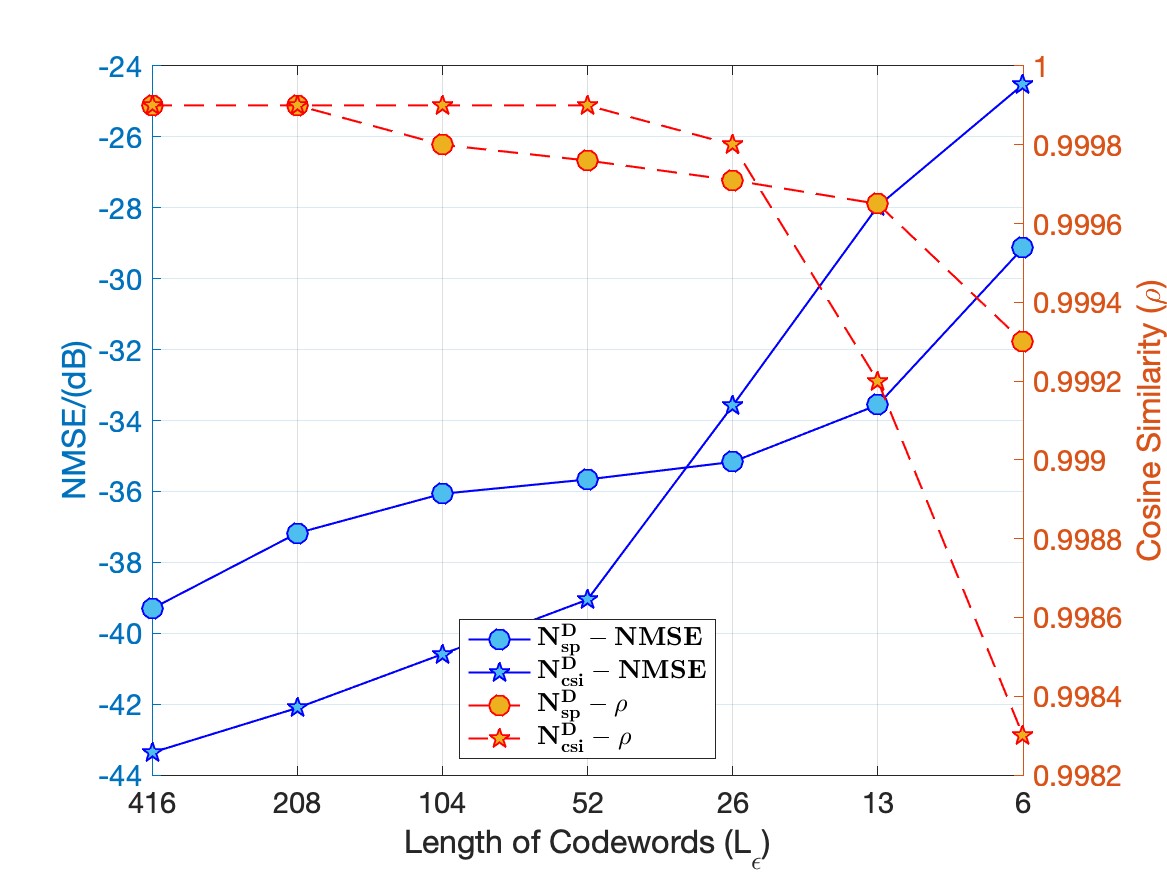}}
	\hspace{0pt}
	\subfloat[CDL-E: $\widehat{\mathbf{V}}$]{
		\label{vemev_E}
		\includegraphics[width=1.6 in]{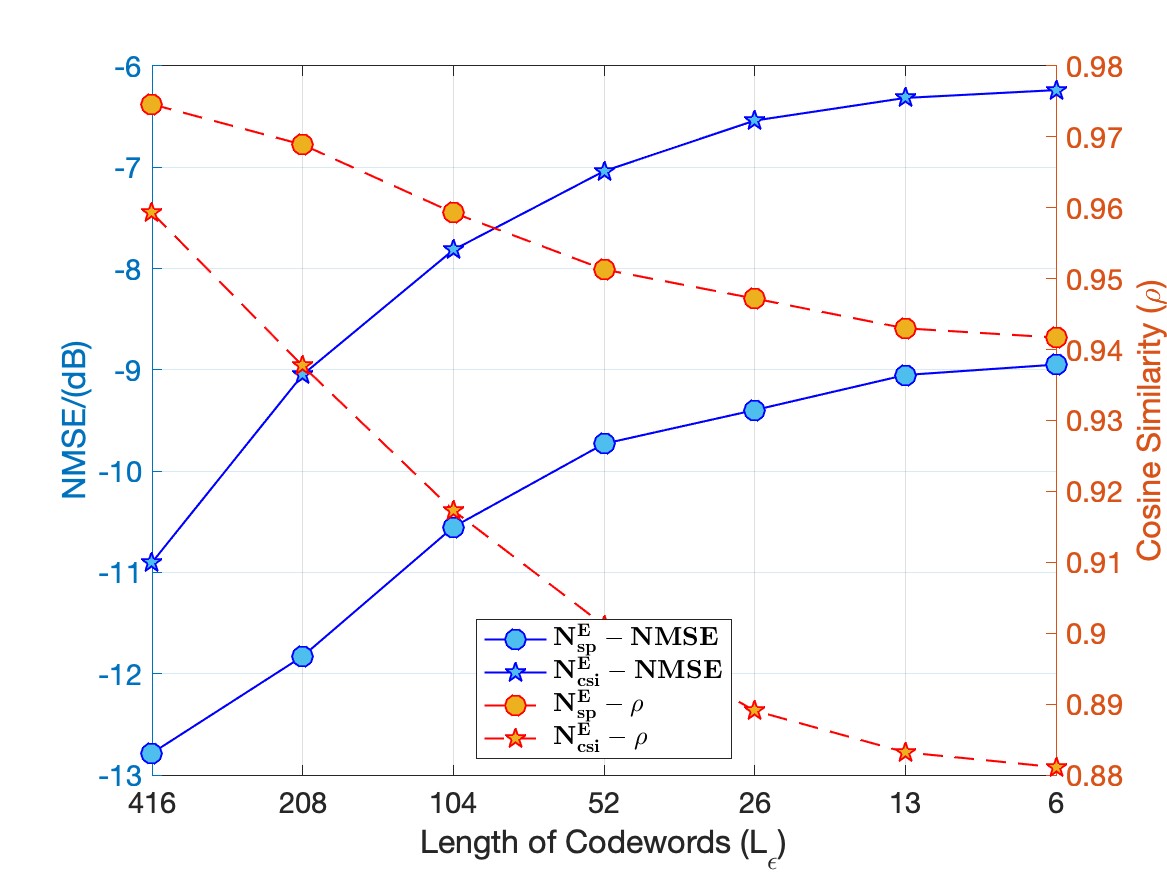}}
	\hspace{0pt}
	\subfloat[CDL-E: $\widehat{\mathbf{S}}$]{
		\label{semev_E}
		\includegraphics[width=1.6 in]{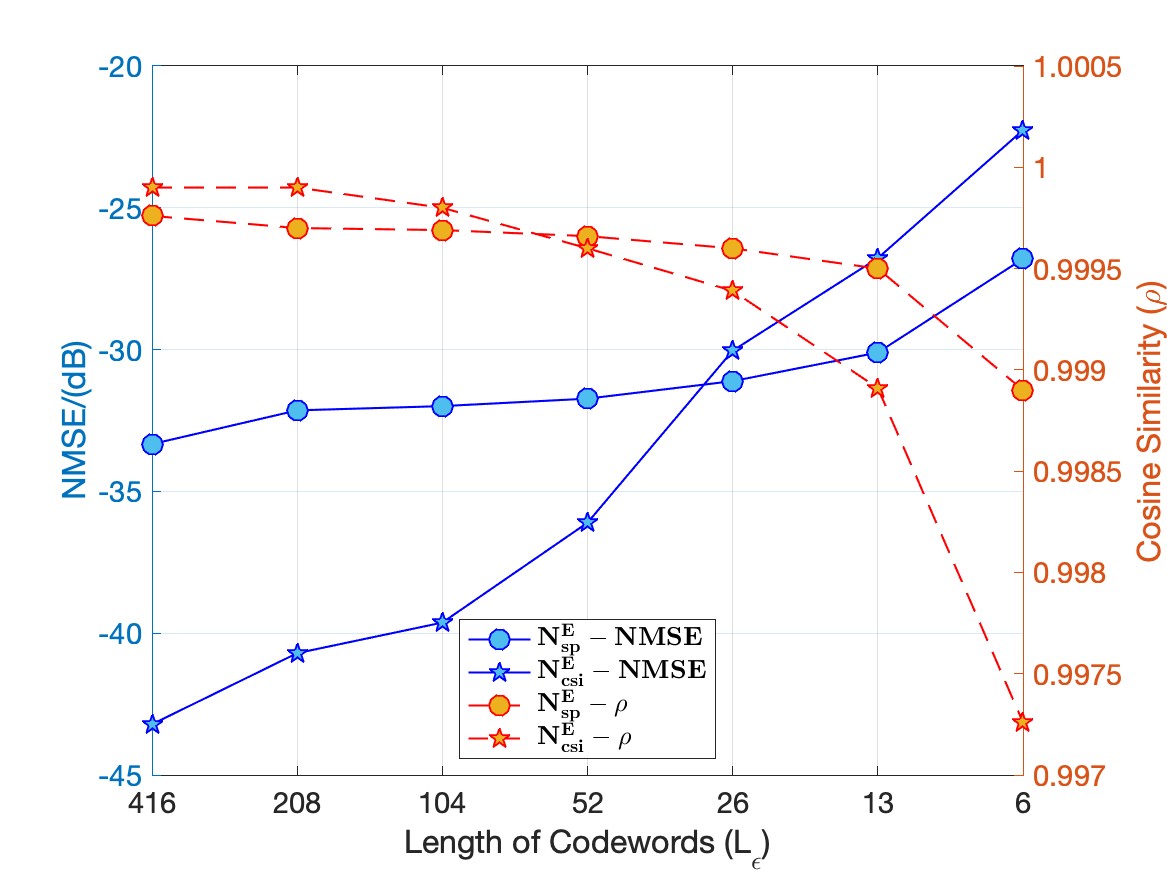}}
	\hspace{0pt}
	\caption{Illustration of robustness experimental results. The x-axis represents different lengths of codewords.  The blue and red y-axes are  NMSE (dB) and cosine similarity ($\rho$).}
	\label{robustness}
\end{figure}
\begin{table*}[htbp]
	\centering
	\caption{The numerical results of simulation experiments carried out and discussed in section \ref{simulation results}. }
	\label{numerical results}
	\begin{tabular}{lcccccccccccc}
		\toprule
		\multicolumn{1}{c}{}                          &                                                   &   & \multicolumn{2}{c}{\bf CDL-A} & \multicolumn{2}{c}{\bf CDL-B} & \multicolumn{2}{c}{\bf CDL-C} & \multicolumn{2}{c}{\bf CDL-D} & \multicolumn{2}{c}{\bf CDL-E}             \\ \hline
		\multicolumn{1}{c}{}                          &                                                   &   &  \scriptsize NMSE (dB)       & $\rho$          & \scriptsize NMSE (dB)          & $\rho$           &  \scriptsize NMSE (dB)         & $\rho$            & \scriptsize NMSE (dB)         & $\rho$            & \scriptsize NMSE (dB)                      & $\rho$        \\ \hline
		\multicolumn{1}{c|}{\multirow{6}{*}{\begin{tabular}[c]{@{}c@{}}$L_{\varepsilon}=416$\\ $\beta_{h}=16$\\$\beta_{emev}=256$\end{tabular}}}   & \multicolumn{1}{c|}{\multirow{2}{*}{\begin{tabular}[c]{@{}c@{}}\textcolor{red}{$\mathbb{N}_{sp}^*$}\\ \textcolor{red}{(proposed)}\end{tabular}}}     & $\widehat{\mathbf{V}}$  & --14.018      & 0.9807     & --9.432       & 0.9481     & --11.721     & 0.9686     & --13.830     & 0.9902      & --12.784                     & 0.9746  \\
		\multicolumn{1}{c|}{}                         & \multicolumn{1}{c|}{}                             & $\widehat{\mathbf{S}}$ & --41.318      & 0.9999     & --24.964      & 0.9979     & --31.566     & 0.9996     & --39.307     & 0.9999      & --33.332                     & 0.9998 \\
		\multicolumn{1}{c|}{}                         & \multicolumn{1}{c|}{\multirow{2}{*}{$\mathbb{N}_{csi}^*$}}      & $\widehat{\mathbf{V}}$ & --12.148      & 0.9695     & --9.362       & 0.9421     & --15.264     & 0.9851      & --10.064     & 0.9579      & --10.902                     & 0.9593  \\
		\multicolumn{1}{c|}{}                         & \multicolumn{1}{c|}{}                             & $\widehat{\mathbf{S}}$  & --42.999      & 0.9999     & --31.292      & 0.9990     & --38.057     & 0.9999      & --43.368     & 0.9999      & --43.211                     & 0.9999  \\
		\multicolumn{1}{c|}{}                         & \multicolumn{1}{c|}{\multirow{2}{*}{$\mathbb{N}_{mix}$}}     & $\widehat{\mathbf{V}}$  & --11.895      & 0.9694     & --9.196       & 0.9448     & --10.206     & 0.9558      & --12.367     & 0.9726      & --11.733                     & 0.9685  \\
		\multicolumn{1}{c|}{}                         & \multicolumn{1}{c|}{}                             & $\widehat{\mathbf{S}}$  & --38.764      & 0.9999     & --22.736      & 0.9970     & --30.065     & 0.9993      & --36.161     & 0.9998      & --29.025                     & 0.9993  \\ \hline
		\multicolumn{1}{c|}{\multirow{6}{*}{\begin{tabular}[c]{@{}c@{}}$L_{\varepsilon}=208$\\ $\beta_{h}=32$\\$\beta_{emev}=512$\end{tabular}}}   & \multicolumn{1}{c|}{\multirow{2}{*}{\begin{tabular}[c]{@{}c@{}}\textcolor{red}{$\mathbb{N}_{sp}^*$}\\ \textcolor{red}{(proposed)}\end{tabular}}}     & $\widehat{\mathbf{V}}$  & --11.657      & 0.9683     & --8.722       & 0.9395     & --10.592     & 0.9601      & --12.398     & 0.9728      & --11.828                     & 0.9689  \\
		\multicolumn{1}{c|}{}                         & \multicolumn{1}{c|}{}                             & $\widehat{\mathbf{S}}$  & --39.512      & 0.9999     & --24.193      & 0.9978     & --31.456     & 0.9996     & --37.178     & 0.9999      & --32.143                     & 0.9997  \\
		\multicolumn{1}{c|}{}                         & \multicolumn{1}{c|}{\multirow{2}{*}{$\mathbb{N}_{csi}^*$}}      & $\widehat{\mathbf{V}}$  & --11.545      & 0.9649     & --7.970       & 0.9202     & --14.113     & 0.9806      & --8.905      & 0.9290      & --9.048                      & 0.9377  \\
		\multicolumn{1}{c|}{}                         & \multicolumn{1}{c|}{}                             & $\widehat{\mathbf{S}}$  & --40.598      & 0.9999     & --23.947      & 0.9977     & --35.133     & 0.9999      & --42.110     & 0.9999      & --40.704                     & 0.9999  \\
		\multicolumn{1}{c|}{}                         & \multicolumn{1}{c|}{\multirow{2}{*}{$\mathbb{N}_{mix}$}}     & $\widehat{\mathbf{V}}$  & --10.909      & 0.9623     & --8.580       & 0.9374     & --9.536      & 0.9491      & --11.607     & 0.9677      & --10.870                     & 0.9619  \\
		\multicolumn{1}{c|}{}                         & \multicolumn{1}{c|}{}                             & $\widehat{\mathbf{S}}$  & --38.111      & 0.9999     & --22.082      & 0.9965     & --29.778     & 0.9992     & --35.734     & 0.9998      & --28.401                     & 0.9992  \\ \hline
		\multicolumn{1}{c|}{\multirow{6}{*}{\begin{tabular}[c]{@{}c@{}}$L_{\varepsilon}=104$\\ $\beta_{h}=64$\\$\beta_{emev}=1024$\end{tabular}}}   & \multicolumn{1}{c|}{\multirow{2}{*}{\begin{tabular}[c]{@{}c@{}}\textcolor{red}{$\mathbb{N}_{sp}^*$}\\ \textcolor{red}{(proposed)}\end{tabular}}}     & $\widehat{\mathbf{V}}$  & --11.202      & 0.9641     & --8.321       & 0.9344     & --9.660      & 0.9514      & --11.217     & 0.9650      & --10.552 & 0.9593  \\
		\multicolumn{1}{c|}{}                         & \multicolumn{1}{c|}{}                             & $\widehat{\mathbf{S}}$  & --37.645      & 0.9999     & --23.765      & 0.9976     & --30.574     & 0.9994      & --36.067     & 0.9998      & --31.993 & 0.9997 \\
		\multicolumn{1}{c|}{}                         & \multicolumn{1}{c|}{\multirow{2}{*}{$\mathbb{N}_{csi}^*$}}      & $\widehat{\mathbf{V}}$  & --10.58       & 0.9562     & --6.943       & 0.8989     & --10.688     & 0.9373      & --7.5421     & 0.9119      & --7.816                      & 0.9173  \\
		\multicolumn{1}{c|}{}                         & \multicolumn{1}{c|}{}                             & $\widehat{\mathbf{S}}$  & --37.476      & 0.9998     & --18.961      & 0.9939     & --28.065     & 0.9996      & --40.592     & 0.9999      & --39.626                     & 0.9998  \\
		\multicolumn{1}{c|}{}                         & \multicolumn{1}{c|}{\multirow{2}{*}{$\mathbb{N}_{mix}$}}     & $\widehat{\mathbf{V}}$  & --9.998       & 0.9540     & --8.101       & 0.9309     & --8.874      & 0.9416      & --10.661     & 0.9603      & --9.983                      & 0.9539  \\
		\multicolumn{1}{c|}{}                         & \multicolumn{1}{c|}{}                             & $\widehat{\mathbf{S}}$  & --37.643      & 0.9998     & --21.568      & 0.9963     & --29.206     & 0.9992      & --35.680     & 0.9998      & --28.225                     & 0.9992  \\ \hline
		\multicolumn{1}{c|}{\multirow{6}{*}{\begin{tabular}[c]{@{}c@{}}$L_{\varepsilon}=52$\\ $\beta_{h}=128$\\$\beta_{emev}=2048$\end{tabular}}}  & \multicolumn{1}{c|}{\multirow{2}{*}{\begin{tabular}[c]{@{}c@{}}\textcolor{red}{$\mathbb{N}_{sp}^*$}\\ \textcolor{red}{(proposed)}\end{tabular}}}     & $\widehat{\mathbf{V}}$  & --9.581       & 0.9504     & --8.011       & 0.9297     & --9.070      & 0.9444      & --10.465     & 0.9584      & --9.7297                     & 0.9513  \\
		\multicolumn{1}{c|}{}                         & \multicolumn{1}{c|}{}                             & $\widehat{\mathbf{S}}$  & --36.016      & 0.9998     & --23.657      & 0.9976     & --29.855     & 0.9993      & --35.663     & 0.9998     & --31.729                    & 0.9997 \\
		\multicolumn{1}{c|}{}                         & \multicolumn{1}{c|}{\multirow{2}{*}{$\mathbb{N}_{csi}^*$}}      & $\widehat{\mathbf{V}}$  & --8.152       & 0.9235     & --6.386       & 0.8851     & --8.646      & 0.9317     & --7.422      & 0.9094      & --7.038                      & 0.9011  \\
		\multicolumn{1}{c|}{}                         & \multicolumn{1}{c|}{}                             & $\widehat{\mathbf{S}}$  & --33.241      & 0.9994     & --13.486      & 0.9899     & --22.293     & 0.9985      & --39.054     & 0.9999      & --36.104                     & 0.9996  \\
		\multicolumn{1}{c|}{}                         & \multicolumn{1}{c|}{\multirow{2}{*}{$\mathbb{N}_{mix}$}} & $\widehat{\mathbf{V}}$  & --9.394       & 0.9475     & --7.877       & 0.9275     & --8.499      & 0.9364      & --10.021     & 0.9542      & --9.507                      & 0.9487  \\
		\multicolumn{1}{c|}{}                         & \multicolumn{1}{c|}{}                             & $\widehat{\mathbf{S}}$  & --36.001      & 0.9998     & --21.547      & 0.9962     & --29.249     & 0.9992      & --35.369     & 0.9997     & --28.046                     & 0.9992 \\ \hline
		\multicolumn{1}{c|}{\multirow{6}{*}{\begin{tabular}[c]{@{}c@{}}$L_{\varepsilon}=26$\\ $\beta_{h}=256$\\$\beta_{emev}=4096$\end{tabular}}}  & \multicolumn{1}{c|}{\multirow{2}{*}{\begin{tabular}[c]{@{}c@{}}\textcolor{red}{$\mathbb{N}_{sp}^*$}\\ \textcolor{red}{(proposed)}\end{tabular}}}     & $\widehat{\mathbf{V}}$  & --9.187       & 0.9448     & --7.721       & 0.9249     & --8.717      & 0.9394      & --9.917      & 0.9532      & --9.397                      & 0.9472  \\
		\multicolumn{1}{c|}{}                         & \multicolumn{1}{c|}{}                             & $\widehat{\mathbf{S}}$  & --35.434      & 0.9997     & --23.559      & 0.9976     & --29.982     & 0.9993      & --35.167     & 0.9997     & --31.112                     & 0.9996  \\
		\multicolumn{1}{c|}{}                         & \multicolumn{1}{c|}{\multirow{2}{*}{$\mathbb{N}_{csi}^*$}}      & $\widehat{\mathbf{V}}$  & --7.020       & 0.9007     & --5.893       & 0.8713     & --6.931      & 0.8986      & --7.143      & 0.9034      & --6.539                      & 0.8891 \\
		\multicolumn{1}{c|}{}                         & \multicolumn{1}{c|}{}                             & $\widehat{\mathbf{S}}$  & --26.173      & 0.9977     & --7.831       & 0.9813     & --17.173     & 0.9926      & --33.582     & 0.9998      & --30.038                     & 0.9994 \\
		\multicolumn{1}{c|}{}                         & \multicolumn{1}{c|}{\multirow{2}{*}{$\mathbb{N}_{mix}$}} & $\widehat{\mathbf{V}}$  & --8.963       & 0.9425     & --7.683       & 0.9249     & --8.274      & 0.9334      & --9.530      & 0.9492      & --9.139                      & 0.9445  \\
		\multicolumn{1}{c|}{}                         & \multicolumn{1}{c|}{}                             & $\widehat{\mathbf{S}}$  & --35.393      & 0.9998     & --20.877      & 0.9956     & --28.506     & 0.9991     & --34.578     & 0.9997     & --27.672                     & 0.9991   \\ \hline
		\multicolumn{1}{c|}{\multirow{6}{*}{\begin{tabular}[c]{@{}c@{}}$L_{\varepsilon}=13$\\ $\beta_{h}=512$\\$\beta_{emev}=8192$\end{tabular}}}  & \multicolumn{1}{c|}{\multirow{2}{*}{\begin{tabular}[c]{@{}c@{}}\textcolor{red}{$\mathbb{N}_{sp}^*$}\\ \textcolor{red}{(proposed)}\end{tabular}}}     & $\widehat{\mathbf{V}}$  & --8.812       & 0.9405     & --7.659       & 0.9233     & --8.303      & 0.9335      & --9.232      & 0.9459      & --9.051                      & 0.9430  \\
		\multicolumn{1}{c|}{}                         & \multicolumn{1}{c|}{}                             & $\widehat{\mathbf{S}}$  & --33.754      & 0.9997     & --22.682      & 0.9969     & --29.037     & 0.9992      & --33.561     & 0.9996     & --30.105                     & 0.9995  \\
		\multicolumn{1}{c|}{}                         & \multicolumn{1}{c|}{\multirow{2}{*}{$\mathbb{N}_{csi}^*$}}      & $\widehat{\mathbf{V}}$  & --6.786       & 0.8952     & --5.711       & 0.8658     & --6.492      & 0.8879      & --6.524      & 0.8887      & --6.315                     & 0.8832 \\
		\multicolumn{1}{c|}{}                         & \multicolumn{1}{c|}{}                             & $\widehat{\mathbf{S}}$  & --17.651      & 0.9943     & --3.233       & 0.9691     & --13.018     & 0.9866      & --27.956     & 0.9992      & --26.774                    & 0.9989 \\
		\multicolumn{1}{c|}{}                         & \multicolumn{1}{c|}{\multirow{2}{*}{$\mathbb{N}_{mix}$}} & $\widehat{\mathbf{V}}$  & --8.356       & 0.9342     & --7.628       & 0.9231     & --7.994      & 0.9288      & --9.055      & 0.9431      & --8.866                      & 0.9408  \\
		\multicolumn{1}{c|}{}                         & \multicolumn{1}{c|}{}                             & $\widehat{\mathbf{S}}$  & --33.717      & 0.9996     & --20.771      & 0.9954     & --28.358     & 0.9990      & --32.944     & 0.9996     & --27.311                     & 0.9990  \\ \hline
		\multicolumn{1}{c|}{\multirow{6}{*}{\begin{tabular}[c]{@{}c@{}}$L_{\varepsilon}=6$\\ $\beta_{h}=1024$\\$\beta_{emev}=16384$\end{tabular}}} & \multicolumn{1}{c|}{\multirow{2}{*}{\begin{tabular}[c]{@{}c@{}}\textcolor{red}{$\mathbb{N}_{sp}^*$}\\ \textcolor{red}{(proposed)}\end{tabular}}}     & $\widehat{\mathbf{V}}$  & --8.298       & 0.9335     & --7.599       & 0.9229     & --7.879      & 0.9272      & --8.969      & 0.9421      & --8.948                      & 0.9417  \\
		\multicolumn{1}{c|}{}                         & \multicolumn{1}{c|}{}                             & $\widehat{\mathbf{S}}$  & --32.833      & 0.9996     & --20.660      & 0.9952     & --27.729     & 0.9989      & --29.132     & 0.9993      & --26.799                     & 0.9989  \\
		\multicolumn{1}{c|}{}                         & \multicolumn{1}{c|}{\multirow{2}{*}{$\mathbb{N}_{csi}^*$}}      & $\widehat{\mathbf{V}}$  & --6.033       & 0.8754     & --5.465       & 0.8579     & --5.971      & 0.8735      & --5.283      & 0.8723      & --6.238                     & 0.8811  \\
		\multicolumn{1}{c|}{}                         & \multicolumn{1}{c|}{}                             & $\widehat{\mathbf{S}}$  & --10.133      & 0.9885     & 2.702        & 0.9611     & --6.120      & 0.9802      & --24.523     & 0.9983      & --22.261                    & 0.9973 \\
		\multicolumn{1}{c|}{}                         & \multicolumn{1}{c|}{\multirow{2}{*}{$\mathbb{N}_{mix}$}} & $\widehat{\mathbf{V}}$  & --7.633       & 0.9233     & --7.355       & 0.9181     & --7.538      & 0.9215      & --8.165      & 0.9352      & --8.125                      & 0.9319  \\
		\multicolumn{1}{c|}{}                         & \multicolumn{1}{c|}{}                             & $\widehat{\mathbf{S}}$  & --32.357      & 0.9997     & --19.231      & 0.9936     & --26.690     & 0.9986      & --28.593     & 0.9992      & --24.871                     & 0.9983 \\
		\bottomrule
	\end{tabular}
\end{table*}
This subsection proves the robustness of  proposed architecture by comparing EMEVNet with the classical CSI feedback scheme CsiNet \cite{wen_csinet}.   In order to narrate conveniently,  $\mathbb{N}_{csi}^{*}$ is defined as the classical CsiNet framework. We distinguish different channels by  superscripts. e.g. $\mathbb{N}_{csi}^{A}$ represents  CsiNet based on CDL-A channel. For fairness, specific datasets with $50,000$ samples  are used in this subsection, and $\mathbb{N}_{sp}^*$ and $\mathbb{N}_{csi}^{*}$ are trained respectively. Similarly, the experiments in this part also verify 5 channel types and 7 different compression ratios set as \textbf{Section \ref{NNsetting}}. It should be noted that the comparison experiment is based on the same length of codewords, that is, the performance is compared with the same feedback overhead. Considering that CsiNet encodes and feedbacks the channel matrix $\mathbf{H}$, we keep the same feedback overhead and carry out SVD transform on the decoded $\widehat{\mathbf{H}}$ at the BS. The testing objects are $\left\lbrace \mathbb{N}_{sp}^{A}, \mathbb{N}_{sp}^{B}, \mathbb{N}_{sp}^{C}, \mathbb{N}_{sp}^{D}, \mathbb{N}_{sp}^{E} \right\rbrace$, and the baselines are set as $\left\lbrace \mathbb{N}_{csi}^{A}, \mathbb{N}_{csi}^{B}, \mathbb{N}_{csi}^{C}, \mathbb{N}_{csi}^{D}, \mathbb{N}_{csi}^{E}\right\rbrace$.

The experimental results can be seen in Fig. \ref{robustness}. The arrangement of different subfigures  is the same as that in Fig. \ref{super}. Each figure has two y-axes of different scales, corresponding to $NMSE (dB)$ and $\rho$ respectively.  In addition, there are four performance curves in each figure, and the solid blue lines show $NMSE$ performance, the dotted red lines represent $\rho$ performance, the circle marked lines represent EMEVNet  $\mathbb{N}_{sp}^{*}$, and the pentagram marked lines represent the baseline, i.e. CsiNet $\mathbb{N}_{csi}^{*}$.   The visualization results   show that the lines marked by the pentagram have  larger gradients. This phenomenon is more obvious in the performance curve of reconstructed $\widehat{\mathbf{S}}$ at the BS.  This can be explained as the baseline $\mathbb{N}_{csi}^{*}$ is greatly affected by the length of codewords $L_{\varepsilon}$.  As for eigenmatrix $\widehat{\mathbf{V}}$,  except  Fig. \ref{vemev_C} shows that the baseline is better than EMEVNet with long $L_{\varepsilon}$, the other results all show the performance of EMEVNet is better with all $L_{\varepsilon}$. As for eigenvector $\widehat{\mathbf{S}}$, we can find  baseline has better performance with long $L_{\varepsilon}$ in all channels. For NLOS  channels (CDL-A, CDL-B and CDL-C), when the length of codewords decreases to $L_{\varepsilon}=104$, the performance of EMEVNet becomes better than that of  baseline. While discussing LOS channels (CDL-D and CDL-E), this threshold is relaxed to $L_{\varepsilon}=26$. After analysis, we believe that this is because there exists  line of sight fading path with strong energy in LOS channels, which cause improvement of  baseline. At the same time, we find that the performance  of baseline attenuates seriously in the case of limited $L_{\varepsilon}$. However,  the performance of EMEVNet is relatively stable. Especially in the case of limit $L_{\varepsilon}$, EMEVNet shows much better performance than baseline. To sum up, the proposed EMEVNet  has good robustness and adaptability. It can be applied to larger system compression rate, that is, it occupies smaller feedback codewords, which can effectively reduce the feedback overhead and improve the spectrum utilization.

\subsection{Numerical Results} \label{num results}
This subsection  shows the numerical results of all simulation experiments in \textbf{Sections \ref{Feasibility Analysis}, \ref{Superioity Analysis}} and \textbf{\ref{Robustness Analysis}}. All numerical results can be found in Tab. \ref{numerical results}. The $\mathbb{N}_{sp}^*$ and $\mathbb{N}_{csi}^*$ in the table correspond to the special EMEVNet and CsiNet trained by large datasets, respectively. And $\mathbb{N}_{mix}$ represents general EMEVNet obtained from mixed datasets. The horizontal comparison is to verify the performance of different channel environments, and the vertical comparison is to verify the impact of different length of codewords. Meanwhile, from the numerical analysis, the reconstruction performance of $\widehat{\mathbf{V}}$ is worse than that of $\widehat{\mathbf{S}}$, which is also consistent with the results in \textbf{Section \ref{Feasibility Analysis}}.  From the analysis of numerical results, we can conclude that $\widehat{\mathbf{S}}$ is almost perfectly reconstructed, while the  performance of $\widehat{\mathbf{V}}$ is worse with the decrease of $L_{\varepsilon}$.

\section{Conclusion}
In this paper, a novel channel feedback architecture for mmWave FDD systems was proposed. The key idea of our architecture was to feed back useful channel information to the BS, instead of the complete CSI matrix.  This paper discussed the beamforming technology based on SVD transformation, and the core design of the architecture was to feed back eigenmatrix and eigenvector. The major technical methods used in this paper include: using SVD transform to extract the effective information; utilizing attention mechanism to design a  dual channel auto-encoder; deploying a channel identification NN at the UE to switch the appropriate specific EMEVNet.
We considered and verified five common CDL channel environments and seven incremental system compression ratios. And all simulations were carried out in mmWave system.  First, we demonstrated the feasibility of our proposed architecture by visualizing some experimental results. Then, we designed two comparison experiments to prove the superiority and robustness of proposed architecture, respectively.  Finally, we showed the numerical results of all simulation experiments to further prove our analysis. This paper provided a new solution to solve the problem that the BS hardly acquires downlink CSI  in FDD wireless communication system. Through extracting and feeding back useful information for BS, the intelligent communication system is able to further improve the performance and reduce overhead.

\bibliographystyle{IEEEtran}
\bibliography{0EMEVfeedback_reference}

\end{document}